\documentclass[a4paper,11pt]{article}

\usepackage{geometry}									
\usepackage[english]{babel}		
\usepackage[utf8]{inputenc}		
\usepackage[T1]{fontenc}		

\usepackage{lmodern}			

\usepackage{amssymb,amsfonts,amsmath,amsthm}
\usepackage{booktabs}
\usepackage{hyphenat}
\usepackage{booktabs}
\usepackage{bbm}
\usepackage{multirow}
\usepackage{graphicx}
\usepackage{setspace}
\usepackage{algpseudocode}
\usepackage{algorithm}
\usepackage{xspace}
\usepackage[dvipsnames]{xcolor}
\usepackage{tikz}
\usetikzlibrary{calc}
\usetikzlibrary{shapes.geometric, arrows}
\usetikzlibrary{positioning}

\makeatletter
\let\oldabs\abs
\def\abs{\@ifstar{\oldabs}{\oldabs*}}
\let\oldnorm\norm
\def\norm{\@ifstar{\oldnorm}{\oldnorm*}}
\makeatother

\definecolor{googleblue}{rgb}{0.2588, 0.5216, 0.9569}
\definecolor{googlered}{rgb}{0.8588, 0.1961, 0.2118}
\definecolor{googlegreen}{rgb}{0.0588, 0.6157, 0.3451}
\definecolor{googleyellow}{rgb}{0.9569, 0.7059, 0}

\tikzstyle{startstop} = [
    rectangle, 
    rounded corners, 
    minimum width=3cm, 
    minimum height=1cm,
    text centered, 
    fill=CarnationPink!80
]
\tikzstyle{recursion} = [
    rectangle, 
    rounded corners, 
    minimum width=3cm, 
    minimum height=1cm,
    text centered, 
    fill=googleyellow!50
]
\tikzstyle{process} = [
    rectangle, 
    rounded corners, 
    minimum width=3cm, 
    minimum height=1cm,
    text centered, 
    fill=googleblue!50
]
\tikzstyle{decision} = [
    diamond, 
    rounded corners, 
    minimum height=1cm, 
    text centered, 
    fill=googlegreen!50,
]
\tikzstyle{arrow} = [thick,->,>=stealth]

\renewcommand{\vec}[1]{\boldsymbol{#1}}
\newcommand{\eg}{\textit{e.g.}\xspace}
\newcommand{\ie}{\textit{i.e.}\xspace}
\newcommand{\wrt}{\textit{w.r.t.}\xspace}

\newcounter{node}
\makeatletter
\newcommand{\customlabel}[2]{%
   \protected@write \@auxout {}{\string \newlabel {#1}{{#2}{\thepage}{#2}{#1}{}} }%
   \hypertarget{#1}{}
}
\tikzset{
  autonumbered node/.style={
    /utils/exec={\stepcounter{node}},
    label={[anchor=north east]above right:{\arabic{node}}}
  }
}
\newcommand{\nodelabel}[1]{%
  \protect\customlabel{#1}{\arabic{node}}
}
\makeatother

\tikzset{
  autonumbered node/.style={
    /utils/exec={\stepcounter{node}},
    label={[anchor=north, xshift=0mm, yshift=0mm]{\fbox{\arabic{node}}}}
  },
  autonumbered decision node/.style={
    /utils/exec={\stepcounter{node}},
    label={[anchor=north, xshift=0mm, yshift=-2mm]{\fbox{\arabic{node}}}}
  }
}

\usepackage{authblk}						

\title{mNARX\textsuperscript{+}: A surrogate model for complex dynamical systems using manifold-NARX and automatic feature selection}
\author[1]{Styfen Schär\thanks{styfen.schaer@ibk.baug.ethz.ch}}
\author[1]{Stefano Marelli\thanks{marelli@ibk.baug.ethz.ch}}
\author[1]{Bruno Sudret\thanks{sudret@ethz.ch}}
\affil[1]{Chair of Risk, Safety and Uncertainty Quantification, ETH Z\"{u}rich, Switzerland}

\date{\today}

\usepackage{microtype}			

\usepackage{indentfirst}		

\usepackage{caption}
\usepackage{subcaption}

\usepackage{natbib}
\usepackage{hyperref}
\hypersetup{
pdftitle = {mNARX\textsuperscript{+}: A surrogate model for complex dynamical systems using manifold-NARX and automatic feature selection},
pdfauthor = {Styfen Sch\"ar, Stefano Marelli, Bruno Sudret},
pdfsubject = {},
pdfkeywords = {NARX, Dynamical systems, Surrogate modeling, Autoregressive modeling, Principal component analysis, Least angle regression},
colorlinks = false,			
}

\usepackage[capitalize]{cleveref}							
\creflabelformat{equation}{#2(#1)#3}						
\crefrangelabelformat{equation}{#3(#1)#4 to #5(#2)#6}		
\labelcrefformat{subequation}{#2(#1)#3}						
\labelcrefrangeformat{subequation}{#3(#1)#4 to #5(#2)#6}	

\graphicspath{{figures}}

\newcommand{\addnew}[1]{#1}
\newcommand{\delete}[1]{}
\newcommand{\replace}[2]{\delete{#1}\addnew{#2}}

\begin{document}

\maketitle

\begin{abstract}
We propose an automatic approach for manifold nonlinear autoregressive with exogenous inputs (mNARX) modeling that leverages the feature-based structure of functional-NARX ($\mathcal{F}$-NARX) modeling.
This novel approach, termed mNARX\textsuperscript{+}, preserves the key strength of the mNARX framework, which is its expressivity allowing it to model complex dynamical systems, while simultaneously addressing a key limitation: the heavy reliance on domain expertise to identify relevant auxiliary quantities and their causal ordering.

Our method employs a data-driven, recursive algorithm that automates the construction of the mNARX model sequence. 
It operates by sequentially selecting temporal features based on their correlation with the model prediction residuals, thereby automatically identifying the most critical auxiliary quantities and the order in which they should be modeled. 
This procedure significantly reduces the need for prior system knowledge.

We demonstrate the effectiveness of the mNARX\textsuperscript{+} algorithm on two case studies: a Bouc-Wen oscillator with strong hysteresis and a complex aero-servo-elastic wind turbine simulator. 
The results show that the algorithm provides a systematic, data-driven method for creating accurate and stable surrogate models for complex dynamical systems.
\end{abstract}

\section{Introduction}\label{sec:introduction}

Modeling the dynamic response of time-dependent systems has received increasing attention across disciplines in recent years. 
This growing interest is driven by the fact that many real-world systems exhibit inherently time-dependent behavior. 
For instance, the response of structures such as buildings and bridges evolves over time under external loads, including wind, wave, and seismic excitations. 
Similarly, the dynamic behavior of wind turbines is influenced not only by environmental loads but also by external stimuli such as control system actions.
Accurate modeling of such systems has therefore become essential for a range of applications, including predictive maintenance \citep{langeron_2021, Samsuri_2023}, fault and defect diagnosis \citep{mattson_2006, Gao_2016}, or optimization of control inputs \citep{levin_1996, Hu_2024}. 
Other applications that require many evaluations of the system for various input scenarios and thus demand particularly efficient modeling include design and optimization \citep{Yu_2023, deshmukh_2017}, uncertainty quantification \citep{mai_2016, bhattacharyya_2020, chandra_2025, cheng_2025}, and reliability analysis \citep{garg_2022, Zhou_2023, Zhang2_2024}. 
In such contexts, surrogate models are often employed as fast-to-evaluate approximations of the underlying dynamical system. 
The development of such surrogate models is at the center of this work.

Despite the already widespread adoption of surrogate models, their construction can be a challenging task, particularly for dynamical systems. 
As an example, systems with coupled subsystems, control units, or nonlinear elements such as springs and dampers can result in highly complex relations between the external input excitations, usually referred to as \emph{exogenous inputs}, and the system responses. 
Additionally, high-dimensional exogenous inputs can exacerbate the curse of dimensionality and thus further complicate the construction of an accurate surrogate.

Difficulties may also arise from the choice of surrogate modeling framework.  
\replace{A widely used approach for modeling dynamical systems with exogenous inputs is the \emph{nonlinear autoregressive with exogenous inputs} (NARX) model \citep{billings_2013, azarhoosh_2025}.}{
Widely used approaches for modeling dynamical systems with exogenous inputs are \emph{recurrent neural networks} such as long short-term memory (LSTM) neural networks \citep{hochreiter1997, zhao2023} and \emph{nonlinear autoregressive with exogenous inputs} (NARX) models \citep{billings_2013, azarhoosh_2025}.
}
These models estimate the system output at a given time based on \replace{its past outputs}{their own past predictions} and the history of the exogenous inputs.  
NARX models\addnew{, which are the central framework used in this work due to their simple structure,} have demonstrated effectiveness across a range of applications, including mooring systems \citep{Yetkin_2017, Zhang_2024} and marine structures \citep{Kim_2015, Poursorkhabi_2023}, which response depends on external wave loading; gas turbines \citep{chiras_2001}, whose behavior is governed by fuel flow; multi-story buildings \citep{spiridonakos_2015, Li_2021b}, which deform in response to ground motions; and geotechnical systems \citep{Wunsch_2018, Dassanayake_2023}, where moisture dynamics are influenced by \eg precipitation and temperature.  
However, achieving stable long-term predictions remains a challenge due to difficulties in selecting appropriate time lags, particularly when using fine time discretizations and when signal downsampling can lead to significant information loss.

To address the problem of lag selection within NARX models, the feature-centric NARX approach was recently introduced in \citet{schaer_fnarx_2025}. Their framework, called \emph{functional-NARX} ($\mathcal{F}$-NARX), calculates model responses based on both exogenous and autoregressive features extracted from a macroscopic time window (the system \textit{memory}), rather than from individual time lags. 
This approach leverages the temporal smoothness and compressibility of real-world processes, making it drastically less sensitive to time-step discretization, and significantly reducing model overparametrization.

While $\mathcal{F}$-NARX does solve the issue of lag selection, it does not directly address the challenges associated with highly nonlinear input-output mappings in complex systems.
To overcome this challenge, \citet{schaer_mnarx_2024} developed an independent approach that leverages prior knowledge about the system, which is available in many applications. 
Their sequential surrogate modeling approach called \emph{manifold-NARX} (mNARX) capitalizes on auxiliary quantities that are related to the underlying physics of the system but are simpler to model.
These auxiliary quantities can either be recorded or calculated from the training data, and serve as additional exogenous inputs to a final NARX model that predicts the quantity of interest (QoI).
Although mNARX can be highly effective in predicting the behavior of complex dynamical systems, its performance relies on the appropriate selection of the auxiliary quantities, a central yet challenging task for complex engineering scenarios. 

Building on the strengths of $\mathcal{F}$-NARX, this work addresses the open challenges of mNARX modeling, namely the selection of the most relevant auxiliary quantities and their prediction using autoregressive models.
In more detail, we propose an automated approach to both select the auxiliary quantities, and to construct a corresponding sequence of $\mathcal{F}$-NARX models. 
The construction of each $\mathcal{F}$-NARX is automated by iteratively including temporal features of both exogenous and autoregressive inputs, based on their correlation with the residual prediction of a constantly updated $\mathcal{F}$-NARX model.
By iteratively refining the model, the proposed method enhances forecast performance and simplifies the construction of each $\mathcal{F}$-NARX model in the modeling chain.
Ultimately, this makes the mNARX approach more widely applicable, even for systems lacking sufficient prior information to identify a suitable time-dependent manifold.
In turn, this increase in time-efficiency and user-friendliness can promote the acceptance and adoption of mNARX in a wider range of practical applications.

The structure of this paper is as follows.  
Section~\ref{sec:state_of_the_art} introduces the $\mathcal{F}$-NARX and mNARX modeling frameworks and defines the relevant notation and terminology.  
Section~\ref{sec:automatic_mnarx_modeling} presents the novel contribution of this work: the automatic construction of mNARX surrogate models using $\mathcal{F}$-NARX modeling.  
Section~\ref{sec:applications} then demonstrates the proposed approach through two case studies: a Bouc-Wen oscillator under random excitation and an aero-servo-elastic onshore wind turbine simulator.  
Finally, Section~\ref{sec:discussion_and_conclusion} summarizes the findings, discusses their implications, and presents concluding remarks.

\section{Functional and manifold-NARX modeling}\label{sec:state_of_the_art}

\subsection{Functional nonlinear autoregressive with exogenous inputs modeling}\label{sec:fnarx_modeling}
Let us consider a dynamical system $\mathcal{M}$ with a deterministic response $y(t) \in \mathbb{R}$ evolving along a time axis $\mathcal{T}$. 
We assume the system starts at an initial state $\vec{\beta}$ and its response is driven by an exogenous input vector $\vec{x}(t) \in \mathbb{R}^M$. This relationship is expressed as:
\begin{equation}\label{eq:dyn_surrogate_modeling_problem}
    y(t) = \mathcal{M}(\vec{x}(\mathcal{T} \leq t), \vec{\beta}),
\end{equation}
where the notation $\bullet(\mathcal{T} \leq t)$ emphasizes that the system response at time $t$ depends on the input history up to and including time $t$.  

In this work, we look at dynamical systems from the perspective of surrogate modeling. Hence, the objective is to construct a computationally efficient mathematical model $\widehat{\mathcal{M}}$ that approximates the expensive-to-evaluate model $\mathcal{M}$. The surrogate model therefore aims to satisfy:
\begin{equation}\label{eq:surr_problem_statement}
    \widehat{y}(t) = 
    \widehat{\mathcal{M}}(\vec{x}(\mathcal{T} \leq t), \vec{\beta}) 
    \approx
    \mathcal{M}(\vec{x}(\mathcal{T} \leq t), \vec{\beta}),
\end{equation}
and it is constructed from a set of full-model evaluations, known as the \emph{experimental design} (ED).
An experimental design typically comprises a relatively small number $N_\text{ED}$ ($\mathcal{O}(10^{1-2})$) of input/output pairs $\left( \vec{x}(t)^{(i)}, \vec{y}(t)^{(i)} \right)$, referred to as \emph{model realizations}.
Note that these realizations are time-series that are available at discrete time steps in practice, as explained below.

A popular surrogate modeling class for dynamical systems is that of \emph{nonlinear autoregressive with exogenous inputs} (NARX) models \citep{billings_2013}. 
NARX models require that the experimental design data is defined on a set of discrete time steps $\{0, \delta t, \dots, (N - 1)\delta t \}$, where $N$ denotes the number of time steps. Note that we assume, without loss of generality, that all the realizations in the experimental design have the same number of time steps.  
NARX models then approximate the system response at the current time step as a parametric function with parameters $\vec{c}$, based on the current and past values of the exogenous input, as well as past values of the output (the autoregressive inputs):
\begin{equation}\label{eq:discrete_prediction}
    \widehat{y}(t) = \widehat{\mathcal{M}}(\vec{\varphi}(t); \vec{c}),
\end{equation}
where the lagged vector $\vec{\varphi}(t) \in \mathbb{R}^n$ reads:
\begin{equation}\label{eq:varphi}
    \begin{split}
        \vec{\varphi}(t) = \{
        &y(t - \delta t), y(t - 2\delta t), \dots, y(t - n_y\delta t), \\
        &x_1(t), x_1(t - \delta t), \dots, x_1(t - n_{x_1}\delta t), \\ 
        &\dots, \\
        &x_M(t), x_M(t - \delta t), \dots, x_M(t - n_{x_M}\delta t)\},
    \end{split}
\end{equation}
and gathers the exogenous input lags $x_j(t-k\delta t)$ and autoregressive lags $y(t-(k+1)\delta t)$ with $k \in \mathbb{N}$.

Despite its popularity, the classical NARX approach faces several notable challenges in the context of surrogate modeling.  
For instance, systems with long memory or complex temporal dependencies can be difficult to model, as they require either careful selection of time lags, or the inclusion of many lags, which significantly increases the problem dimensionality.
Selecting individual lags is in itself both time-consuming and somewhat arbitrary, while high dimensionality typically results in a large number of parameters $\vec{c}$ that must be constrained from the available training data.
To address these issues, \citet{schaer_fnarx_2025} proposed an alternative to classical NARX models, termed \emph{functional-NARX} ($\mathcal{F}$-NARX).  
The key idea behind $\mathcal{F}$-NARX is that the underlying dynamical system is typically continuous in time, and exhibits temporal regularity and smoothness, with time discretization being merely an artifact of digital data assimilation and computer-based simulation.  
$\mathcal{F}$-NARX therefore exploits the temporal compressibility of the process by not using the lagged input vector $\vec{\varphi}(t)$ directly, but rather a set of temporal features extracted from it.  
While $\mathcal{F}$-NARX is conceptually based on a continuous-time assumption, for the sake of conciseness we introduce here only its discrete-time formulation relevant for this work (see \citet{schaer_fnarx_2025} for a general presentation).

Let $\vec{\varphi}_j(t) \in \mathbb{R}^{n_j}$ denote the portion of the lagged vector $\vec{\varphi}(t)$ in Eq.~\eqref{eq:varphi} that corresponds to the $j$-th variable (i\ie, one of $\vec{x}_1, \dots, \vec{x}_M, \vec{y}$).  
Consequently, it is assumed that the memory of the system \wrt the $j$-th variables is at least $n_j\delta t$.
By leveraging the assumed temporal compressibility of the sliding window $\vec{\varphi}_j(t)$, a set of temporal features $\vec{\xi}_j(t) \in \mathbb{R}^{\tilde{n}_j}$ is extracted as
\begin{equation}
    \vec{\xi}_j(t) = \mathcal{K}_j(\vec{\varphi}_j(t)),
\end{equation}
where $\mathcal{K}_j: \mathbb{R}^{n_j} \to \mathbb{R}^{\tilde{n}_j}$ denotes a feature extraction operator, with the goal of achieving $\tilde{n}_j \ll n_j$.  

When dealing with time-discrete signals, we can adopt principal component analysis (PCA) \citep{Jolliffe_2002} as the feature extraction method, which allows us to formulate the feature extraction process as a simple linear transformation:
\begin{equation}\label{eq:pca_mapping}
    \vec{\xi}_j(t) = \vec{\varphi}_j(t) \vec{V}_j,
\end{equation}
where $\vec{V}_j \in \mathbb{R}^{n_j \times \tilde{n}_j}$ denotes the PCA projection matrix, obtained via eigen-decomposition of the covariance matrix of the signal.

An important advantage of PCA is that the extracted features (namely the \emph{principal components} (PCs)) are mutually orthogonal, and thus uncorrelated.  
This is in strong contrasts with classical NARX models, where temporally adjacent time lags, are typically highly correlated. 
Such correlation can introduce numerical instability and bias toward recent autoregressive lags during the model fitting, both of which are strongly mitigated by the orthogonal feature basis of PCA.
The PCs can also be naturally ranked by their eigenvalues: the first component of $\vec{\xi}_j(t)$ captures the largest proportion of variance in $\vec{\varphi}_j(t)$, with subsequent components accounting for progressively smaller contributions.  
By retaining only the first $\tilde{n}_j \ll n_j$ components of $\vec{\xi}_j(t)$, a small number of features can be obtained with often minimal loss of information.
Furthermore, this PCA-based representation is parametrized only by a single $\tilde{n}_j$ for each variable instead of arbitrarily selecting multiple lags individually.

Applying this procedure to each variable yields the \emph{feature vector} $\vec{\xi}(t)$, which serves as a compressed representation of the full lagged vector $\vec{\varphi}(t)$:
\begin{equation}\label{eq:feature_vector}
    \vec{\xi}(t) = \{ \vec{\xi}_{y}(t), \vec{\xi}_{x_1}(t), \dots, \vec{\xi}_{x_M}(t) \}.
\end{equation}
Finally, the $\mathcal{F}$-NARX model approximates the response $y(t)$ based on the features $\vec{\xi}(t)$ as follows:
\begin{equation}\label{eq:fnarx_discrete_prediction}
    \widehat{y}(t) = \widehat{\mathcal{M}}(\vec{\xi}(t); \vec{c}).
\end{equation}
The fitting process of an $\mathcal{F}$-NARX model remains analogous to that of a classical NARX model, and reduces to a standard regression problem.  
For the $i$-th realization in the experimental design, we construct the \emph{feature matrix} $\vec{\Xi}^{(i)}$ by stacking the feature vectors at each time step.  
Similarly, stacking the corresponding response values yields the response vector $\vec{y}^{(i)}$:
\begin{equation}
\label{eq:feature_matrix}
    \vec{\Xi}^{(i)} = \begin{pmatrix}
       \vec{\xi}^{(i)}(t_0) \\
       \vec{\xi}^{(i)}(t_0+\delta t) \\
       \vdots \\
       \vec{\xi}^{(i)}((N-1)\delta t)
    \end{pmatrix},
    \quad
    \vec{y}^{(i)} = \begin{pmatrix}
       y^{(i)}(t_0) \\
       y^{(i)}(t_0+\delta t) \\
       \vdots \\
       y^{(i)}((N-1)\delta t)
    \end{pmatrix}.
\end{equation}
Note that $t_0 = \max(n_y, n_{x_1}, \dots, n_{x_M}) \delta t$ accounts for the use of lagged values.
The final regression matrix and output vector are obtained by stacking the feature matrices and response vectors across all realizations:
\begin{equation}\label{eq:feature_matrix_exp_design}
    \vec{\Xi} = \begin{pmatrix}
        \vec{\Xi}^{(1)} \\
        \vdots \\
        \vec{\Xi}^{(N_\text{ED})}
    \end{pmatrix}, 
    \quad
    \vec{y} = \begin{pmatrix}
        \vec{y}^{(1)} \\ 
        \vdots \\
        \vec{y}^{(N_\text{ED})} 
    \end{pmatrix},
\end{equation}
and the optimal set of model parameters $\hat{\vec{c}}$ is then obtained by minimizing a specified loss function $\mathcal{L}$:
\begin{equation}\label{eq:loss_minimization}
    \hat{\vec{c}} = \mathop{\arg\min}_{\vec{c}} \mathcal{L}\left(\vec{y}, \widehat{\mathcal{M}}(\vec{\Xi}; \vec{c})\right).
\end{equation}
  
This work employs linear-in-the-parameters NARX models, specifically the polynomial NARX formulation from \citet{schaer_fnarx_2025}. 
Accordingly, at each time step $t$, given the full feature vector $\vec{\xi}(t)$, we construct a monomial $\mathcal{P}_{\vec{\alpha}}(\vec{\xi}(t))$ as follows:
\begin{equation}\label{eq:monomial}
    \mathcal{P}_{\vec{\alpha}}(\vec{\xi}(t)) = \prod_{j=1}^{\tilde{n}} \xi_j(t)^{\alpha_j},
\end{equation}
where $\xi_j(t)$ denotes the $j$-th element of $\vec{\xi}(t)$, and $\vec{\alpha} \in \mathbb{N}^{\tilde{n}}$ is a multi-index of nonnegative integers.  
The response is then approximated as a weighted sum of monomials:
\begin{equation}
    \widehat{y}(t) = \sum_{\vec{\alpha} \in \mathcal{A}} c_{\vec{\alpha}} \mathcal{P}_{\vec{\alpha}}(\vec{\xi}(t)),
\end{equation}
where $c_{\vec{\alpha}}$ are real-valued coefficients to be learned from the experimental design data.  
The multi-index set $\mathcal{A}$ is truncated to control the complexity of the model.  
In this work, we adopt a hyperbolic truncation strategy, originally introduced for polynomial chaos expansion in \citet{Blatman_2010}, characterized by a truncation parameter $q$ and a maximum polynomial degree $d$.  
The resulting multi-index set is therefore defined as:
\begin{equation}
    \mathcal{A}^{\tilde{n}, d, q} = \left\{ \vec{\alpha} \in \mathbb{N}^{\tilde{n}} \;:\; ||\vec{\alpha}||_q \le d \right\},
\end{equation}
where $||\vec{\alpha}||_q = \left(\sum_{j=1}^{\tilde{n}} \alpha_j^q \right)^{1/q}$ for $0 < q \le 1$.

The use of the linear-in-the-parameters polynomial NARX model allows use to use the modified least-angle regression (LARS) method proposed in \citet{schaer_fnarx_2025}.
In the traditional LARS algorithm \citep{Efron_2004}, the loss function $\mathcal{L}$ is given by:
\begin{equation}\label{eq:lars_loss}
    \mathcal{L}(\vec{c}) = \frac{1}{2}\left\lVert\vec{y} - {\mathcal{P}}(\vec{\Xi})\vec{c}\right\rVert_2^2 + \gamma\left\lVert\vec{c}\right\rVert_1,
\end{equation}
where ${\mathcal{P}}(\vec{\Xi})$ denotes the regression matrix formed by applying the chosen polynomial basis functions to $\vec{\Xi}$, and $\gamma$ is a penalty term that enforces sparsity in $\vec{c}$. 
In the modified version, sparsity is incorporated under the consideration that the model forecast, rather than its one-step-ahead prediction error, is being optimized.
This is accomplished by evaluating the forecast error at various steps along the LARS path.

\subsection{Manifold nonlinear autoregressive with exogenous inputs modeling}\label{sec:mnarx_modeling}
Classical NARX models are recognized as powerful tools for modeling dynamical systems and $\mathcal{F}$-NARX has shown promise to extend them in surrogate modeling settings due to its accuracy and long-term stability (see Section~\ref{sec:fnarx_modeling} and \citet{schaer_fnarx_2025}).
However, they both encounter notable limitations when dealing with highly complex systems.
One of the main limitations lies in their expressive power, which is often directly linked to the size of their parameter vector $\vec{c}$ (see Eq.~\eqref{eq:loss_minimization}).
Consequently, approximating highly nonlinear input/output mappings can quickly become problematic, particularly when working with small experimental designs, since a large parameter vector can become difficult to fit.
To this end, \citet{schaer_mnarx_2024} introduced manifold-NARX (mNARX) as a way to model complex dynamical systems when data is scarce, but prior knowledge of the system dynamics is available.
mNARX leverages this knowledge by sequentially constructing auxiliary quantities that lie along a known causal or physical pathway to the quantity of interest.  
The construction of each auxiliary quantity typically involves a simpler mapping than the original problem of predicting the final quantity of interest.  
Therefore, mNARX essentially trades a single highly nonlinear mapping for several simpler ones.

This idea is formalized by replacing the original mapping:
\begin{equation}
    \widehat{\mathcal{M}}: \vec{x}(\mathcal{T} \le t) \to y(t)
\end{equation}
with a simplified mapping:
\begin{equation}
    \widehat{\mathcal{M}}: \vec{\zeta}(\mathcal{T} \le t) \to y(t)
\end{equation}
where $\vec{\zeta}(t) \in \mathbb{R}^{M_\zeta}$ is a time-dependent input manifold derived from the original exogenous input, and additional intermediate response quantities related to the quantity of interest $y(t)$. 
The manifold has two functions: to guarantee that the input maintains manageable dimensionality, and to incorporate prior knowledge about the system dynamics, thereby reducing the complexity of the mapping $\widehat{\mathcal{M}}$.
The manifold is constructed iteratively by introducing so-called \emph{auxiliary quantities} denoted as $z_i(t)$. They are either obtained through processing of the exogenous inputs $\vec{x}(t)$, or by introducing physically meaningful \emph{intermediate response quantities}, related \eg to control systems or single components, which are key to incorporating prior information about the system. 
This iterative construction can be written as:
\begin{equation}\label{eq:aux_quantity}
    \begin{split}
        {\vec{z}}_{1}(t) &= \mathcal{F}_1(\vec{x}(\mathcal{T} \le t), \vec{z}_{1}(\mathcal{T} <t)) \\
        {\vec{z}}_{2}(t) &= \mathcal{F}_2(\vec{x}(\mathcal{T} \le t), {\vec{z}}_{1}(\mathcal{T} \le t), \vec{z}_{2}(\mathcal{T} <t)) \\
        \vdots \\
        {\vec{z}}_{M_{\vec{z}}}(t) &= \mathcal{F}_{M_{\vec{z}}}(\vec{x}(\mathcal{T} \le t), {\vec{z}}_{1}(\mathcal{T} \le t), \dots, {\vec{z}}_{M_{\vec{z}}-1}(\mathcal{T} \le t), \vec{z}_{M_{\vec{z}}}(\mathcal{T} <t)),
    \end{split}
\end{equation}
where the $\mathcal{F}_i$ can represent arbitrarily complex functions, \eg cosine transforms, NARX models (as used in \citet{schaer_mnarx_2024}) or even $\mathcal{F}$-NARX model as we will adopt in this work. 
The final input manifold is then defined as $\vec{\zeta}(t)= \{ z_1(t), \dots, z_{M_{\vec{z}}}(t), \vec{x}(t) \}$, and it replaces the sole raw exogenous input $\vec{x}(t)$ in the classical NARX formulation from Eq.~\eqref{eq:surr_problem_statement}. 
In contrast to the original surrogate modeling problem from Eq.~\eqref{eq:dyn_surrogate_modeling_problem}, the manifold-enriched problem therefore reads:
\begin{equation}
    \widehat{y}(t) = 
    \widehat{\mathcal{M}}(\vec{\zeta}(\mathcal{T}\le t), \vec{\beta}).
\end{equation}

As an example, in the wind turbine case study in \citet{schaer_mnarx_2024}, the authors demonstrate the use of mNARX by first applying a discrete cosine transform (DCT) to reduce the dimensionality of the original exogenous input (a three-dimensional transient wind field) and retain only the low-frequency components. These DCT modes are considered auxiliary quantities as they are obtained by preprocessing the raw exogenous input.
Subsequently, a series of NARX models is built to approximate the turbine control system responses (\eg, blade pitch) and the higher harmonics of the rotor speed as intermediate response quantities, based on the prior knowledge on the dynamics of a wind turbine. 
All the auxiliary quantities are then gathered and used to surrogate the flapwise blade root bending moment as the primary quantity of interest.

Note that the mNARX approach assumes that the auxiliary quantities are available in the experimental design data, or can be extracted/postprocessed from either the simulation inputs or their outputs.
Therefore, the training process of the mNARX surrogate follows a similar methodology as for $\mathcal{F}$-NARX models, with the main difference lying in that mNARX may need to also train surrogates for some of the auxiliary quantities $\mathcal{F}_1, \dots, \mathcal{F}_{M_{\vec{z}}}$, instead of a single $\mathcal{F}$-NARX model.
During prediction on unseen data, only the exogenous inputs $\vec{x}(t)$ are available, and the auxiliary quantities $\vec{z}_1(t), \dots, \vec{z}_{M_{\vec{z}}}(t)$ must be computed sequentially in the same order as in Eq.~\eqref{eq:aux_quantity}:
\begin{equation}\label{eq:aux_quantity_prediction}
    \begin{split}
        \widehat{\vec{z}}_{1}(t) &= \mathcal{F}_1(\vec{x}(\mathcal{T} \le t), \widehat{\vec{z}}_{1}(\mathcal{T} < t)) \\
        \widehat{\vec{z}}_{2}(t) &= \mathcal{F}_2(\vec{x}(\mathcal{T} \le t), \widehat{\vec{z}}_{1}(\mathcal{T} \le t), \widehat{\vec{z}}_{2}(\mathcal{T} < t)) \\
        \vdots \\
        \widehat{\vec{z}}_{{M_{\vec{z}}}}(t) &= \mathcal{F}_{M_{\vec{z}}}(\vec{x}(\mathcal{T} \le t), \widehat{\vec{z}}_{1}(\mathcal{T} \le t), \dots, \widehat{\vec{z}}_{{M_{\vec{z}}}-1}(\mathcal{T} \le t), \widehat{\vec{z}}_{{M_{\vec{z}}}}(\mathcal{T} < t)).
    \end{split}
\end{equation}
After the auxiliary quantities are constructed, the final prediction of the quantity of interest is calculated as:
\begin{equation}
    \widehat{y}(t) = \widehat{\mathcal{M}}(\widehat{\vec{\zeta}}(\mathcal{T} \le t), \widehat{\vec{y}}(\mathcal{T} < t), \vec{\beta}),
\end{equation}
where $\widehat{\vec{\zeta}}(\mathcal{T} \le t) = \{ \widehat{\vec{z}}_{1}(\mathcal{T} \le t), \dots, \widehat{\vec{z}}_{n}(\mathcal{T} \le t), \vec{x}(\mathcal{T} \le t) \}$.

The authors of \citet{schaer_mnarx_2024} demonstrated that this incremental approach can dramatically improve NARX performance and long-term stability on rather complex problems, even with relatively scarce training data, as each individual modeling step is much simpler than the original full problem.

\section{Automatic mNARX modeling}\label{sec:automatic_mnarx_modeling}
A key weakness of the mNARX approach presented in Section~\ref{sec:mnarx_modeling}, is that its construction often relies on specific domain expertise. 
Specifically, building the input manifold requires prior knowledge to identify relevant auxiliary quantities, which may not always be available or complete. 
Even when sufficient insight exists to identify potential manifold components, determining the optimal construction order (\ie, the sequence of predictive models for the intermediate response quantities) can still be challenging.

Therefore, in this section we develop a data-driven automated approach for constructing the input manifold with limited prior information.
This automated approach leverages the temporal feature-based structure of $\mathcal{F}$-NARX modeling, detailed in Section~\ref{sec:fnarx_modeling}, to identify both the manifold components, and the order they need to be fitted.
We refer to this method as mNARX\textsuperscript{+} to reflect its automated nature and its role as an extension of the original mNARX framework.

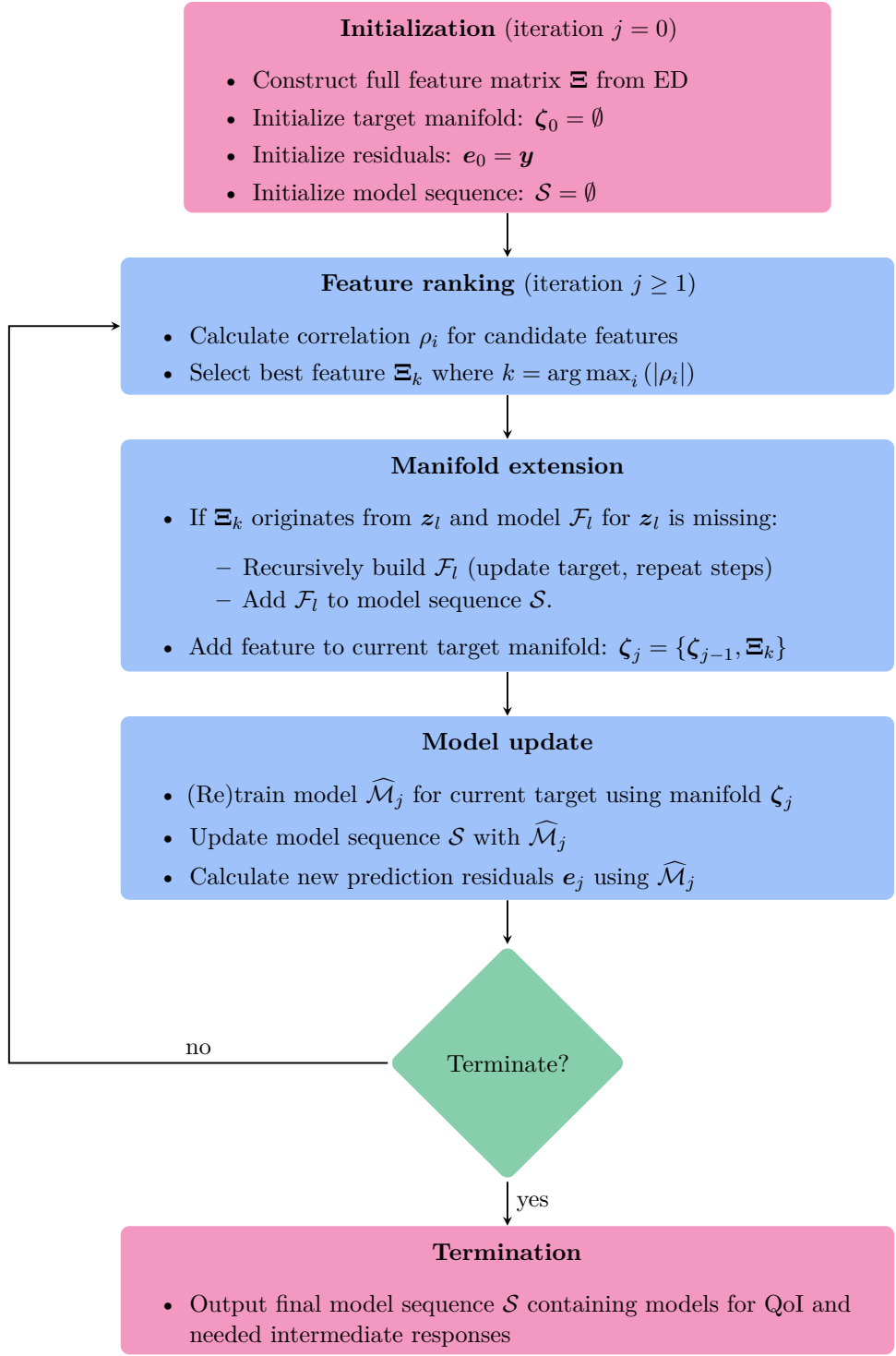
\begin{figure}[htb!]
    \centering
    \def\nodedistance{0.7cm}
    \def\boxwidth{12cm}
    \def\itemizesep{-0.1em}
    \begin{tikzpicture}
        \node (start) [startstop, text width=10cm, align=flush left] {
            \vspace{-1em}
            \begin{center}
                \textbf{Initialization} (iteration $j=0$)
            \end{center}
            \vspace{-1em}
            \begin{itemize}
                \setlength\itemsep{\itemizesep}
                \item Construct full feature matrix $\vec{\Xi}$ from ED
                \item Initialize target manifold: $\vec{\zeta}_{0} = \emptyset$
                \item Initialize residuals: $\vec{e}_{0} = \vec{y}$
                \item Initialize model sequence: $\mathcal{S} = \emptyset$
            \end{itemize}
        };
        
        \node (feature_ranking) [process, below=\nodedistance of start.south, anchor=north, text width=\boxwidth, align=flush left] {
            \vspace{-1em}
            \begin{center}
                \textbf{Feature ranking} (iteration $j \ge 1$)
            \end{center}
            \vspace{-1em}
            \begin{itemize}
                \setlength\itemsep{\itemizesep}
                \item Calculate correlation $\rho_i$ for candidate features
                \item Select best feature $\vec{\Xi}_k$ where $k = \mathop{\arg\max}_{i} \left( |\rho_i| \right)$
            \end{itemize}
        };
        \draw [arrow] (start) -- (feature_ranking);
    
        \node (manifold_extension) [process, below=\nodedistance of feature_ranking.south, anchor=north, text width=\boxwidth, align=flush left] {
            \vspace{-1em}
            \begin{center}
                \textbf{Manifold extension}
            \end{center}
            \vspace{-1em}
            \begin{itemize}
                \setlength\itemsep{\itemizesep}
                \item If $\vec{\Xi}_k$ originates from $\vec{z}_l$ and model $\mathcal{F}_l$ for $\vec{z}_l$ is missing: 
                \begin{itemize}
                    \setlength\itemsep{\itemizesep}
                    \item Recursively build $\mathcal{F}_l$ (update target, repeat steps)
                    \item Add $\mathcal{F}_l$ to model sequence $\mathcal{S}$.
                \end{itemize}
                \item Add feature to current target manifold: $\vec{\zeta}_{j} = \{ \vec{\zeta}_{j-1}, \vec{\Xi}_k \}$
            \end{itemize}
        };
        \draw [arrow] (feature_ranking) -- (manifold_extension);
    
        \node (model_update) [process, below=\nodedistance of manifold_extension.south, anchor=north, text width=\boxwidth, align=flush left] {
            \vspace{-1em}
            \begin{center}
                \textbf{Model update}
            \end{center}
            \vspace{-1em}
            \begin{itemize}
                \setlength\itemsep{\itemizesep}
                \item (Re)train model $\widehat{\mathcal{M}}_{j}$ for current target using manifold $\vec{\zeta}_{j}$
                \item Update model sequence $\mathcal{S}$ with $\widehat{\mathcal{M}}_{j}$
                \item Calculate new prediction residuals $\vec{e}_{j}$ using $\widehat{\mathcal{M}}_{j}$
            \end{itemize}
        };
        \draw [arrow] (manifold_extension) -- (model_update);
    
        \node (decision_converged) [decision, below=\nodedistance of model_update.south, anchor=north, text width=3cm, align=center] {
            Terminate?
        };
        \draw [arrow] (model_update) -- (decision_converged);
        \draw [arrow] (decision_converged.west) -- ++ (-6.0,0) node[midway, above] {no} |- (feature_ranking.west);
    
        \node (termination) [startstop, below=\nodedistance of decision_converged.south, anchor=north, text width=\boxwidth, align=flush left] {
            \vspace{-1em}
            \begin{center}
                \textbf{Termination}
            \end{center}
            \vspace{-1em}
            \begin{itemize}
                \setlength\itemsep{\itemizesep}
                \item Output final model sequence $\mathcal{S}$ containing models for QoI and needed intermediate responses
            \end{itemize}
        };
        \draw [arrow] (decision_converged) -- node[midway, right] {yes} (termination); 
    \end{tikzpicture}
    \caption{Flowchart illustrating the proposed \addnew{mNARX\textsuperscript{+}} algorithm to automatically construct an mNARX surrogate model leveraging the temporal feature-based $\mathcal{F}$-NARX structure.}\label{fig:auto_mnarx}
\end{figure}

We propose a top-down sequential feature selection approach, that starts from an initial extensive list of possible features extracted from the available experimental design, and ends in a final composite input manifold with manageable dimension.
It can be summarized in the following steps, which are also graphically presented in Figure~\ref{fig:auto_mnarx}:
\begin{enumerate}
    \item \textit{Initialization and exhaustive feature extraction} -- an extensive list of temporal features is extracted from the experimental design. 
    This can be a large number of time-dependent features (often $\mathcal{O}(10^2)$), extracted from input characteristics (Fourier coefficients, principal components, etc.) as well as QoIs and intermediate responses (stresses, strains, displacements, etc.).
    The residual error is initialized to the final desired QoI, $y(t)$.
    \item \textit{Feature ranking and selection} -- all of the extracted features are ranked according to their correlation with the current residual error. 
    \item \textit{Manifold extension} -- if the selected feature is available at this stage of the algorithm execution, it is directly added to the current manifold, and the execution goes to the next step of the algorithm. 
    If the selected feature is not available (because it belongs to a not-yet-calculated intermediate response), an $\mathcal{F}$-NARX model for the intermediate response is calculated before adding the selected feature and progressing to the next step. 
    In practice, this calculation is achieved via a recursive call of the entire algorithm, using the not-yet-calculated intermediate response as the QoI.
    \item \textit{Model and residual update} -- Once the manifold has been expanded, a new $\mathcal{F}$-NARX model is fitted, and a new residual error is calculated and updated.
    If a convergence criterion is met, the algorithm is stopped, and the manifold construction sequence is returned. 
    Otherwise, the algorithm repeats from Step~2.
\end{enumerate}

Note that the feature selection at Step~2 selects only \textit{single features} from either input or response time series.
This allows the algorithm to parsimoniously select only a relatively small set of relevant features, while ignoring input and output signal characteristics that are not relevant for modeling the QoI. 
This has the effect of drastically reducing the problem dimension, and it is made possible by the feature-based representation of the system inputs and outputs provided by $\mathcal{F}$-NARX.

Due to the complexity of the algorithm, and the extensive bookkeeping it involves due to its recursive nature, each step in \replace{this automatic mNARX}{the mNARX\textsuperscript{+}} construction process is described in one of the following Sections~\ref{sec:initialization}-\ref{sec:convergence_check}. 
The flowchart provided in Figure~\ref{fig:auto_mnarx} is intended as a map to help the reader visualize the interconnections between the various steps, as well as the main symbols and terminology used throughout. 
A detailed description of the implementation of the algorithm is available in \ref{app:implementation_details} for reproducibility purposes.

\subsection{Initialization}\label{sec:initialization}
Let $\vec{x} = \{ \vec{x}^{(i)} \}_{i=1}^{N_\text{ED}}$ denote the set of exogenous input time series across the $N_\text{ED}$ realizations in the experimental design (ED). 
Similarly, let $\vec{z} = \{ \vec{z}^{(i)} \}_{i=1}^{N_\text{ED}}$ be the set of \emph{candidate} auxiliary quantity time series available in the training data, and $\vec{y} = \{ \vec{y}^{(i)} \}_{i=1}^{N_\text{ED}}$ be the corresponding time series for the quantity of interest (QoI). 
By \emph{candidate} auxiliary quantities we mean both arbitrary transforms of the exogenous inputs (\eg, moving averages, dimensionality reduction coefficients, etc.) and additional available computational model responses directly related to the quantity of interest $y(t)$ (\eg, stresses or strains at different nodes, gradients, control system states, etc.). 
We subsequently use the symbol $\vec{s}(t)$ to denote intermediate responses which require a predictive model (\ie, an $\mathcal{F}$-NARX model in our case) for the function $\mathcal{F}_l$ and we use $\tilde{\vec{x}}(t)$ to denote the remaining auxiliary quantities with predefined transforms $\mathcal{F}_l$.
These quantities collectively form the vector $\vec{z}^{(i)} = \{ \tilde{\vec{x}}^{(i)}, \vec{s}^{(i)} \}$.
The full experimental design thus consists of the data tuples $\{ \vec{x}^{(i)}, \vec{z}^{(i)}, \vec{y}^{(i)} \}_{i=1}^{N_\text{ED}}$. 
This setup aligns with classical mNARX modeling, where auxiliary quantities $\vec{z}^{(i)}$ are assumed to be available during training (either measured or computed) but are generally unavailable during prediction on new, unseen data (where only $\vec{x}$ is provided). 
Consequently, this automated approach still assumes a minimal level of prior knowledge---namely, the ability to identify or provide a set of \emph{candidate} auxiliary quantities. 
Neverthless, unlike the original mNARX \citep{schaer_mnarx_2024}, no prior knowledge regarding the importance, physical meaning, or causal ordering of these candidates is required.

The automated construction process begins by assembling the full $\mathcal{F}$-NARX feature matrix $\vec{\Xi}$ in Eq.~\eqref{eq:feature_matrix_exp_design} (see Section~\ref{sec:fnarx_modeling}). 
This matrix incorporates temporal features extracted from all available time series data in the experimental design, including the exogenous inputs $\vec{x}$, the potential auxiliary quantities $\vec{z}$, and the target quantity $\vec{y}$:
\begin{equation}
   {\vec{\Xi}} = \begin{pmatrix}
      {\vec{\xi}}^{(1)}(t_0) \\
      \vdots \\
      {\vec{\xi}}^{(1)}((N-1)\delta t) \\
      \vdots \\
      {\vec{\xi}}^{(N_\text{ED})}(t_0) \\
      \vdots \\
      {\vec{\xi}}^{(N_\text{ED})}((N-1)\delta t)
   \end{pmatrix},
\end{equation}
where each row vector $\vec{\xi}^{(i)}(t_k)$ gathers the temporal features at time step $t_k$ for realization $i$:
\begin{equation}
   \vec{\xi}^{(i)}(t_k) = \{ 
      \vec{\xi}_{y}^{(i)}(t_k),
      \;
      \vec{\xi}_{x_1}^{(i)}(t_k),
      \dots, 
      \vec{\xi}_{x_M}^{(i)}(t_k), 
      \;
      \vec{\xi}_{z_1}^{(i)}(t_k),
      \dots, 
      \vec{\xi}_{z_{M_{z}}}^{(i)}(t_k)
   \}.
\end{equation}

This matrix $\vec{\Xi}$ represents the complete pool of candidate features derived from the $M$ exogenous inputs, $M_z$ potential auxiliary quantities, and the autoregressive component $y$.
The goal of the algorithm is to iteratively select features (columns) from $\vec{\Xi}$ that are most predictive of the target quantity (initially $y$, but potentially shifting to some auxiliary quantity $z_j$ during the recursive process). 
These selected features will form the input manifold $\vec{\zeta}$ for the final $\mathcal{F}$-NARX model. 

We initialize the manifold for the primary target $y$ as empty at iteration $j=0$: $\vec{\zeta}_{y,0} = \emptyset$. 
Note that we use subscript $y$ here to denote the manifold for the final QoI; additional manifolds for auxiliary quantities might be constructed during the process.

The feature selection process, detailed in Section~\ref{sec:feature_ranking}, relies on ranking features based on their relationship with the current prediction residuals $\vec{e}$. 
Initially, we assume a trivial model which predicts: $\widehat{y}(t) = 0 \; \forall \; t \in \mathcal{T}$.
Therefore, at iteration $j=0$, the residuals are initialized with the stacked values of the quantity of interest $\vec{y}$ from the ED:
\begin{equation}
   \vec{e}_{0} = 
   \begin{pmatrix} 
      \vec{y}^{(1)} \\ 
      \vdots \\
      \vec{y}^{(N_\text{ED})} 
   \end{pmatrix}.
\end{equation}

\subsection{Feature ranking}\label{sec:feature_ranking}
To select the most informative feature at each iteration $j \ge 1$, candidate features (columns $\vec{\Xi}_i$ from the full feature matrix $\vec{\Xi}$ that are not yet included in the current manifold) are ranked based on their expected contribution to improving the prediction of the current target quantity. 
We assume an \emph{assessment function}, subsequently denoted by $\mathcal{H}$, that returns a score $\rho_i \in \mathbb{R}$ quantifying the relevance of the $i$-th candidate feature $\vec{\Xi}_i$ with respect to the prediction residuals $\vec{e}_{j-1}$:
\begin{equation}
   \rho_i = \mathcal{H}(\vec{\Xi}_i, \vec{e}_{j-1}).
\end{equation}
In this work, we use a correlation-based measure, the Kendall rank correlation coefficient \citep{kendall_1938}, known for its robustness to nonlinear relationships. 
The feature $\vec{\Xi}_k$ selected at iteration $j$ is the one that maximizes the absolute value of this correlation, indicating the strongest monotonic relationship (positive or negative) with the residuals:
\begin{equation}
   k = \mathop{\arg\max}_{i} \left( |\rho_i| \right).
\end{equation}

Note that this iterative selection based on residuals is inspired by common compressive-sensing algorithms such as LASSO or LARS \citep{Tibshirani_1996_LASSO,Efron_2004}, with the main difference lying in the choice of a more stable similarity measure.

\subsection{Manifold extension}\label{sec:manifold_extension}
Suppose that feature $\vec{\Xi}_k$ (the $k$-th column of the full feature matrix $\vec{\Xi}$) has been identified at iteration $j$ as the most promising candidate according to the ranking in Section~\ref{sec:feature_ranking}. 
We now consider its origin to determine how it affects the manifold construction $\vec{\zeta}_{j}$ for the current target quantity $\vec{y}$:
\begin{itemize}
    \item Scenario 1: The feature originates from the autoregressive component ($\vec{\Xi}_k \in \vec{\Xi}_{y}$) or from an exogenous input ($\vec{\Xi}_k$ corresponds to some $\vec{x}_i$).
    \item Scenario 2: The feature originates from a potential auxiliary quantity ($\vec{\Xi}_k$ corresponds to some $\vec{z}_l$).
\end{itemize}

In Scenario 1, the selected feature $\vec{\Xi}_k$ can be directly added to the input manifold for the current target model. 
This is because features derived from $\vec{y}$ (autoregressive) or $\vec{x}$ (exogenous) can always be computed during prediction: autoregressive features are calculated from the surrogate's own past predictions, and exogenous features are calculated from the known exogenous inputs. 
The manifold is then updated as follows: $\vec{\zeta}_{y,j} = \{ \vec{\zeta}_{y,j-1}, \vec{\Xi}_k \}$.

In Scenario 2, the feature $\vec{\Xi}_k$ is derived from an auxiliary quantity $\vec{z}_l$. 
This feature is available during training, but is \emph{not} during prediction on unseen data, when only the exogenous input $\vec{x}$ is provided. 
Including $\vec{\Xi}_k$ in the manifold $\vec{\zeta}_{y,j}$ for predicting $\vec{y}$ implies that a function $\mathcal{F}_l$ for the auxiliary quantity $\vec{z}_l$ must either exist, or needs to be constructed as part of the overall mNARX sequence. 
This ensures that $\widehat{\vec{z}}_l$, and consequently its features $\vec{\Xi}_{\vec{z}_l}$ (including $\vec{\Xi}_k$), can be computed recursively during prediction using available inputs (from $\vec{x}$ and potentially other previously predicted auxiliary quantities $\widehat{\vec{z}}_{i}$). 
We must therefore manage the construction of these necessary auxiliary functions $\mathcal{F}_l$.

For the basic auxiliary quantities $\widetilde{\vec{x}}(t)$, these functions $\mathcal{F}_l$ are predefined and already available.
However, for intermediate responses $\vec{s}(t)$, this function $\mathcal{F}_l$ is an $\mathcal{F}$-NARX model that we have to construct first.
To achieve this, we adopt a recursive strategy, that involves temporarily switching the target from $\vec{y}$ to $\vec{z}_l$, and recursively applying the feature selection and model update steps until $\mathcal{F}_l$ is built. 
The main difficulty resides in identifying and respecting all the causal dependencies between auxiliary quantities. 
\addnew{
This is achieved by temporarily excluding from selection those features that depend on quantities still under construction, as detailed in \ref{app:implementation_details}.
}
Once the prerequisite model $\mathcal{F}_l$ exists, the selected feature $\vec{\Xi}_k$ is added to the manifold for the primary target $\vec{y}$: $\vec{\zeta}_{y,j} = \{ \vec{\zeta}_{y,j-1}, \vec{\Xi}_k \}$.
While this task is not particularly complex in itself, it involves significant bookkeeping to avoid circular dependencies.
This in turn results in rather cumbersome notation, and therefore 
the full recursion and dependency management is detailed in \ref{app:implementation_details}.

\addnew{
It is worth noting that when no auxiliary quantities or intermediate responses are available, the mNARX\textsuperscript{+} reduces to a standard $\mathcal{F}$-NARX model with automated feature selection.
This is still favorable, \eg in the case of a PCA-based $\mathcal{F}$-NARX model, which requires a manual choice of the number of retained components. 
In contrast, mNARX\textsuperscript{+} automatically selects only the relevant features from all available ones.
}

\subsection{Model update}\label{sec:model_update}
Once the input manifold $\vec{\zeta}_{y,j}$ for the primary target $\vec{y}$ (or temporarily for an auxiliary target $\vec{z}_l$) has been extended with the selected feature $\vec{\Xi}_k$, the corresponding $\mathcal{F}$-NARX surrogate model $\widehat{\mathcal{M}}_{j}$ is (re)trained. This training uses the extended feature set in $\vec{\zeta}_{y,j}$ as predictors, and the target quantity ($\vec{y}$ or $\vec{z}_l$) as the response. 
Using this updated surrogate $\widehat{\mathcal{M}}_{j}$, the forecast is recomputed on the experimental design data. 
This allows the prediction residuals $\vec{e}_{j}$ relative to the actual target values to be recalculated, preparing for the next feature ranking iteration ($j+1$).

\subsection{Algorithm termination}\label{sec:convergence_check}
Based on the updated model $\widehat{\mathcal{M}}_{j}$ and the corresponding residuals $\vec{e}_{j}$, a decision is made at each iteration whether to terminate the algorithm or continue adding features. 
Termination occurs under one or more of the following conditions:
\begin{itemize}
    \item There are no features left that show significant correlation with the residuals $\vec{e}_j$. That is, the highest observed correlation is below a given threshold value $\theta_\rho$, \ie: $\max(|\vec{\rho}|) < \theta_\rho$.
    \item The norm of the current residuals $\vec{e}_j$ reaches a predefined target threshold, \ie: $||\vec{e}_j|| < \theta_e$, where $||\cdot||$ denotes a vector norm (typically the Euclidean norm).
    \item A predefined maximum number of features or iterations is reached, or resource limits (such as maximum computation time), are exceeded.
\end{itemize}
Upon termination, the algorithm yields the final sequence of functions $\mathcal{F}_l$ for the auxiliary quantities $\vec{z}_l$ and the final $\mathcal{F}$-NARX model for the current target quantity.

\subsection{Algorithm parameters}
The proposed \replace{automated mNARX}{mNARX\textsuperscript{+}} modeling framework is designed to simplify the mNARX construction process by minimizing the number of user-defined parameters and user interaction. 
Therefore, once the set of candidate auxiliary quantities and the primary quantity of interest (QoI) are provided, the user is primarily concerned with the selection of a few key parameters at both the overarching algorithm level and the individual $\mathcal{F}$-NARX model level.

At the algorithm level, the main parameters are:
\begin{itemize}
    \item Assessment function ($\mathcal{H}$): This function is used to rank the candidate features based on their relationship with the current residuals (see Section~\ref{sec:feature_ranking}). While $\mathcal{H}$ is technically a tunable parameter, standard measures, such as Kendall's tau, Spearman's rank correlation, or Pearson's correlation coefficient, are well-suited for this purpose, thereby simplifying this selection. 
    \item The stopping criterion: A simple but effective criterion is the use of a threshold score $\theta_{\rho}$. This threshold determines when to stop the feature selection process (Section~\ref{sec:convergence_check}). The selection of $\theta_{\rho}$ involves a trade-off: an excessively high value may terminate the algorithm prematurely, potentially omitting important features and leading to a suboptimal surrogate model. Conversely, a value that is too low might lead to the inclusion of non-relevant features, resulting in overly complex models that could overfit the training data. 
    The optimal choice for $\theta_{\rho}$ can be influenced by the size of the experimental design, with larger datasets potentially allowing smaller threshold values. 
    In the subsequent applications in Section~\ref{sec:applications}, we use a representative value of $\theta_\rho=0.2$.
\end{itemize}

For the individual $\mathcal{F}$-NARX models, which are constructed for the intermediate responses and the final QoI (as per Section~\ref{sec:fnarx_modeling}), the primary considerations are:
\begin{itemize}
    \item Model memory: The memory length defines the extent of past information (number of time steps) used for temporal feature extraction in each $\mathcal{F}$-NARX model. As suggested in \citet{schaer_fnarx_2025}, a uniform memory length across all exogenous inputs and the autoregressive component can be adopted, reducing the selection to a single value per $\mathcal{F}$-NARX model. 
    It was also noted that $\mathcal{F}$-NARX model performance is relatively robust to moderately oversized memory lengths, offering some flexibility in this selection.
    \item $\mathcal{F}$-NARX implementation-specific parameters: Depending on the chosen $\mathcal{F}$-NARX implementation, additional parameters may need to be specified. 
    For instance, the PCA-based polynomial $\mathcal{F}$-NARX models utilized in this work (see Section~\ref{sec:fnarx_modeling}) require the selection of the maximum polynomial degree and the hyperbolic truncation index. 
    It is crucial to highlight that the \replace{automated mNARX}{mNARX\textsuperscript{+}} algorithm obviates the need for manual selection of principal components; PCA serves as an automated feature extraction mechanism, and the algorithm itself identifies and selects the most relevant components based on the chosen assessment function $\mathcal{H}$.
\end{itemize}

\addnew{
While the parameters at the algorithmic level are invariant to the problem dimensionality, the number of model memories to be selected can scale linearly with the number of exogenous inputs. 
However, we found that a uniform memory across all inputs generally performs well. 
This observation was also reported by \citet{schaer_fnarx_2025}.
The runtime of the algorithm is therefore primarily dictated by the intrinsic complexity of the problem rather than the observed problem dimensionality. 
That is, it increases if a large number of temporal features is required, requiring many iterations, or if the model requires the construction of numerous intermediate responses, potentially leading to repeated or even deep recursion.
}

\section{Applications}\label{sec:applications}
In this section we demonstrate the performance of the \replace{automatic mNARX}{mNARX\textsuperscript{+}} algorithm from Section~\ref{sec:automatic_mnarx_modeling} on two case studies.
\begin{itemize}
    \item In the first case study we investigate a Bouc-Wen oscillator with strong hysteresis under random ground-motion-like excitation. 
    This case study is relevant, since the hysteretic displacement is normally available as an output of the simulator, and can therefore be used as an auxiliary response for the training phase of the mNARX model.
    \item In the second case study we re-visit the aero-servo-elastic wind turbine case study from \citet{schaer_mnarx_2024}, in which the flapwise blade root bending moment of a realistic wind turbine was considered. 
    This highly complex simulation provides many possible auxiliary quantities and is of particular interest: significant prior knowledge on the system was needed to construct the original mNARX model in \citet{schaer_mnarx_2024}. 
    Further, the authors reported challenges in finding suitable lags for the intermediate and final NARX models that would result in stable and accurate long-term surrogate predictions. 
    Our goal here is to demonstrate how the required prior knowledge on the system and the manual work to tune multiple intermediate autoregressive models is drastically reduced with the proposed \replace{automatic-mNARX}{mNARX\textsuperscript{+}} approach.
\end{itemize}
\addnew{For both case studies, we additionally compare the performance of the mNARX\textsuperscript{+} model with long short-term memory (LSTM) neural networks. LSTM models have demonstrated excellent performance in numerous surrogate modeling applications for dynamical systems and are widely regarded as state-of-the-art in this domain \citep{zhao2023}.
}

\subsection{Bouc-Wen oscillator}\label{sec:bouc_wen}
\subsubsection{Problem statement}
\begin{figure}[H]
    \centering
    \includegraphics[width=.9\linewidth]{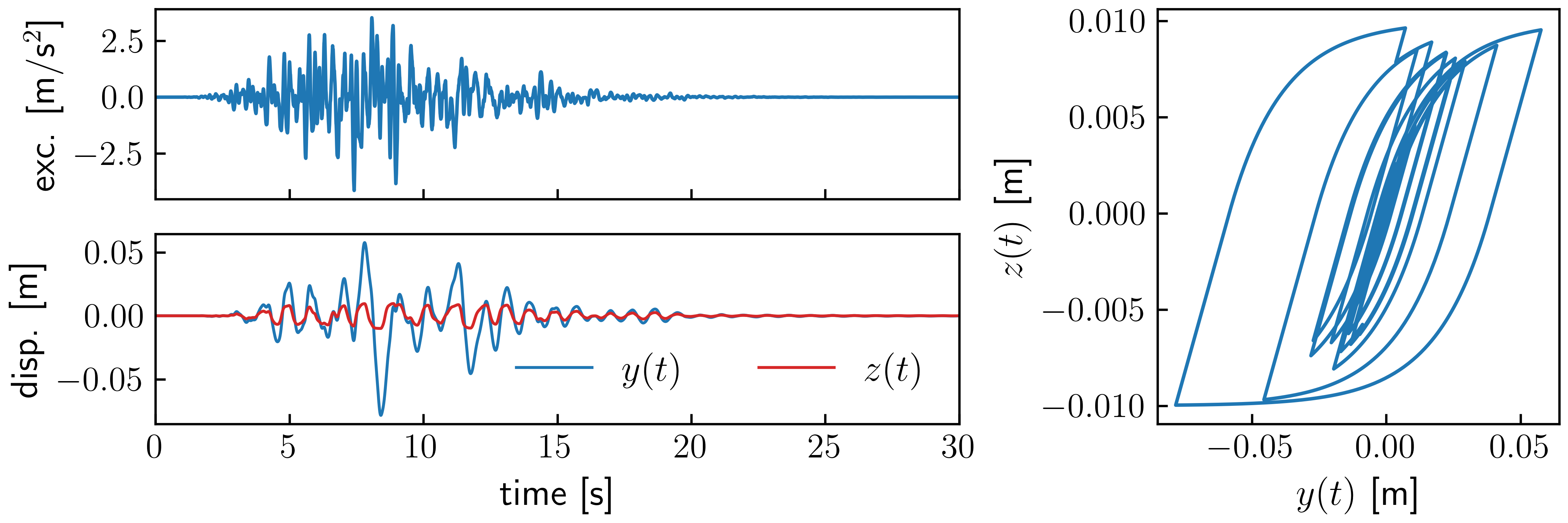}
    \caption{
        (Top left) Synthetic ground motion acceleration.
        (Bottom left) Corresponding Bouc-Wen displacement $y(t)$ and hysteretic displacement $z(t)$.
        (Right) Illustration of the corresponding nonlinear behavior. 
    }
    \label{fig:bouc_wen_example}
\end{figure}

In this first application, we consider a Bouc-Wen oscillator with displacement $y(t)$ determined by the following equations of motion:
\begin{equation}\label{eq:bouc_wen}
    \begin{cases}
        & \ddot{y}(t) + 2\zeta\omega\dot{y}(t) + \omega^2(\rho y(t)+(1-\rho)z(t))=-\ddot{x}(t), \\
        & \dot{z}(t) = \gamma\dot{y}(t) - \alpha|\dot{y}(t)||z(t)|^{n-1}z(t)-\beta\dot{y}(t)|z(t)|^n.
    \end{cases}
\end{equation}
The oscillator is parametrized in terms of the damping ratio $\zeta$, the fundamendal frequency $\omega$ and the post-yield to pre-yield stiffness ratio $\rho$. 
The variable $z(t)$ is a hidden hysteretic displacement which depends on the four parameters $\gamma$, $\alpha$, $\beta$ and $n$, which in turn determine the shape of the hysteretic loops.
The numerical values for the Bouc-Wen model parameters are given in Table~\ref{tab:bouc_wen_parameters} and an example trace along with the corresponding hysteretic loop is shown in Figure~\ref{fig:bouc_wen_example}.

The oscillator is subject to the random excitation $\ddot{x}(t)$, which mimics ground motion acceleration. 
More precisely, we adopted the approach presented in \citet{mai_2016b}, and condition the parametric stochastic Der~Kiureghian ground motion model \citep{rezaeian_2010} on the component 090 of the Northridge earthquake recorded at the LA~00 station. The resulting Der Kiureghian ground motion model parameters are listed in Table~\ref{tab:ground_motion_model}.
An example excitation is shown in Figure~\ref{fig:bouc_wen_example} with the corresponding model Bouc-Wen model response shown in Figure~\ref{fig:bouc_wen_example}.
        
\begin{table}[H]
    \centering
    \caption{Bouc-wen model parameters}
    \begin{tabular}{@{}lcc@{}}
    \toprule
    Parameter & \multicolumn{1}{l}{Unit} & \multicolumn{1}{l}{Value} \\ \midrule
    $\zeta$   & -                        & 0.02                      \\
    $\omega$  & rad/s                    & 10                        \\
    $\rho$    & -                        & 0.2                       \\
    $\gamma$  & -                        & 0.5                       \\
    $\alpha$  & 1/m                      & 25                        \\
    $\beta$   & -                        & 25                        \\
    $n$       & -                        & 1                         \\ \bottomrule
\end{tabular}
    \label{tab:bouc_wen_parameters}
\end{table}

\begin{table}[H]
    \centering
    \caption{Der Kiureghian ground motion model parameters}
    \begin{tabular}{@{}lcc@{}}
    \toprule
    Parameter & Unit & Value \\ \midrule
    $I_a$ & s.g & $0.109$ \\
    $D_{5-95}$ & s & $7.96$ \\
    $t_{mid}$ & s & $7.78$ \\
    $\omega_{mid}$ & Hz & $4.66 \cdot 2\pi$ \\
    $\omega'$ & Hz & $-0.09 \cdot 2\pi$ \\
    $\zeta_f$ & - & $0.24$ \\ \bottomrule
\end{tabular}
    \label{tab:ground_motion_model}
\end{table}

Using the described ground motion model, we generated a total of $10{,}000$ excitations and corresponding Bouc-Wen model responses.
The simulations have a duration of $30$~seconds each and are sampled at $100$~Hz.
We randomly selected $100$ simulations for the training set and the remaining $9{,}900$ for the out-of-sample test dataset. 

\subsubsection{mNARX construction}
Following the algorithm presented in Section~\ref{sec:automatic_mnarx_modeling}, we created an mNARX surrogate to predict the Bouc-Wen model response $y(t)$.
We include the hysteretic displacement $z(t)$ as a candidate intermediate response and the ground motion acceleration ($\ddot{x}(t)$), velocity ($\dot{x}(t)$) and displacement (${x}(t)$) as exogenous inputs.
As a stopping criterion, we use a correlation threshold $\theta_\rho=0.2$. That is, we stop the algorithm if no feature with a Kendall's $\tau$ correlation coefficient above $0.2$ remains.
For the $\mathcal{F}$-NARX surrogates $\widehat{\mathcal{M}}$, we use a polynomial structure, trained by least-angle regression (see Section~\ref{sec:fnarx_modeling}).
To select the model parameters (memory lengths and the parameters defining the polynomials), we used a manual search approach in which we select the parameters that minimize the mean forecast error, which is evaluated on the experimental design: 
\begin{equation}\label{eq:mean_forecast_error}
    \overline{\epsilon} = \frac{1}{N_\text{ED}} \sum_{i=1}^{N_\text{ED}} \epsilon^{(i)},
\end{equation}
where $\epsilon^{(i)}$ is the mean squared error on the $i$-th trace in the ED normalized by its variance:
\begin{equation}\label{eq:forecast_error}
    \epsilon^{(i)} = \frac{1}{N} \frac{
        {\displaystyle \sum_{k=0}^{N-1}} \left( y^{(i)}(k \delta t) - \widehat{y}^{(i)}(k \delta t) \right)^2
    }{
        \text{Var}(\vec{y}^{(i)})
    }.
\end{equation}
\addnew{Note that since the error used to select the parameters is computed on the training data, an automated structure search can also be employed, as demonstrated in \citet{schaer_fnarx_2025}.}
This process resulted in a maximum polynomial degree of $3$, which enables the model to capture nonlinear relationships, and a hyperbolic truncation index of $0.8$, which constrains the interaction between different polynomial terms. 
The optimal memory lengths for the $\mathcal{F}$-NARX models were found to be $40$ time steps ($0.4$~seconds) for the Bouc–Wen model response $y(t)$ and $120$ time steps ($1.2$~seconds) for the intermediate response $z(t)$.

\begin{figure}
    \centering
    \includegraphics[width=\linewidth]{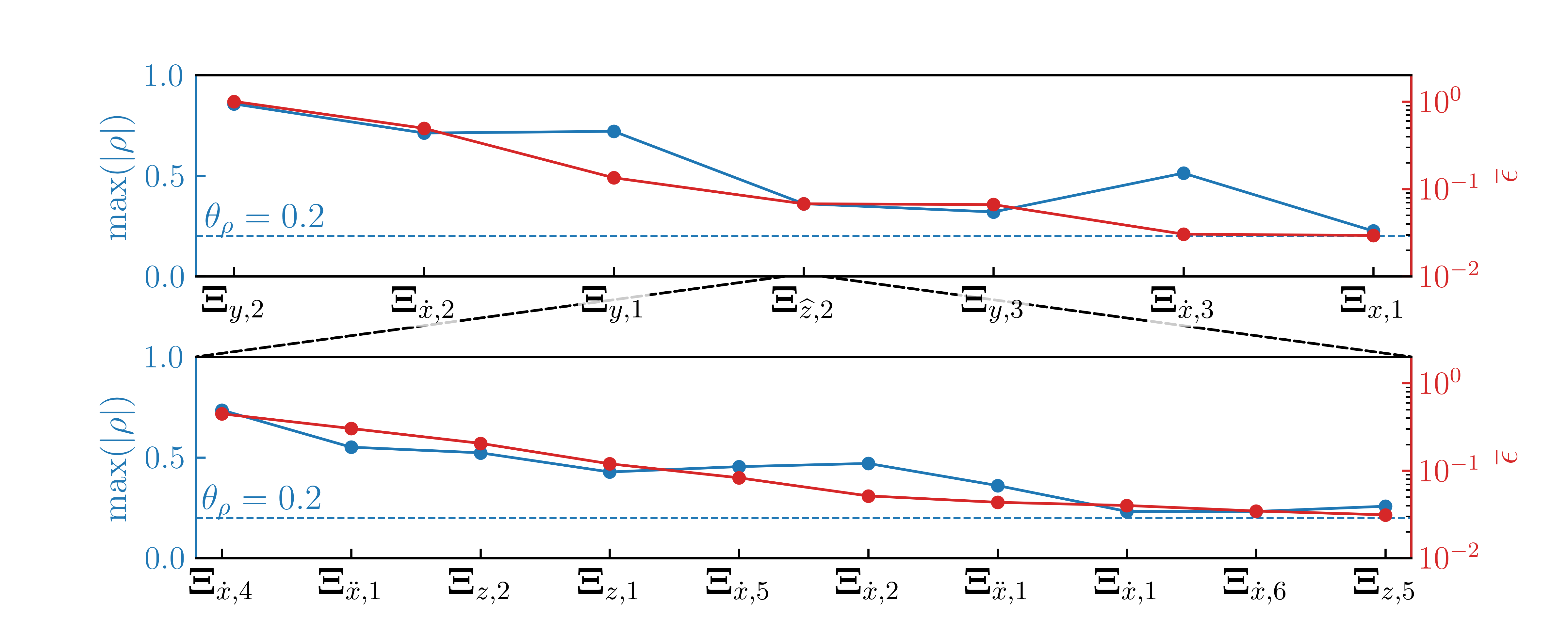}
    \caption{
        (Top) Sequence of selected features for the Bouc-Wen displacement $y(t)$. On the left axis the maximum absolute correlation $\max(|\vec{\rho}|)$ of the selected features is shown. On the right axis the mean forecast error $\overline{\epsilon}$ (see Eq.~\eqref{eq:forecast_error}) is shown.
        The dashed horizontal line indicates the correlation threshold ($\theta_\rho$) at which the selection process ends.
        (Bottom) Sequence of selected features for the auxiliary quantity $z(t)$ with the corresponding correlation and error values.
    }
    \label{fig:bouc_wen_feature_selection}
\end{figure}

The temporal features that were selected while running the algorithm with the best configuration are shown in Figure~\ref{fig:bouc_wen_feature_selection} and are ordered on the abscissa from left to right.
For example, the first feature selected is $\vec{\Xi}_{\dot{x},2}$, which corresponds to the second principal component of the ground motion velocity of the sliding window $\vec{\varphi}(t)$ from Eq.~\eqref{eq:varphi}.  
This is illustrated in the top panel of Figure~\ref{fig:bouc_wen_feature_selection}.
The maximum absolute correlation of the feature at its selection point is shown in blue, while the corresponding mean forecast error on the ED of the surrogate after including the feature is shown in red. Note that the error is shown in a logarithmic scale.
It is also worth highlighting that once a feature of the hysteretic displacement $z(t)$ is selected, $z(t)$ becomes an intermediate response. 
The algorithm therefore recurses with $z(t)$ as a temporary target, resulting in the corresponding evolution of selected features and forecast error shown in the bottom panel of the same figure.

Interestingly, while by design the error $\epsilon$ is strictly decreasing, this does not necessarily hold for $\max(|\vec{\rho}|)$. 
Adding a new feature such as $\vec{\Xi}_{y,3}$ (see the top subplot of Figure~\ref{fig:bouc_wen_feature_selection}) can change the surrogate forecast, resulting in residuals that are more correlated with the remaining available features. 
It can also be observed that the first selected feature is not necessarily the first principal component $\vec{\Xi}_{\bullet, 1}$, as the ranking depends on the correlation with the current residuals, rather than the ordering of the components.

\addnew{
\subsubsection{LSTM construction}\label{sec:lstm_architecture_bouc_wen}
The LSTM used for comparison with the mNARX\textsuperscript{+} model is trained to predict the Bouc–Wen response $y(t)$ directly from the exogenous input $x(t)$, without exploiting auxiliary information such as the hysteretic displacement. 
The network architecture consists of a single hidden LSTM layer with $100$ units. 
The $100$ traces available for training were split into $75$ training traces and $25$ validation traces. 
Training was performed with a batch size of $5$, a learning rate of $0.001$, and the Adam optimizer. 
The training process was stopped after $1{,}000$ epochs once convergence was reached. 
The choice of these hyperparameters was inspired by \citet{atila2025}, which addressed a surrogate modeling problem of comparable complexity.
}

\subsubsection{Results}
\begin{figure}[H]
    \centering
    \includegraphics[width=.9\textwidth]{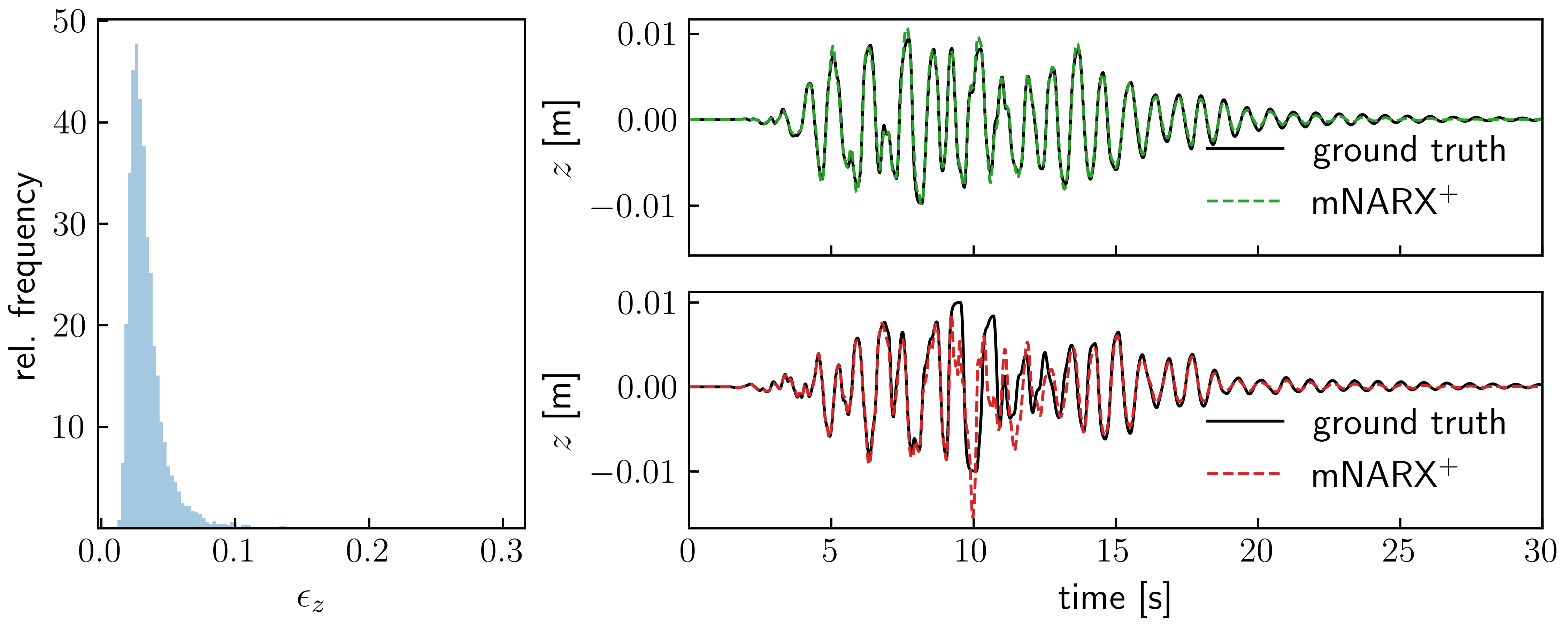}
    \caption{
        \addnew{mNARX\textsuperscript{+} results.}
        (Left) Distribution of the forecast error (see Eq.~\eqref{eq:forecast_error}) for the histeretic displacement $z(t)$.
        (Right) Trace with the highest forecast error shown in the top subplot in red and trace with the lowest forecast error shown in green in the bottom subplot. 
    }
    \label{fig:results_z}
\end{figure}

\begin{figure}[H]
    \centering
    \includegraphics[width=.9\textwidth]{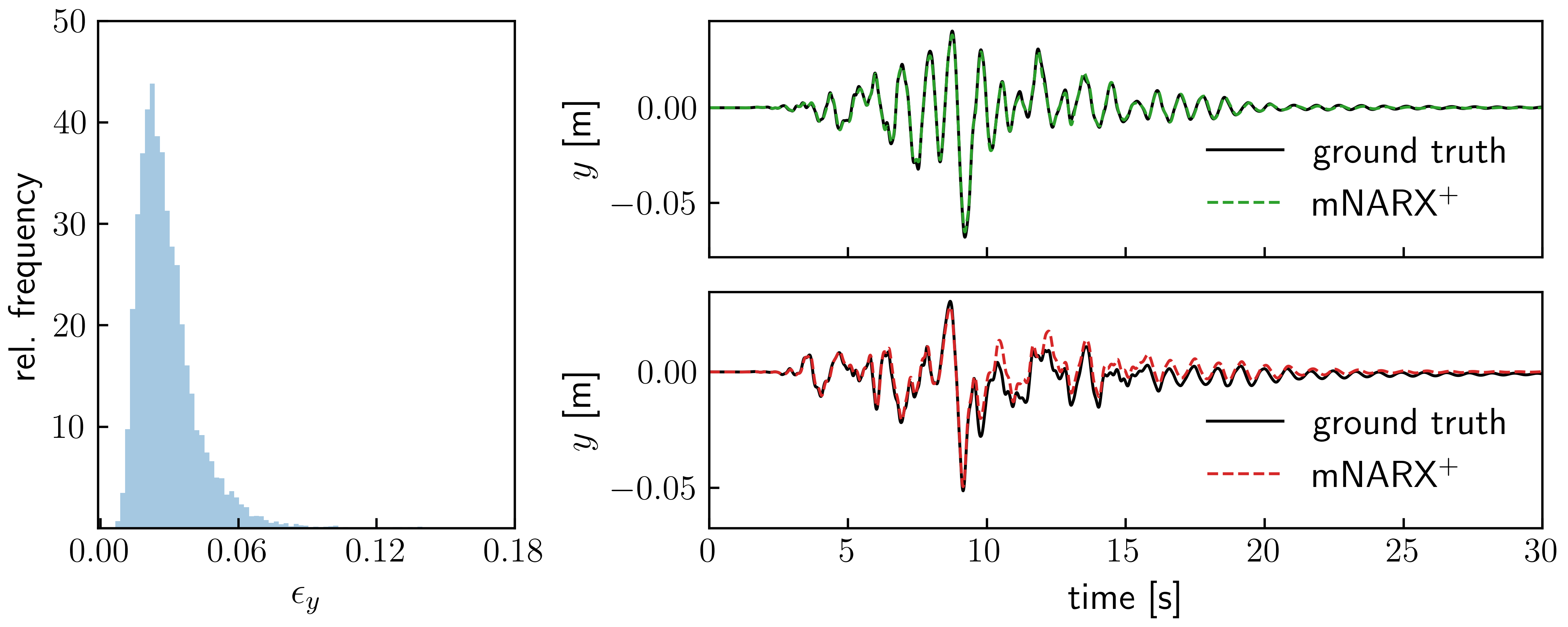}
    \caption{
        \addnew{mNARX\textsuperscript{+} results.}
        (Left) Distribution of the forecast error (see Eq.~\eqref{eq:forecast_error}) for the Bouc-Wen displacement $y(t)$.
        (Right) Trace with the highest forecast error shown in the top subplot in red and trace with the lowest forecast error shown in green in the bottom subplot. 
    }
    \label{fig:results_y}
\end{figure}

The performance of the mNARX surrogate is shown in Figure~\ref{fig:results_z} and Figure~\ref{fig:results_y}.
Figure~\ref{fig:results_z} shows the performance of the $\mathcal{F}$-NARX model for the intermediate response quantity $z(t)$.
On the left subplot we show the distribution of the forecast error as defined in Eq.~\eqref{eq:forecast_error} over the out-of-sample test dataset. 
In the right panels we show the traces corresponding to the highest error $\epsilon_{z,\text{max}}=0.30$ (red) and the lowest error $\epsilon_{z,\text{min}}=0.012$ (green) out of $9{,}900$ test traces. 
While there are a few traces with relatively large error, overall the surrogates shows good accuracy. 
It is noteworthy that, even for the trace with the highest error, the surrogate provides a stable prediction and only deviates from the ground truth for a few seconds.

Similarly, in Figure~\ref{fig:results_y} we show the error distribution for the Bouc-Wen displacement $y(t)$ and the traces corresponding to the highest error ($\epsilon_{y,\text{max}} \approx 0.16$) and the lowest error ($\epsilon_{y,\text{min}} \approx 0.0065$).
These errors are considerably smaller than the ones for the intermediate response. 
This indicates that the $\mathcal{F}$-NARX model relies strongly on the exogenous inputs $\ddot{x}(t)$, $\dot{x}(t)$ and ${x}(t)$ to predict $y(t)$ and that the model learned to compensate for the inaccuracy in the prediction of the intermediate response. 
This aligns well with the results from \citet{schaer_mnarx_2024} where it was also observed that remarkably good forecasts can be achieved despite imperfect predictions on the intermediate responses.

\addnew{

The performance of the LSTM model in predicting the displacement $y(t)$ is shown in Figure~\ref{fig:lstm_results_y}. 
The left subplot shows the distribution of the prediction error $\epsilon_{y}$ over the test dataset, which ranges between $\epsilon_{y,\text{min}} \approx 0.005$ and $\epsilon_{y,\text{max}} \approx 0.1$. 
The top-right subplot illustrates the trace corresponding to the best-case prediction, while the bottom-right subplot shows the worst-case prediction.

Both the mNARX\textsuperscript{+} and the LSTM provide accurate predictions of the displacement $y(t)$. 
The LSTM generally achieves lower errors across best-case, worst-case, and average scenarios, which is consistent with its established strength as a universal function approximator. 
Nonetheless, a more nuanced assessment of the performance of both the mNARX\textsuperscript{+} and LSTM models is given in Figure~\ref{fig:lstm_vs_mnarx_scatter}. 
This figure shows how well the two models can reconstruct peak displacements $\max(|\vec{y}|)$ across the validation dataset. 
It becomes apparent that the LSTM consistently underestimates  the upper tail of the distribution of maximum displacements. 
In contrast, the mNARX\textsuperscript{+} model provides more consistent predictions throughout the response range, including the tails, at the cost of a larger dispersion for middle maximum displacements. 
This bias reduction in the tails  is particularly important in engineering scenarios such as reliability or fragility analysis.

\begin{figure}[H]
    \centering
    \includegraphics[width=.9\textwidth]{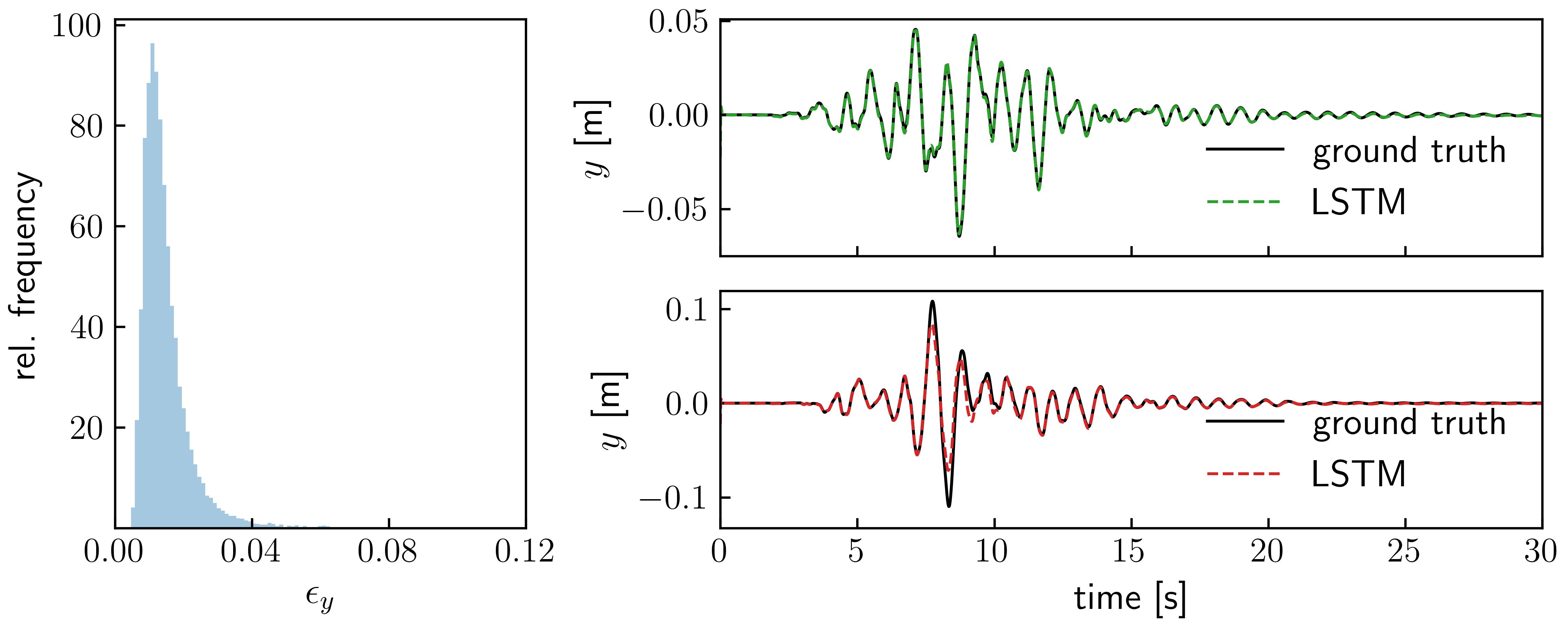}
    \caption{
        \addnew{LSTM results.
        (Left) Distribution of the forecast error (see Eq.~\eqref{eq:forecast_error}) for the Bouc-Wen displacement $y(t)$.
        (Right) Trace with the highest forecast error shown in the top subplot in red and trace with the lowest forecast error shown in green in the bottom subplot. 
        }
    }
    \label{fig:lstm_results_y}
\end{figure}


\begin{figure}[H]
    \centering
    \includegraphics[width=.9\textwidth]{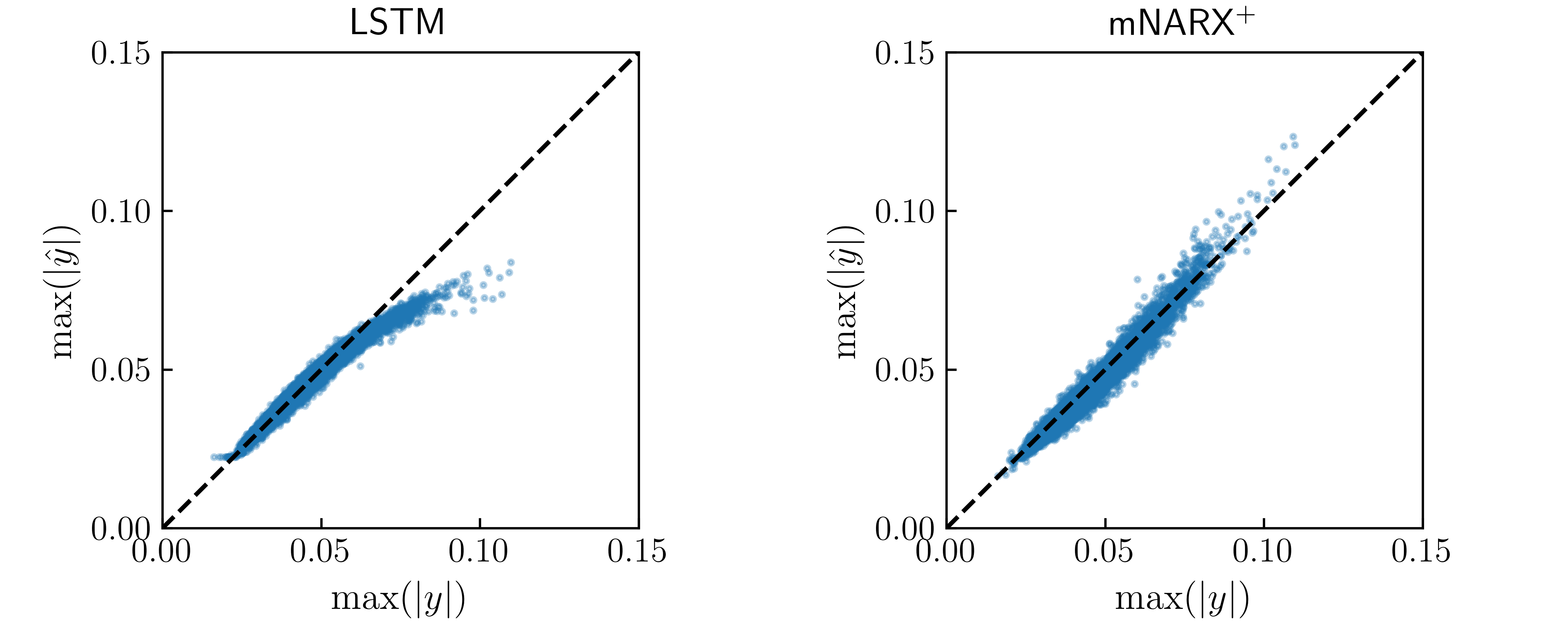}
    \caption{
        \addnew{
        (Left) Scatter plot of the absolute peak response obtained from the LSTM model prediction $\max(|\hat{\vec{y}}|)$ versus the true peak response $\max(|\vec{y}|)$.  
        (Right) Absolute peak response obtained from the mNARX\textsuperscript{+} model prediction versus the true peak response.
        }
    }
    \label{fig:lstm_vs_mnarx_scatter}
\end{figure}

}

\subsection{Aero-servo-elastic wind turbine simulator}\label{sec:wind_turbine}
\subsubsection{Problem statement}
In the second application, we consider an aero-servo-elastic (ASE) wind turbine simulator \citep{openfast_2021}. This simulator comprises both a complex, high-dimensional exogenous input in the form of turbulent wind, and an active control system. 
The interplay between this turbulent wind input and the active control system makes this a highly nonlinear dynamical system.

\begin{figure}[H]
    \centering
    \hspace{.09\linewidth}
    \includegraphics[width=0.5\linewidth, trim={130pt 0pt 80pt 0pt}, clip]{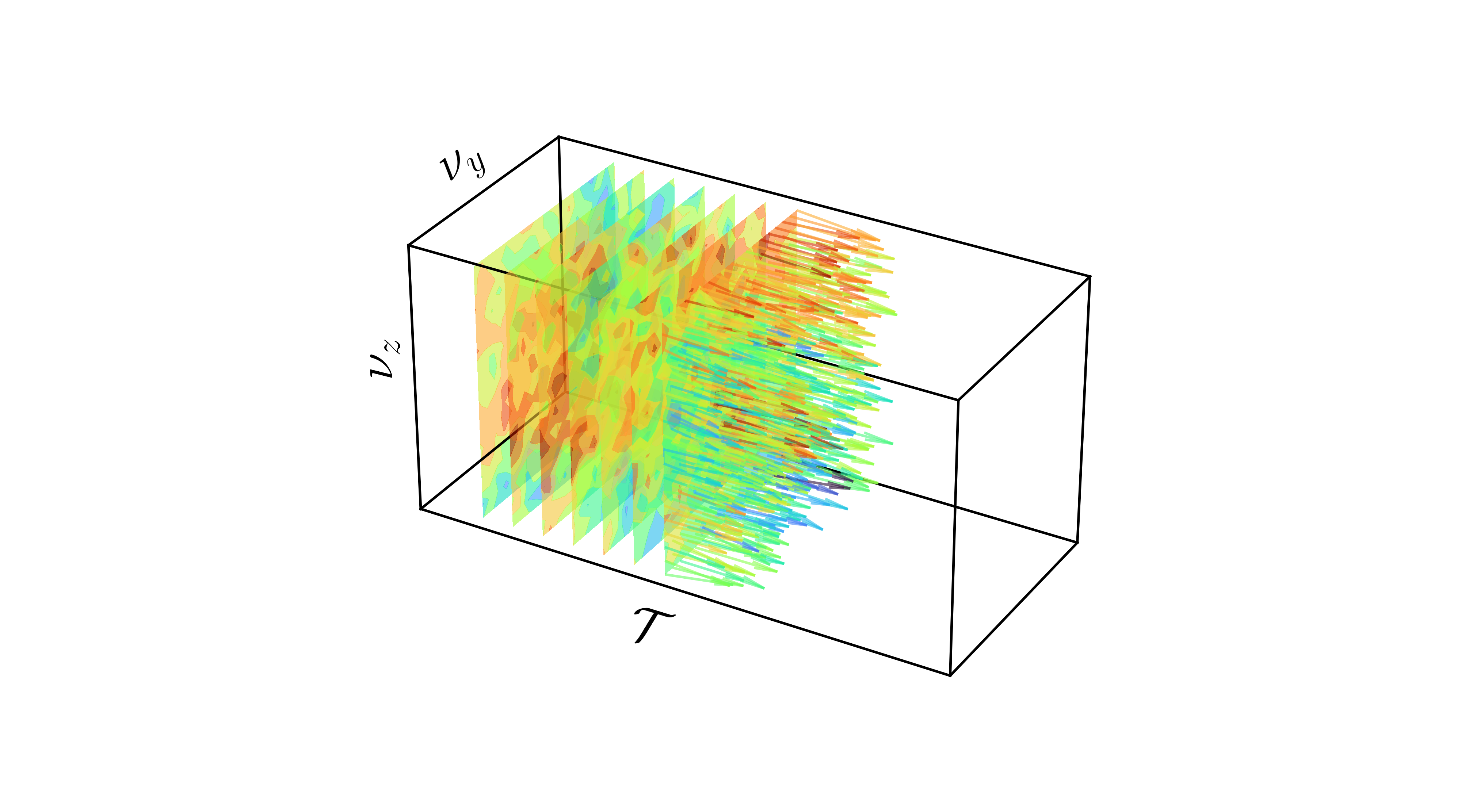}
    \hfill
    \includegraphics[width=0.3\linewidth]{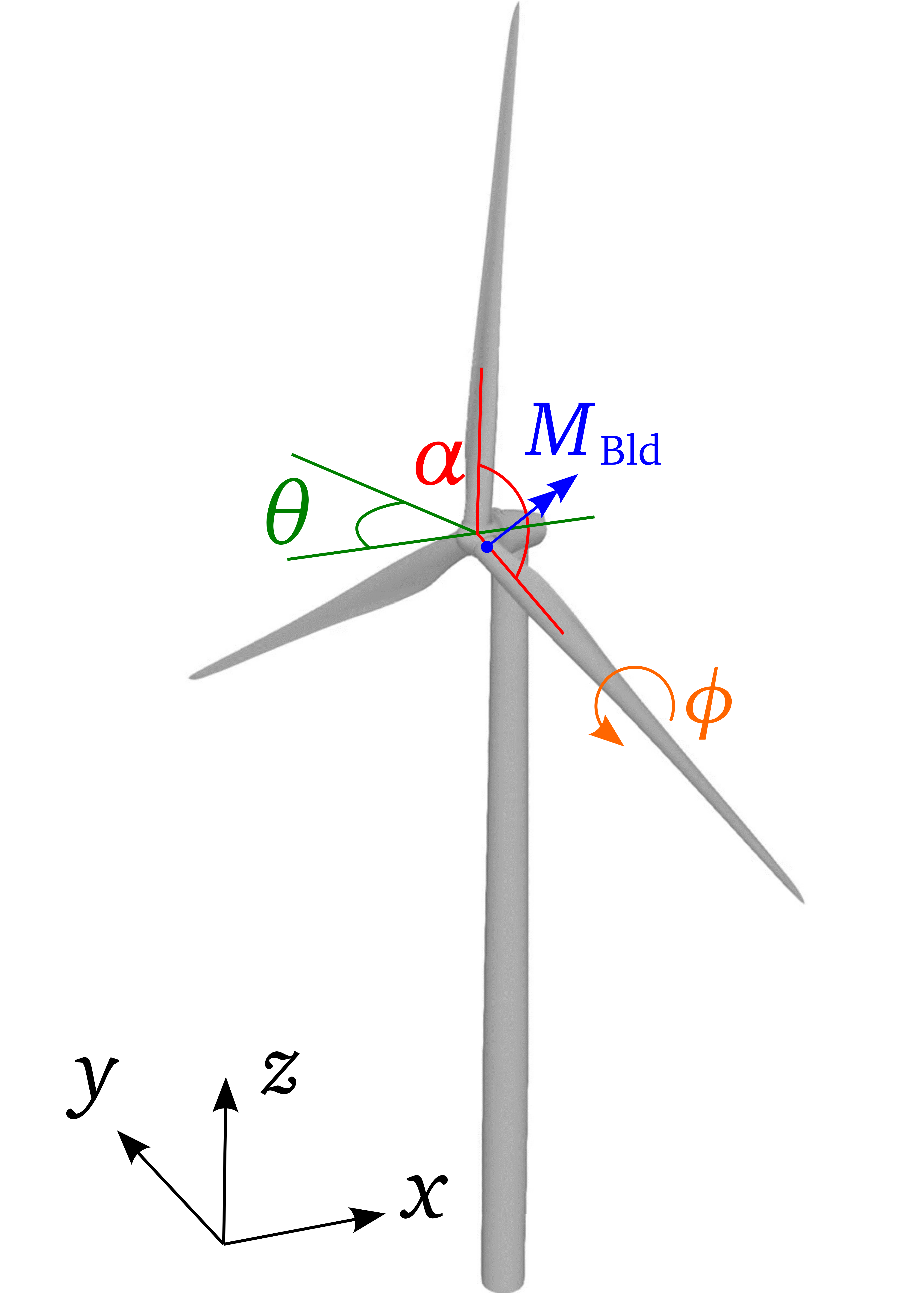}
    \hspace{.09\linewidth}
    \caption{
        (Left) Random wind field evolving along the time axis $\mathcal{T}$.
        (Right) Scheme of an onshore wind turbine with the rotor azimuth $\alpha$ depicted in red, the blade pitch $\phi$ in orange and the yaw angle $\theta$ in green. Additionally, we show the blade root bending moment $M_\text{Bld}$ in blue.
    }
    \label{fig:wind_turbine}
\end{figure}

The simulator input is represented by a random velocity field $\vec{v}$ evolving along the time axis $\mathcal{T}$, and consisting of wind speed values in the longitudinal ($x$), transversal ($y$) and vertical ($z$) directions at each of the $\nu_y \times \nu_z$ spatial grid points:
\begin{equation}
    \vec{v}: \mathcal{T} \to \mathbb{R}^{3 \times \nu_y \times \nu_z}.
\end{equation}
A detailed description on how we generate such wind fields is given in Section~\ref{sec:input_wind_field} and an illustration of an example wind field is shown in the left panel of Figure~\ref{fig:wind_turbine}.
The outputs of the simulator for a given input wind field are a set of univariate time series $f_i$ representing the turbine responses, such as rotor speed, power output and bending moments:
\begin{equation}
    f_i: \mathcal{T} \to \mathbb{R}.
\end{equation}
A comprehensive description of the wind turbine model used and the output quantities of interest are provided in Section~\ref{sec:output_quantities}.

The goal in this case study is to emulate the wind turbine simulator, meaning that we aim at predicting multiple simulator outputs, namely the blade pitch angle, the rotor speed and the blade root bending moment $M_\text{Bld}$, based solely on unseen input wind fields. 
The main challenges associated to this problem are the very high-dimensional exogenous input ($\mathcal{O}(10^{3})$ wind speed values at every time step), the varying geometry (\eg changes in the rotor position over time), as well as the change in the turbine response behavior due to the real-time active control of the blade pitch angle. 
An illustration of an onshore wind turbine and its degrees of freedom is given in the right panel of Figure~\ref{fig:wind_turbine}.

\subsubsection{Input wind fields}\label{sec:input_wind_field}
To generate the input wind fields (also often referred to as turbulence boxes) we use the open-source software Turbsim \citep{turbsim_2009}.
To generate the spatio-temporal random fields representing turbulence boxes, TurbSim takes a set of scalar environmental parameters which we list in Table~\ref{tab:turbsim_specs}. 
Note that two of the parameters, namely the reference wind speed and the turbulence standard deviation are sampled randomly from distributions defined by the IEC~61400-1 design standard for onshore wind turbines of the class B~I \citep{iec_2005}. Further, we use the Kaimal spectral model and an exponential coherence model as suggested by the standard.
The spatial dimension of the turbulence boxes is set to $\nu_y=\nu_z=19$ and the boxes are generate at a sampling rate of $20$~Hz or a time discretization of $\delta t=0.05$~s, respectively.

\begin{table}[H]
    \centering
    \caption{Parameters for wind field generation (according to \citet{iec_2005})}
    \begin{tabular}{@{}lcc@{}}
    \toprule
    Parameter & Unit & Value \\ \midrule
    Reference wind speed  ($V_\text{hub}$) & m/s & Rayleigh distribution \\
    Turbulence standard deviation ($\sigma_1$) & m/s & Lognormal distribution conditional on $V_\text{hub}$ \\
    Wind shear ($\alpha$) & - & $0.2$  \\
    Air density ($\rho$) & kg/m$^3$ & $1{,}225$  \\
    \bottomrule     
\end{tabular}
    \label{tab:turbsim_specs}
\end{table}

\subsubsection{Wind turbine model and output quantities}\label{sec:output_quantities}
The input turbulence boxes are processed using the open-source ASE simulator OpenFAST \citep{openfast_2021} in which we model the well-known reference 5~MW NREL onshore wind turbine \citep{nrel_onshore_2009} controlled by the NREL reference open-source controller (ROSCO) \citep{rosco_2022}. The turbine specifications are detailed in Table~\ref{tab:turbine_specs}.

Each of the $12$~minutes long simulations is carried out at a sampling rate of $160$~Hz ($\delta t=0.00625$~s) which means that the input wind field is interpolated in time during the simulation.
While this is important for numerical stability of the solver, during the surrogate modeling we downsample back to $20$~Hz, which is sufficient to preserve virtually all of the signal information, while significantly reducing computational demands.
The output quantities recorded during the simulation are the blade pitch angle $\phi(t)$, the rotor speed $\omega(t)$, the rotor azimuth $\alpha(t)$ and the flapwise blade root bending moment $M_\text{Bld}$.

\begin{table}[H]
    \centering
    \caption{NREL 5MW reference turbine model specifications}
    \begin{tabular}{@{}ll@{}}
    \toprule
    Category                          & Specification                   \\ \midrule
    Rated power                       & 5 MW                               \\
    Rotor orientation, configuration  & upwind, 3 blades                   \\
    Control                           & variable speed, collective pitch   \\
    Drivetrain                        & high speed, multiple-stage gearbox \\
    Rotor, hub diameter               & 126 m, 3 m                         \\
    Hub Height                        & 90 m                               \\
    Cut-in, rated, cut-out wind speed & 3 m/s, 11.4 m/s, 25 m/s            \\
    Cut-in, rated rotor speed         & 6.9 rpm, 12.1 rpm                   \\ \bottomrule
\end{tabular}
    \label{tab:turbine_specs}
\end{table}

\subsubsection{Training and test data}\label{sec:training_and_test_data}
In this case study we specifically model the behavior of the wind turbine under high wind speed conditions. 
To achieve this, we began with a total of $5{,}000$ simulations, each $12$~minutes in duration. Following standard practice, the initial $2$~minutes of each simulation were truncated to exclude the start-up phase of the turbine.

From this initial set, we selected the $1,009$ simulations that maintained the turbine within or above its rated operating range, characterized by consistently high wind speeds and active blade pitch control. 
From this dataset, a subset of $100$ simulations was used for the training of our surrogate, with the remaining simulations reserved for out-of-sample testing. 

\begin{figure}[H]
    \centering
    \includegraphics[width=0.9\linewidth]{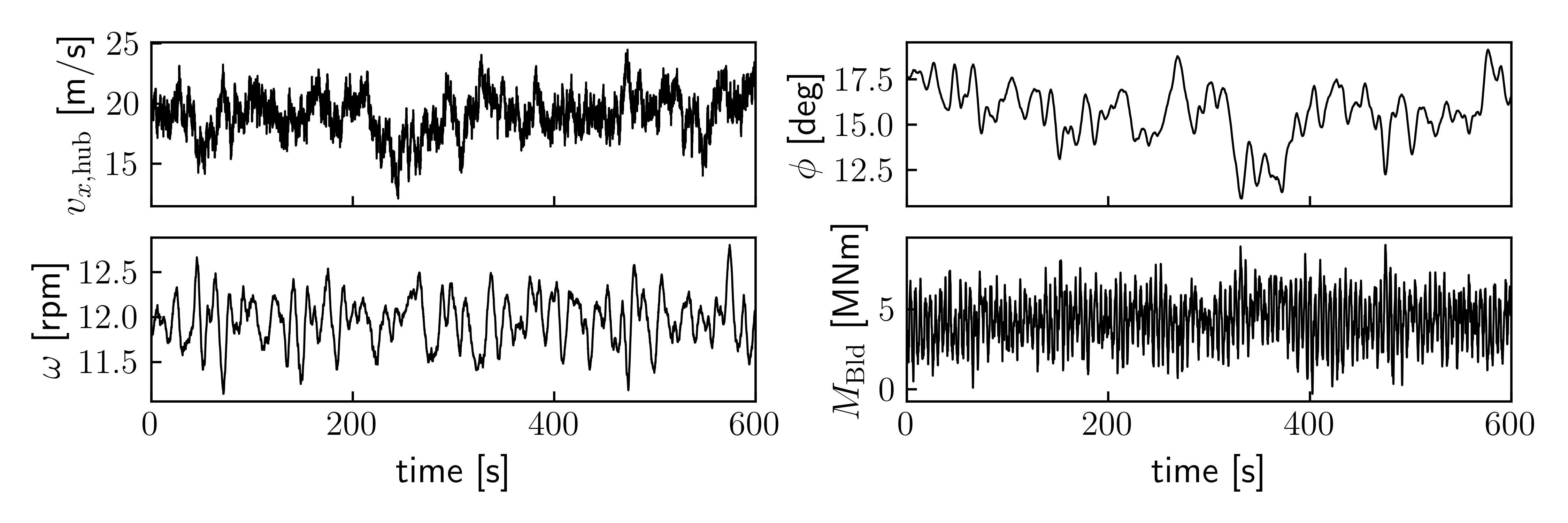}
    \caption{
        \delete{
            Example data of an aero-servo-elastic simulation.
            (Top left) Longitudinal wind speed at the rotor hub $v_{x, \text{hub}}$.
            (Top right) Evolution of the rotor speed $\omega(t)$.
            (Bottom left) Evolution of the blade pitch angle $\phi(t)$.
            (Bottom right) Evolution of the flapwise blade root bending moment $M_\text{Bld}$.
        }
        \addnew{
            Example data of an aero-servo-elastic simulation.
            (Top left) Longitudinal wind speed at the rotor hub $v_{x, \text{hub}}$.
            (Top right) Evolution of the blade pitch angle $\phi(t)$.
            (Bottom left) Evolution of the rotor speed $\omega(t)$.
            (Bottom right) Evolution of the flapwise blade root bending moment $M_\text{Bld}$.
        }
    }
    \label{fig:wind_turbine_example_traces}
\end{figure}

An example simulation is shown in Figure~\ref{fig:wind_turbine_example_traces}. This figure shows the wind speed at the rotor hub and the corresponding pitch control output, as well as the evolution of the rotor speed and the flapwise blade root bending moment.

\subsubsection{mNARX construction}
Our main objective is to automatically create an mNARX surrogate using the algorithm described in Section~\ref{sec:automatic_mnarx_modeling}, to emulate the flapwise blade root bending moment $M_\text{Bld}$.

As candidate auxiliary quantities, we use the blade pitch $\phi(t)$, the rotor speed $\omega(t)$, and the first four harmonics of the rotor azimuth $\{ \sin(k \alpha(t)), \cos(k \alpha(t)) \}$ for $k=1, \dots, 4$, where we obtain the rotor azimuth through the integration of the rotor speed: $\alpha(t) = \left( \int_0^t \omega(\tau)\text{d}\tau \right) \pmod{2\pi}$.
For the exogenous inputs, one could use the raw wind speed values $\vec{v}(t)$. The disadvantage of this approach is that the wind speed values carry localized spatial information of the turbulence box, whereas we expect that the turbine response is mostly driven by the average longitudinal wind speed and certain frequencies of the latter.
Instead of using the raw wind speed values, we therefore follow the approach from \citet{schaer_mnarx_2024} and compute spatial 2D discrete cosine transform (DCT) modes of the longitudinal wind speed component $\vec{v}_x$ instead.
For each time step, we compute the DCT coefficient $\eta^{(i, j)}(t)$ of the slice $\vec{v}_x(t)$ as follows:
\begin{equation}\label{eq:discrete_cosine_transform}
    v_x^{(\kappa, \ell)}(t) = \sum_{i=0}^{\nu_y-1}\sum_{j=0}^{\nu_z-1} \eta^{(i,j)}(t)
    \cos\left[ \frac{\pi}{\nu_y} \left( i+\frac{1}{2} \right)\kappa \right]
    \cos\left[ \frac{\pi}{\nu_z} \left( j+\frac{1}{2} \right)\ell \right].
\end{equation}
We then pass the DCT modes $\eta^{(i,j)}(t)$ for $i,j \in \{0, \dots, 9\}$ as the exogenous inputs to the algorithm. 

Similar to the first case study in Section~\ref{sec:bouc_wen}, we use a correlation threshold of $\theta_\rho = 0.2$ to terminate the algorithm.
As in Section~\ref{sec:fnarx_modeling}, we again employ polynomial $\mathcal{F}$-NARX models, with their structure determined through a manual search approach.
This resulted in a polynomial degree of $d = 3$ and a hyperbolic truncation index of $q = 1.0$.
The memory lengths for the $\mathcal{F}$-NARX models were chosen as $2.0$~seconds for the rotor speed, $4.0$~seconds for the blade pitch, and $4.0$~seconds for the blade moment.

\begin{figure}[htb]
    \centering
    \includegraphics[width=\linewidth,trim=30pt 0pt 15pt 0pt,clip]{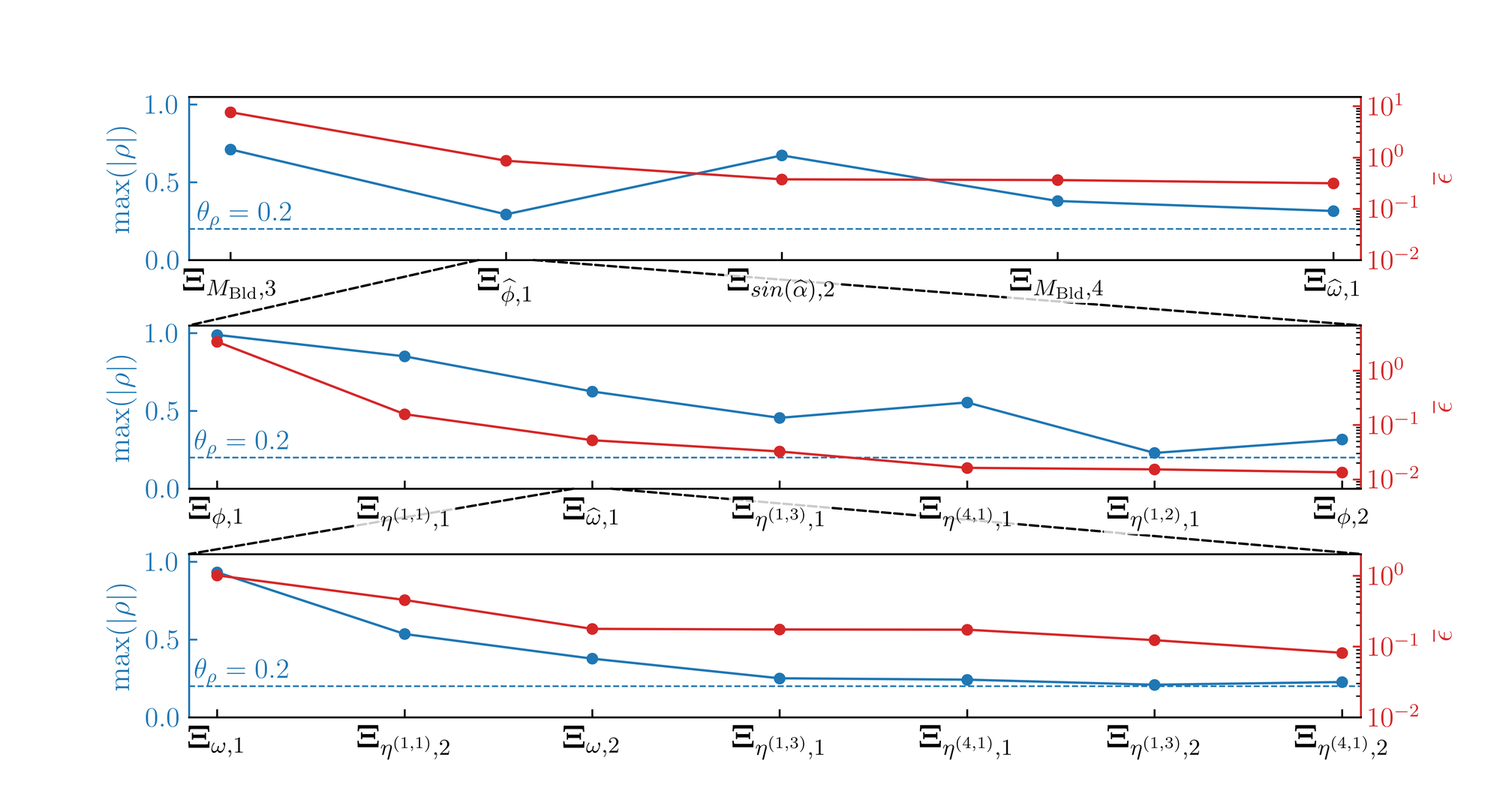}
    \caption{
        (Top) Sequence of selected features for the flapwise blade root bending moment $M_\text{Bld}$. 
        On the left axis the maximum absolute correlation $\max(|\vec{\rho}|)$ of the selected features is shown. 
        On the right axis the mean forecast error $\overline{\epsilon}$ (see Eq.~\eqref{eq:forecast_error}) is shown.
        The dashed horizontal line indicates the correlation threshold ($\theta_\rho$) at which the selection process ends.
        (Middle) Sequence of selected features for the blade pitch angle $\phi(t)$ with the corresponding correlation and error values.
        (Bottom) Sequence of selected features for the rotor speed $\omega(t)$ with the corresponding correlation and error values.
    }
    \label{fig:wind_turbine_feature_selection}
\end{figure}

The full selection process is illustrated in Figure~\ref{fig:wind_turbine_feature_selection}. 
We denote the $i$-th PCA feature of the blade root moment as $\vec{\Xi}_{M_{\text{Bld}}, i}$, that of the blade pitch as $\vec{\Xi}_{\phi, i}$, of the rotor speed as $\vec{\Xi}_{\omega, i}$, and for the $k$ harmonics as $\vec{\Xi}_{\bullet(k\alpha), i}$. 
For the DCT features, the $i$-th temporal PCA feature of the spatial DCT component $(k,l)$ is denoted as $\vec{\Xi}_{\eta^{(k, l)}, i}$.

When processing the blade moment $M_\text{Bld}$, the algorithm selects in the first iteration a PCA feature that originates from the blade pitch $\phi$, prompting a recursion with $\phi$ as a new quantity of interest. 
During this recursion, a PCA feature of the rotor speed $\omega$ is selected, which triggers another recursion and increases the recursion depth to two. 
In this second recursion, only autoregressive features and exogenous inputs are identified. 

The second recursion stops with a total of $7$ PCA features after no features with sufficient correlation remain and the first recursion stops after the selection of $7$ PCA features.
After completing both recursion steps, the algorithm selects a PCA feature derived from the higher rotor harmonics in the third iteration step. 
Since the rotor speed has already been predicted in the second recursion step, its harmonics can now be computed and therefore no additional recursion is required. 
In total, only $5$ PCA features are selected to emulate the blade moment before no features with sufficient correlation remain.

\addnew{
This selection process differs fundamentally from the original mNARX approach, which relies on full auxiliary quantities chosen based on presumed physical relevance.
As a result, the final mNARX\textsuperscript{+} manifold typically differs from that of mNARX, as only subsets of temporal features extracted from the auxiliary quantities are included prior to regression.
Combined with its largely automated construction, this makes mNARX\textsuperscript{+} a distinct and significantly more flexible approach.
}

\addnew{
\subsection{LSTM construction}
Analogous to the Bouc-Wen case study, we construct a long short-term memory (LSTM) neural network as a competitor to the mNARX\textsuperscript{+} surrogate. 
This LSTM model is trained to approximate the flapwise blade root bending moment $M_\text{Bld}(t)$ directly from the spatial DCT modes $\eta^{(i,j)}(t)$, without exploiting additional simulator outputs. 
The $100$ traces available for training were split into a training set of $75$ traces and a validation set of $25$ traces. 
The network employs the same hyperparameters as in Section~\ref{sec:lstm_architecture_bouc_wen}, namely a single hidden layer with $100$ units and a learning rate of $0.001$ using the Adam optimizer. 
Training was stopped after $250$ epochs once convergence was reached.
}

\subsubsection{Results}\label{sec:wind_turbine_results}
\begin{figure}
    \begin{subfigure}{0.49\linewidth}
        \centering
        \includegraphics[width=0.65\linewidth]{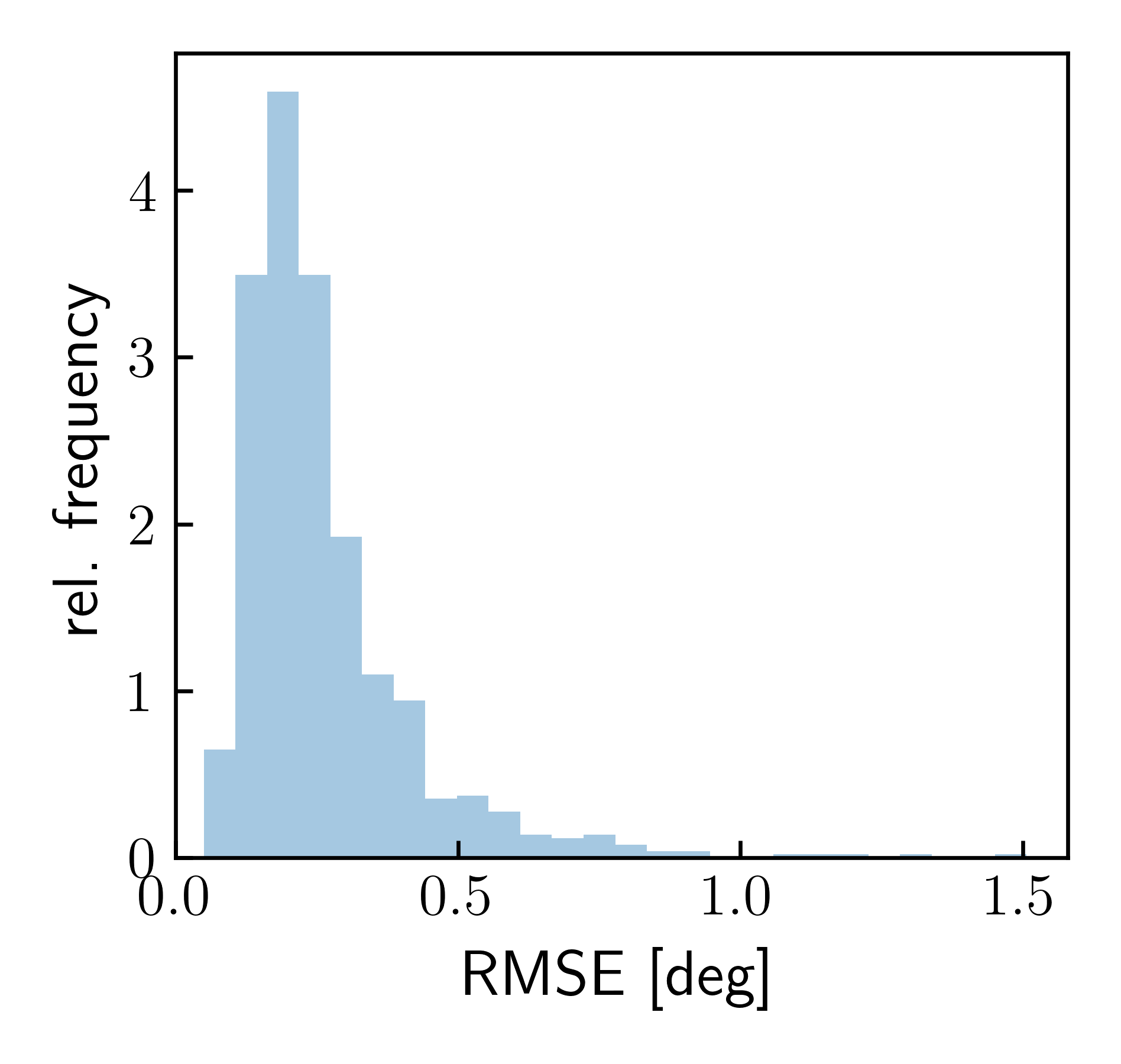}\\
        \includegraphics[width=\linewidth, trim=0 20 0 0]{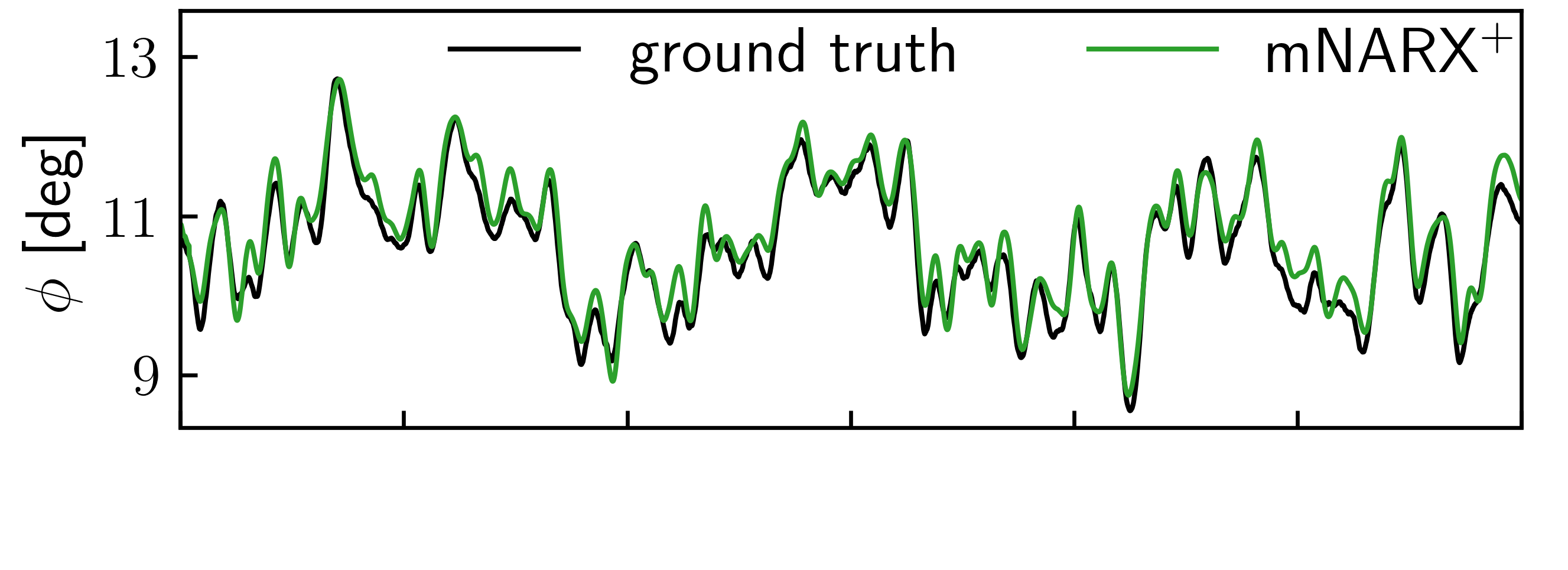}\\
        \includegraphics[width=\linewidth]{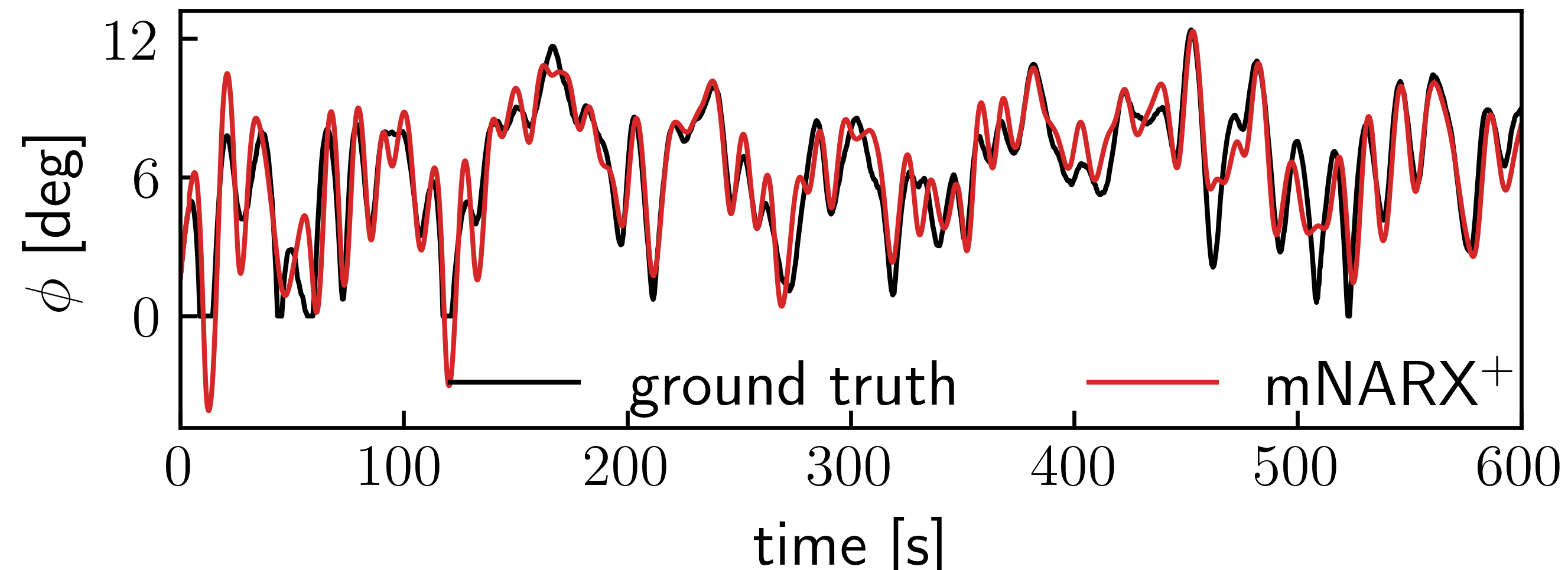}
    \end{subfigure}
    \hfill
    \begin{subfigure}{0.49\linewidth}
        \centering
        \includegraphics[width=0.65\linewidth]{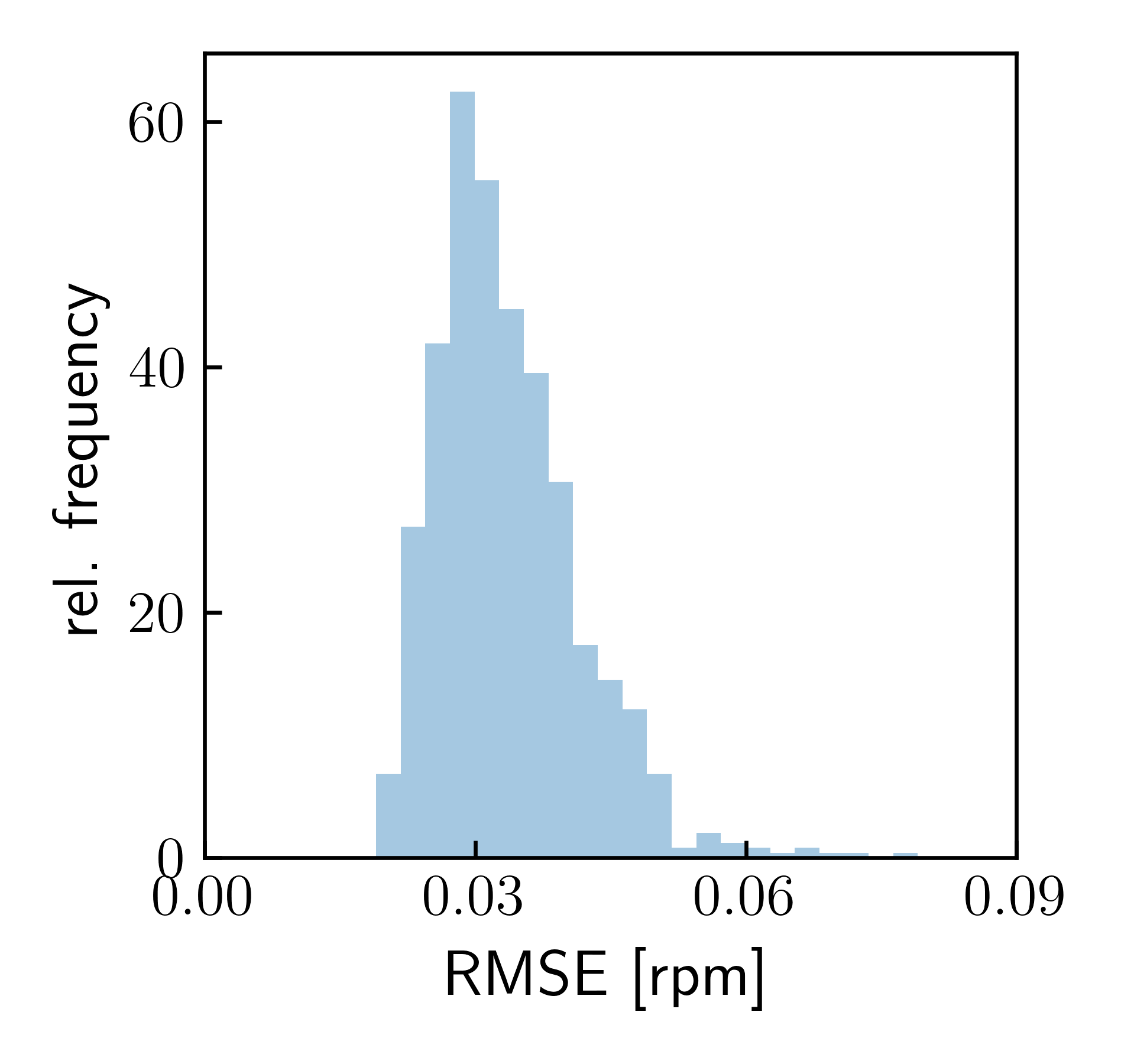}\\
        \includegraphics[width=\linewidth, trim=0 20 0 0]{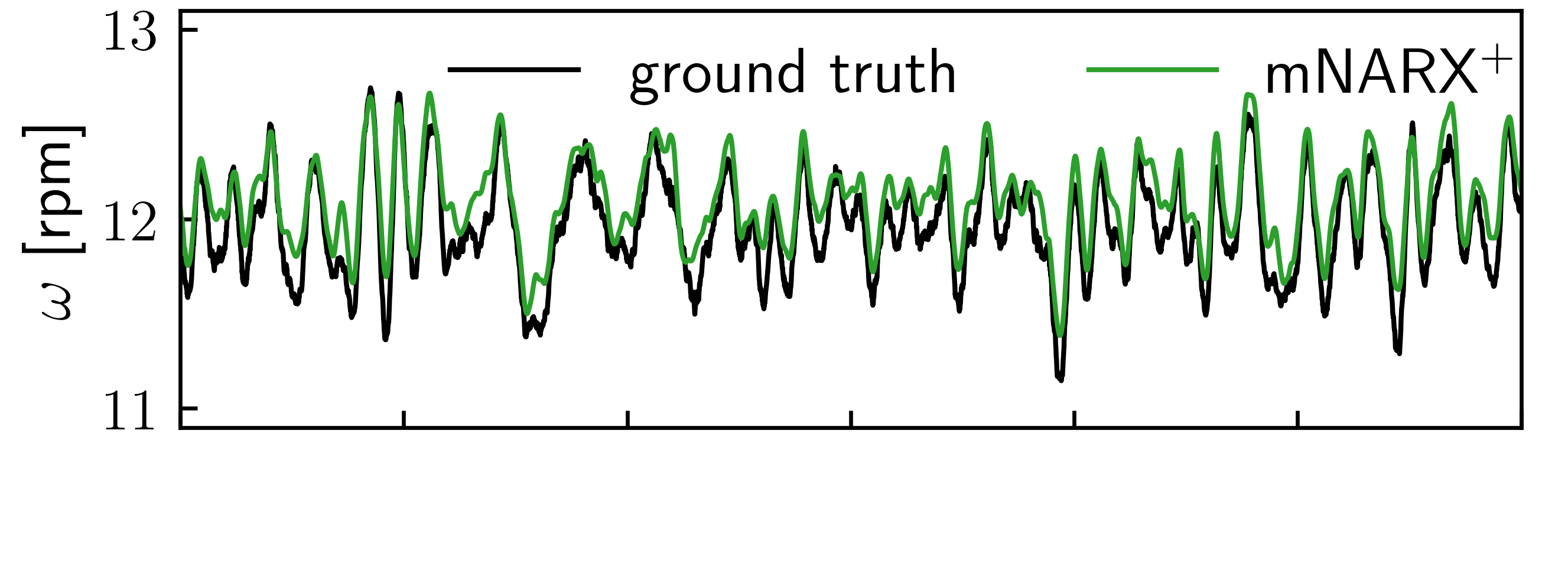}\\
        \includegraphics[width=\linewidth]{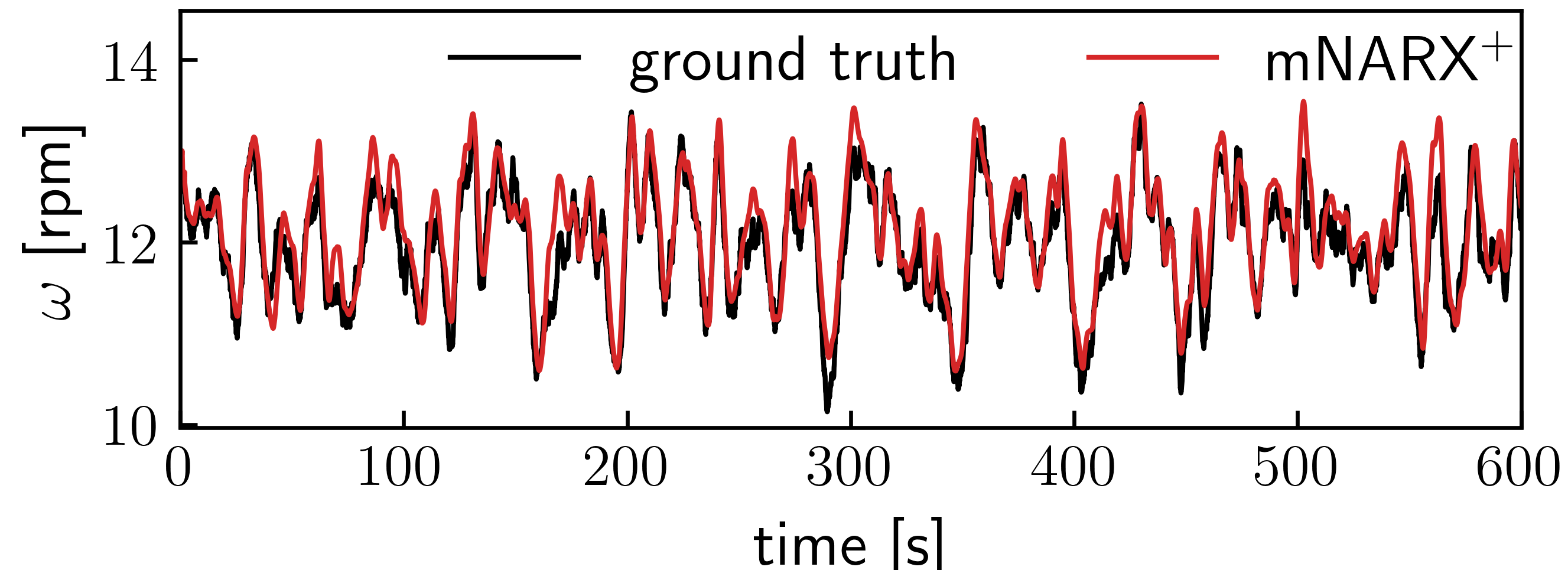}
    \end{subfigure}
    \caption{
        \addnew{mNARX\textsuperscript{+} results.}
        (Top left) Histogram of the root-mean-squared error (RMSE) in degrees for the \textbf{blade pitch} prediction.
        (Bottom left) \textbf{Blade pitch} traces with the lowest (green) and highest (red) RMSE.
        (Top right) Histogram of the root-mean-squared error (RMSE) in revolutions per minute for the \textbf{rotor speed} prediction.
        (Bottom right) \textbf{Rotor speed} traces with the lowest (green) and highest (red) RMSE.
    }
    \label{fig:blade_pitch_and_rotor_speed_results}
\end{figure}

\begin{figure}
    \centering
    \begin{minipage}{0.49\linewidth}
        \centering
        \includegraphics[width=0.65\linewidth]{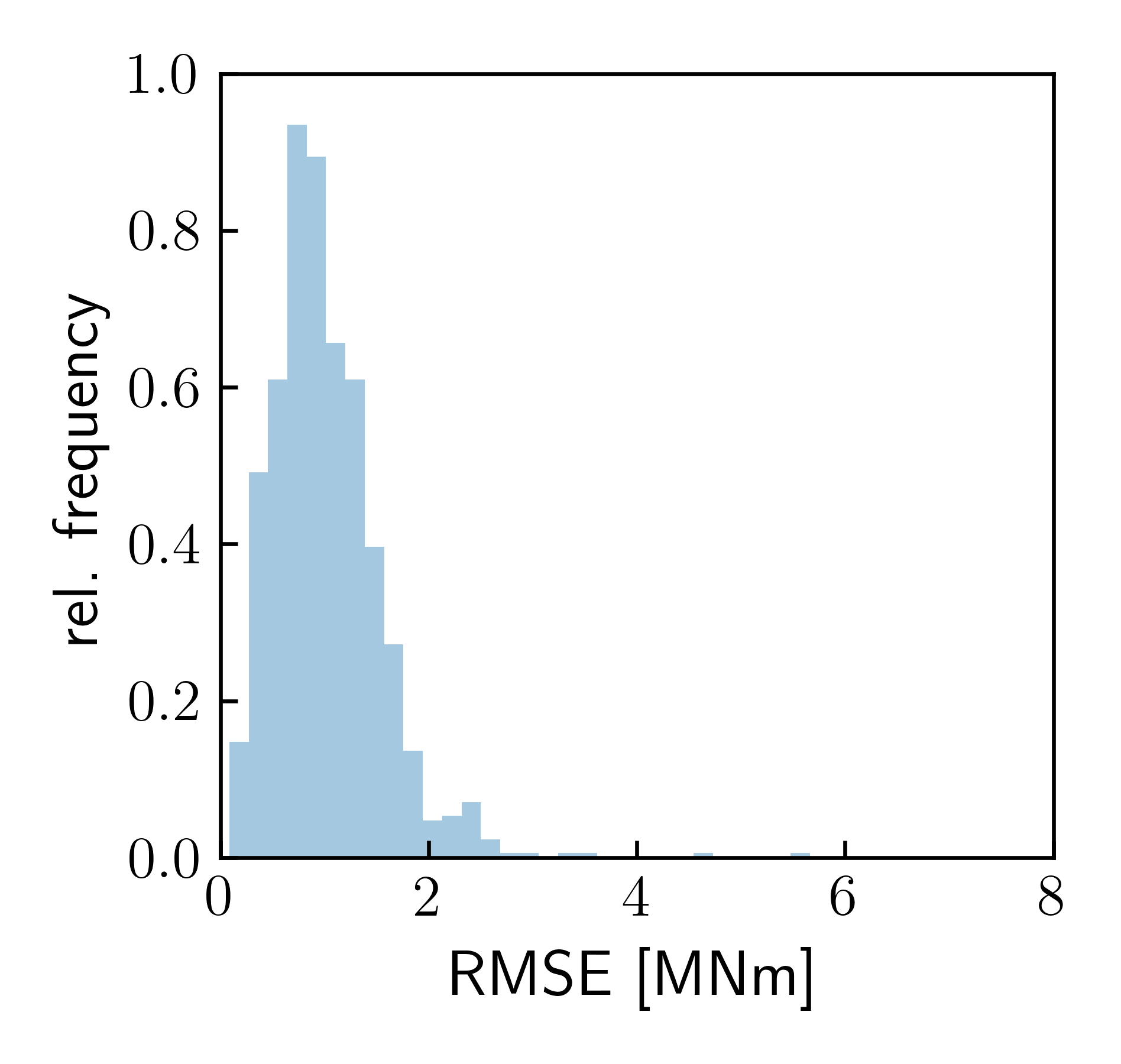}\\
        \includegraphics[width=\linewidth]{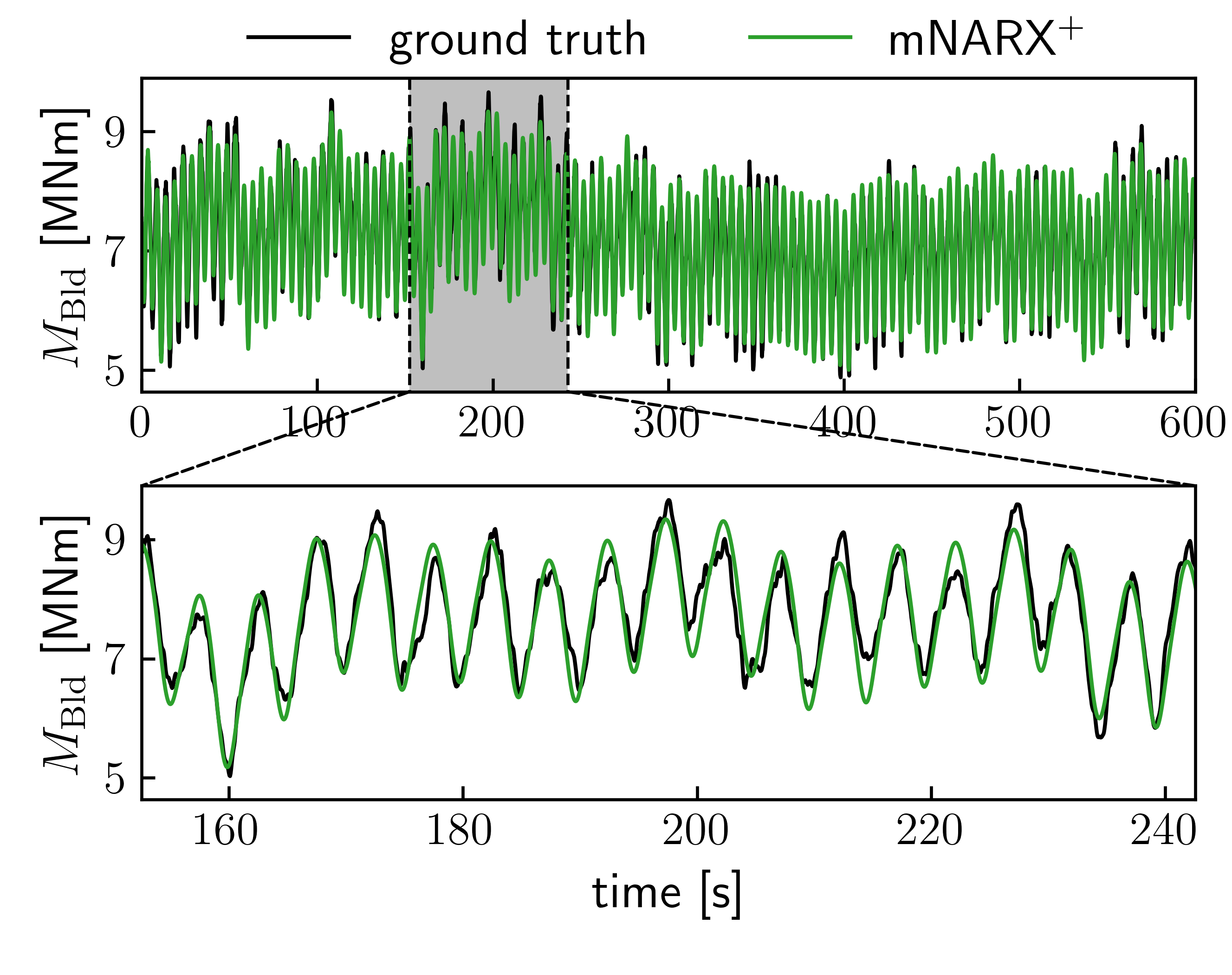}
    \end{minipage}
    \begin{minipage}{0.49\linewidth}
        \centering
        \includegraphics[width=0.65\linewidth]{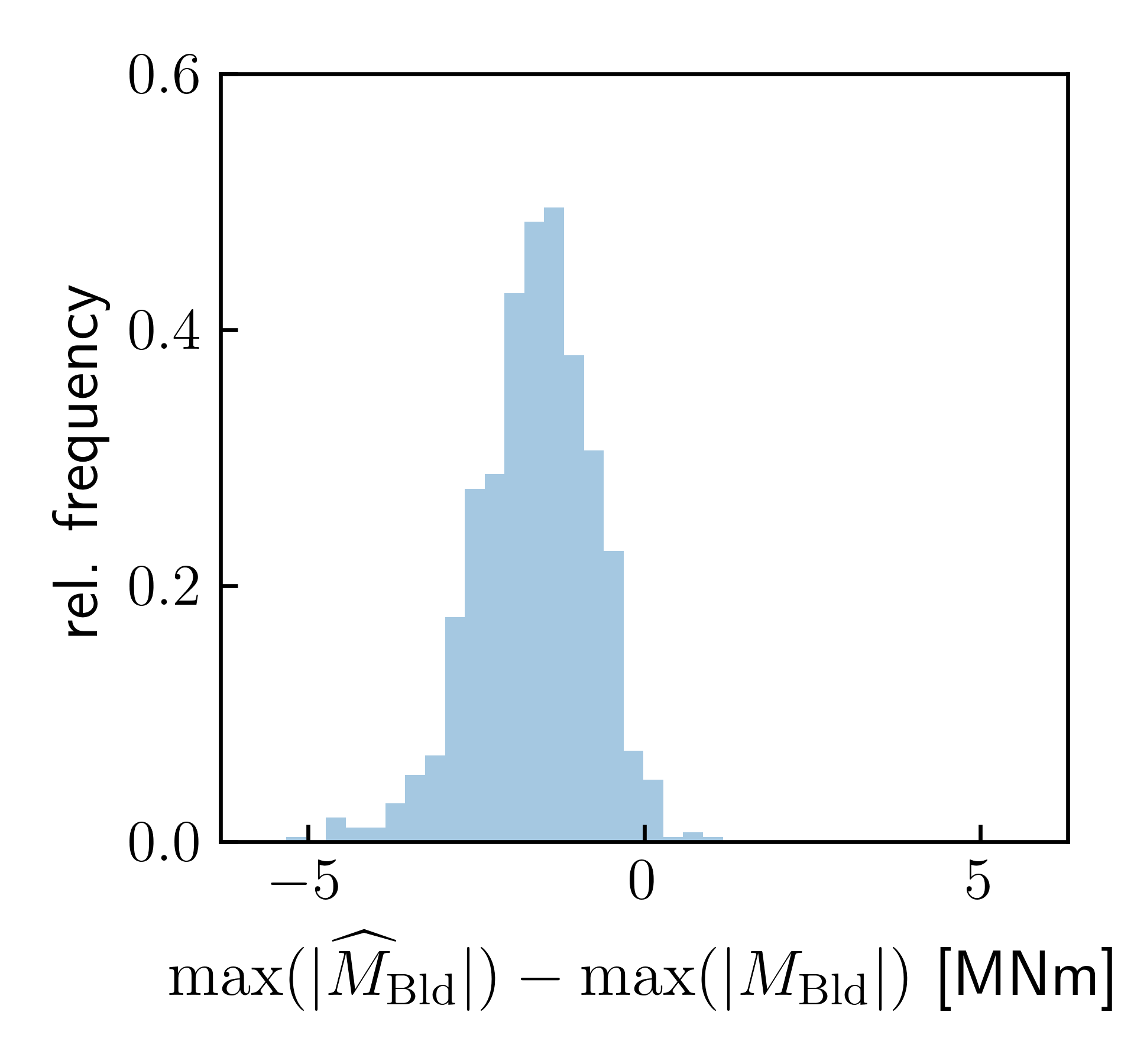}\\
        \includegraphics[width=\linewidth]{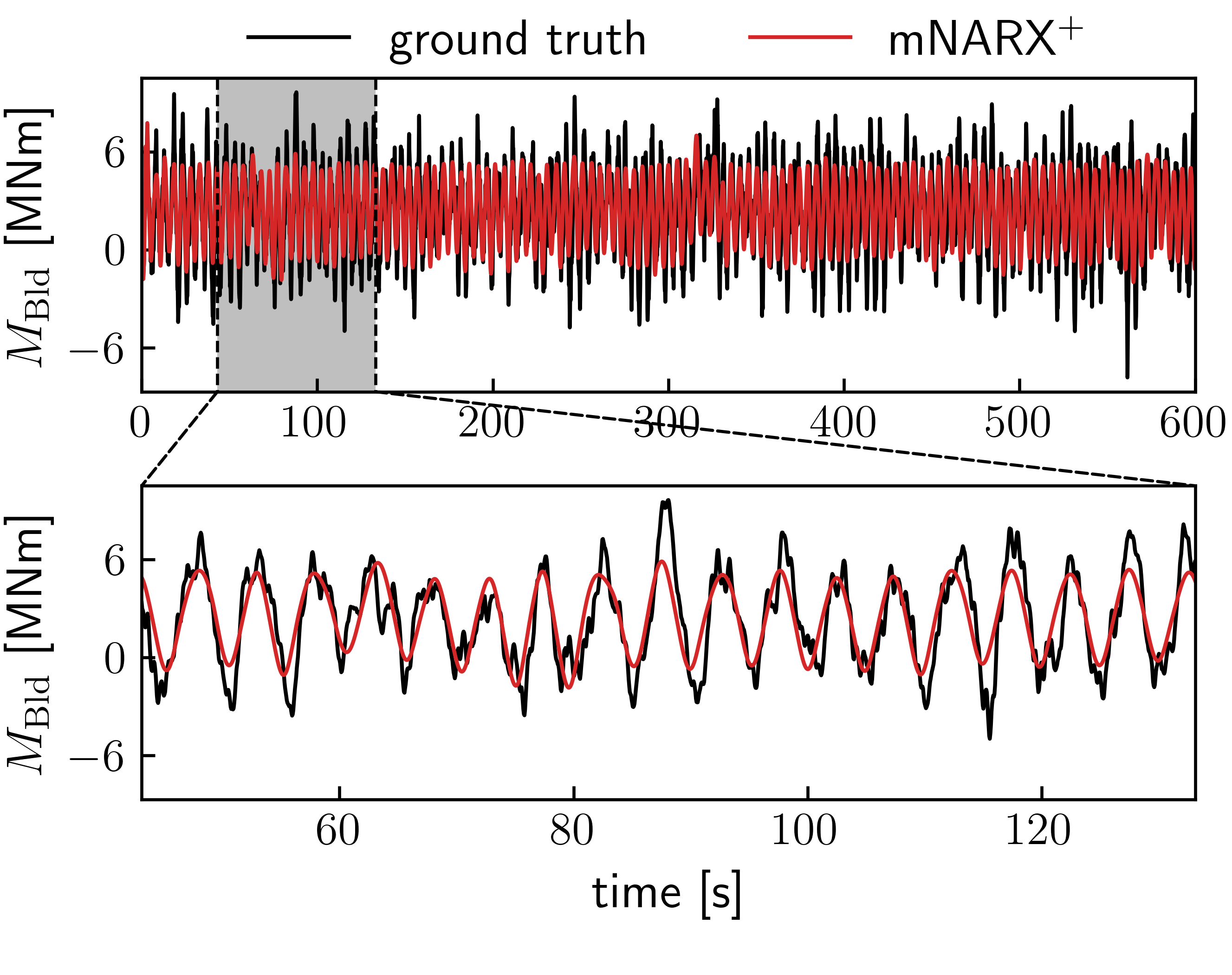}
    \end{minipage}
    \caption{
        \addnew{mNARX\textsuperscript{+} results.}
        (Top left) Histogram of the root-mean-squared error (RMSE) in MNm for the \textbf{blade root bending moment} prediction
        (Top right) Histogram of the difference between the true and predicted peak absolute \textbf{blade root bending moment}.
        (Bottom left) Trace corresponding to the lowest RMSE. 
        (Bottom right) Trace corresponding to the highest RMSE. 
    }
    \label{fig:blade_moment_results}
\end{figure}

The performance of the surrogate for the intermediate responses (rotor speed and blade pitch) and the main quantity of interest (flapwise blade root bending moment) is presented in the Figures~\ref{fig:blade_pitch_and_rotor_speed_results}--\ref{fig:blade_moment_results}. 

For each of the two intermediate responses, the distribution of the root-mean-squared error (RMSE) computed over all traces in the test dataset is shown in the top subplots in Figure~\ref{fig:blade_pitch_and_rotor_speed_results}. 
For both quantities, the RMSE is very low, with the worst-case RMSE being approximately $1.5$ degrees for the blade pitch and $0.1$ revolutions per minute for the rotor speed.

The good accuracy indicated by the RMSE histograms is further supported by the example traces shown in the bottom-left and bottom-right subplots in Figure~\ref{fig:blade_pitch_and_rotor_speed_results}. 
The displayed traces correspond to those with the lowest and highest RMSE out of $909$ test traces, respectively. 
While the blade pitch prediction with the lowest RMSE is extremely accurate, the prediction with the highest error shows moderate deviations from the ground truth.
A similar pattern can be seen in the results for the rotor speed. 
While the prediction with the lowest error is remarkably accurate, the prediction with the highest error visibly deviations from the ground truth but remains stable.

The results for the flapwise blade root bending moments are presented in Figure~\ref{fig:blade_moment_results}.
In the top-left subplot, the histogram of the RMSE on the test dataset is shown.
The top-right subplot displays the distribution of the discrepancy in the peak blade moment: $\max(|\widehat{M}_\text{Bld}|) - \max(|M_\text{Bld}|)$.
While these two plots indicate that the blade moment is generally predicted well, the top-left plot also reveals that the surrogate is relatively inaccurate in a few cases.
The discrepancy in the peak moment further suggests a tendency of the surrogate to underestimate the peak values.

The bottom subplots of Figure~\ref{fig:blade_moment_results} show the traces with the lowest and highest RMSE.
While the prediction corresponding to the lowest error shows very good agreement with the true output, the prediction with the highest error clearly deviates from the ground truth.
In the worst case, the surrogate appears unable to capture both the strong low-frequency oscillations and the higher-frequency signal components at the same time.
While the inability to represent the full extent of the low-frequency oscillation may be mitigated by a larger experimental design, the failure to capture the higher-frequency content suggests that the model lacks relevant information.
The seven temporal features included in this model (see Figure~\ref{fig:wind_turbine_feature_selection}) may be insufficient, and additional higher-frequency features might be required to improve accuracy.

\addnew{

The results for the LSTM are presented in Figure~\ref{fig:blade_moment_results_lstm}, following the same format as in Figure~\ref{fig:blade_moment_results}: the top-left subplot shows the histogram of the RMSE error on the test dataset, while the top right subplots shows the distribution of the error in the peak blade moment. 
The traces corresponding to the lowest and highest RMSE are depicted in the bottom-left and bottom-right subplots, respectively. 

Figures~\ref{fig:blade_moment_results_lstm} demonstrates that, while the LSTM successfully learns the general trend in the data, it fails to capture most of the higher-frequency oscillatory behavior. 
These oscillations arise from the interplay between wind speeds, rotor orientation and blade pitch control. 
Such interplay is difficult to learn directly from the raw input, even with a powerful learner such as the LSTM model. 
In contrast, the mNARX\textsuperscript{+} framework can leverage the auxiliary model responses to enable even a simple learner such as polynomials to achieve remarkably better performance.

\begin{figure}
    \centering
    \begin{minipage}{0.49\linewidth}
        \centering
        \includegraphics[width=0.65\linewidth]{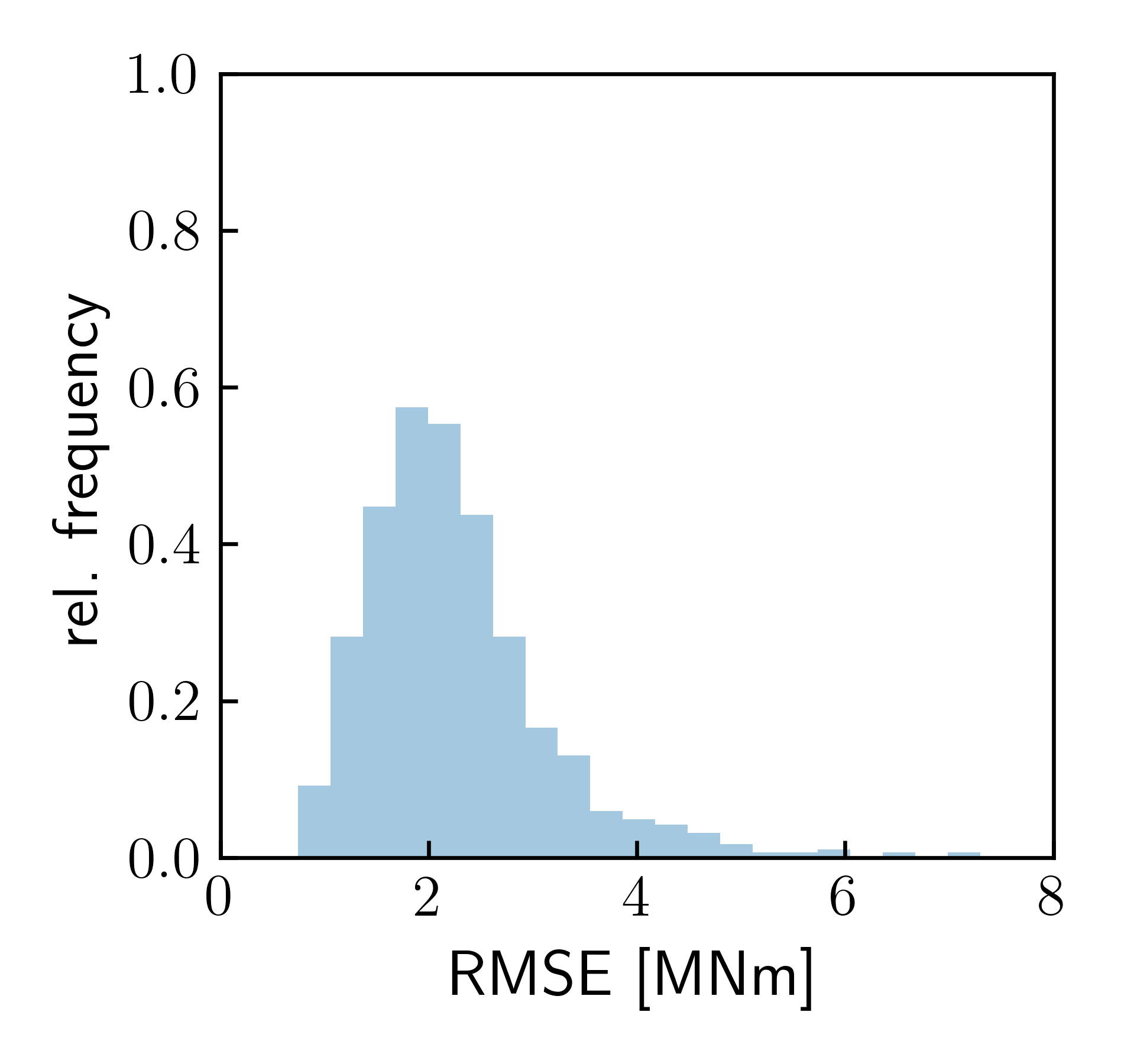}\\
        \includegraphics[width=\linewidth]{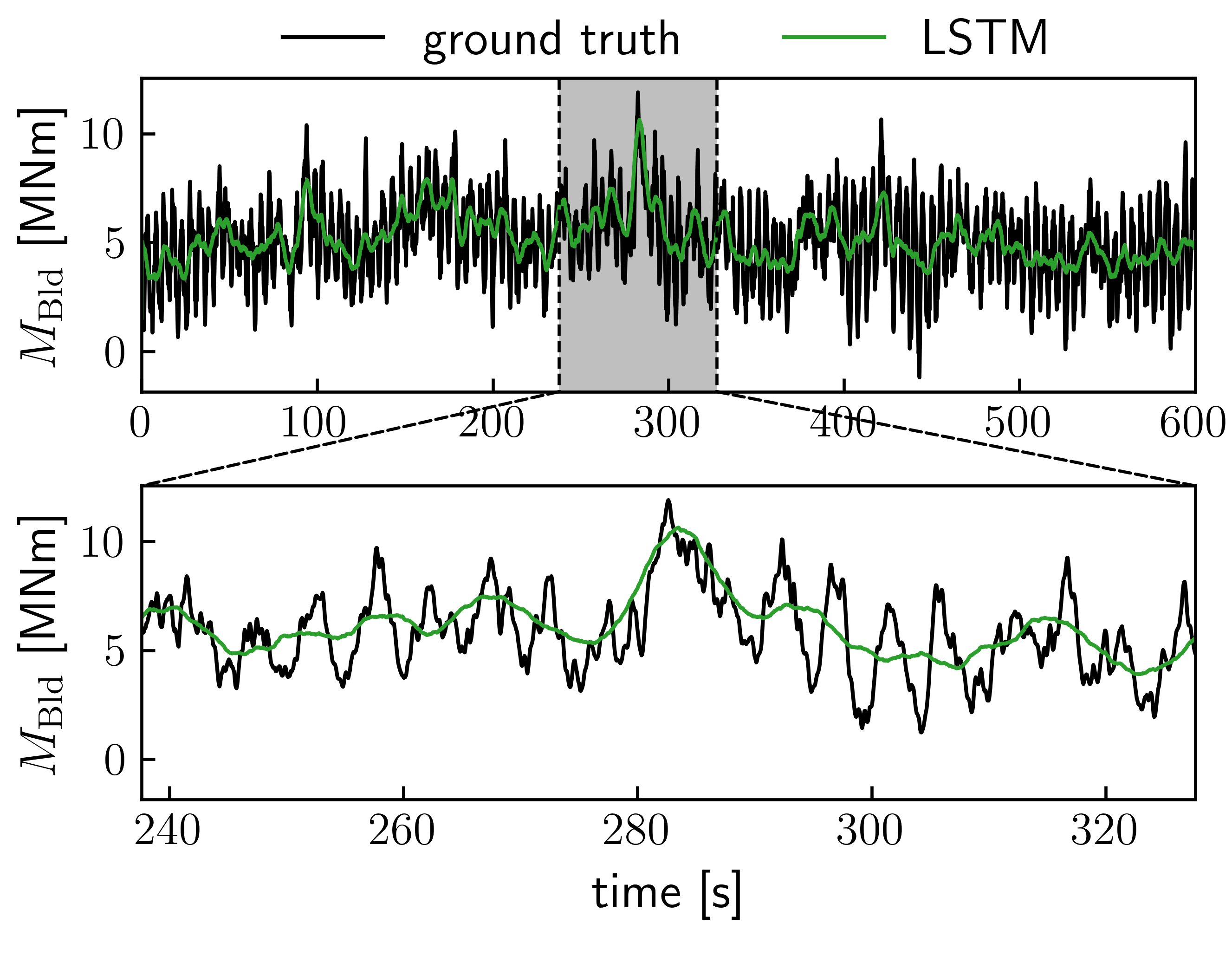}
    \end{minipage}
    \begin{minipage}{0.49\linewidth}
        \centering
        \includegraphics[width=0.65\linewidth]{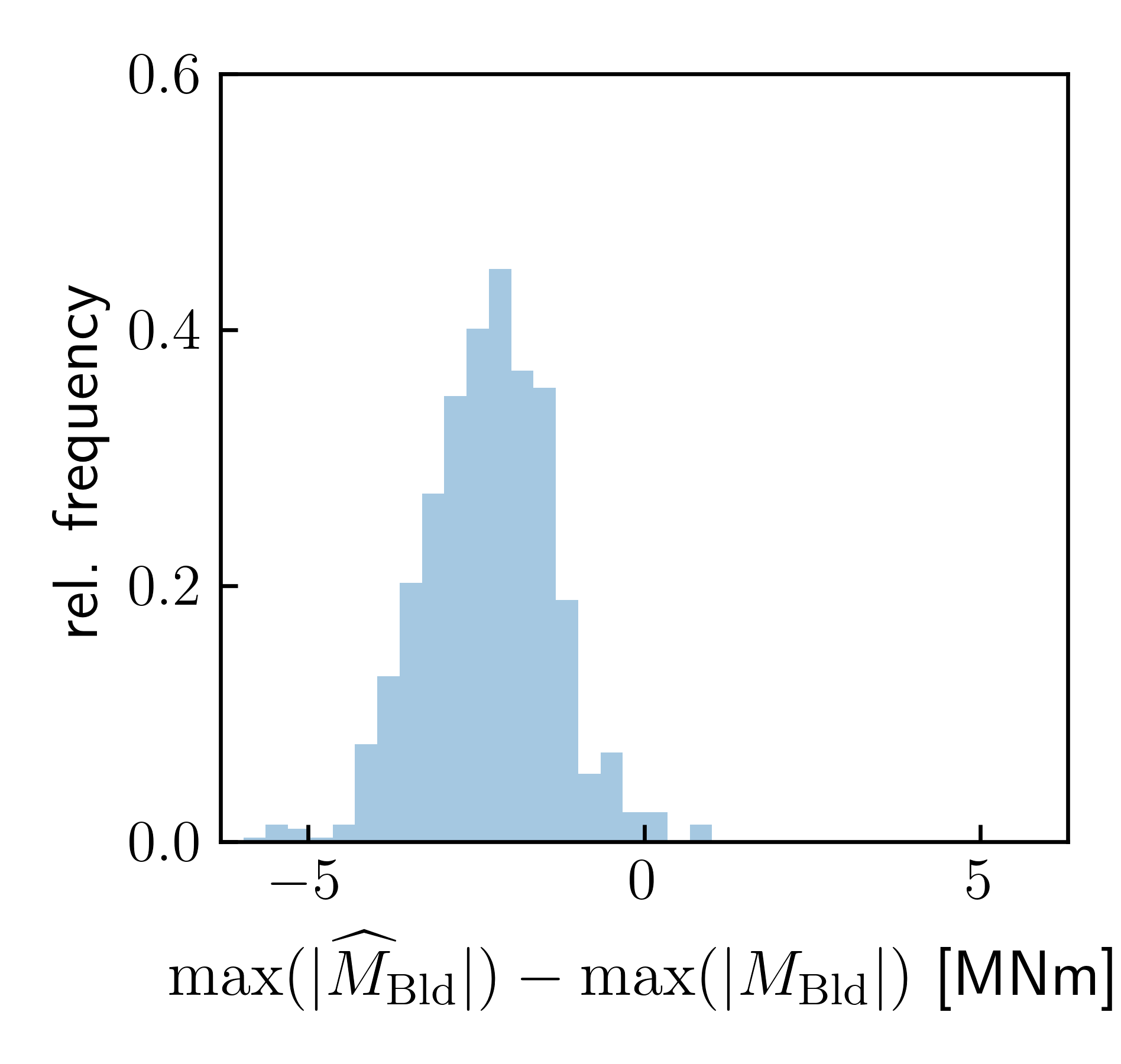}\\
        \includegraphics[width=\linewidth]{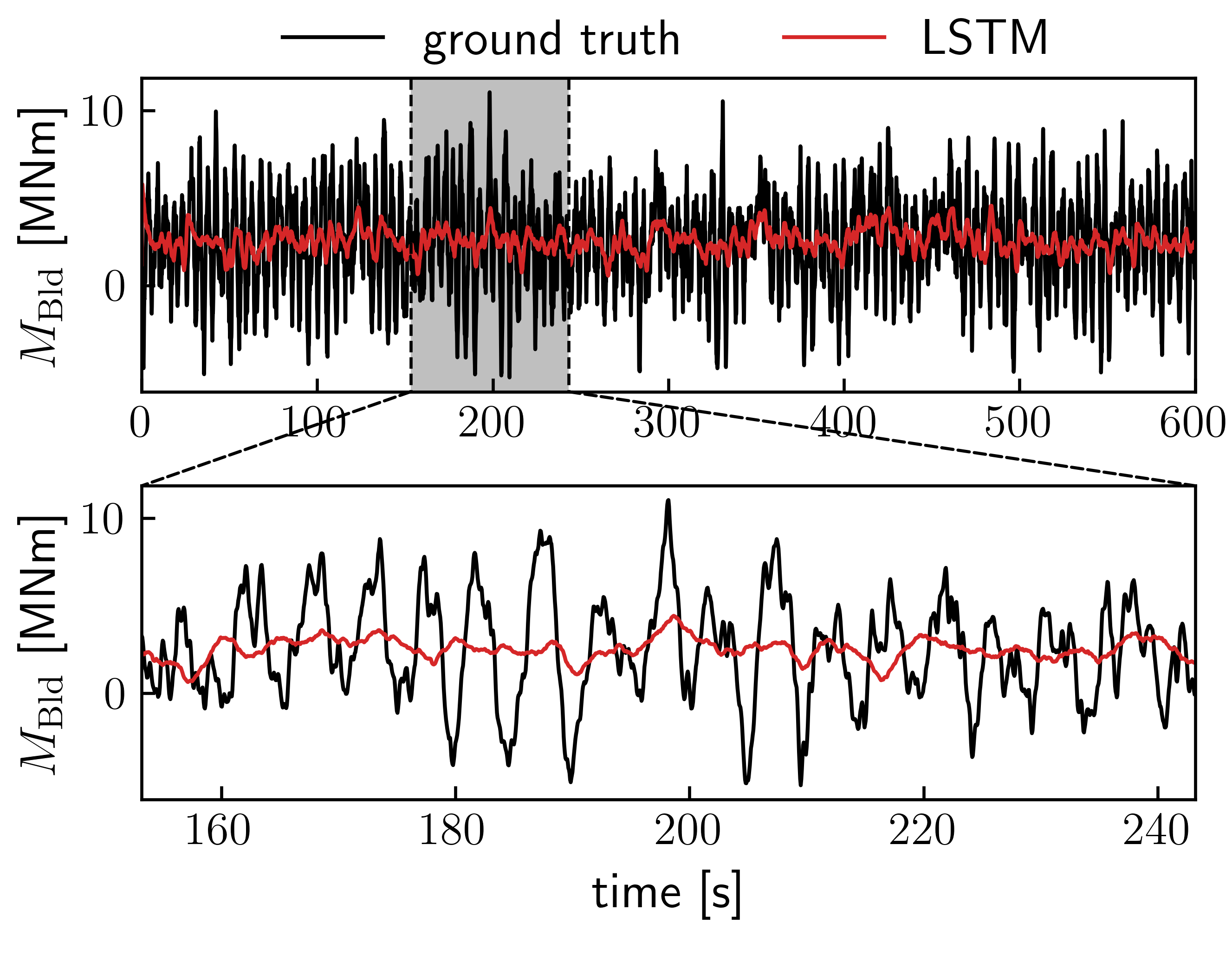}
    \end{minipage}
    \caption{
        \addnew{LSTM results.}
        \addnew{(Top left) Histogram of the root-mean-squared error (RMSE) in MNm for the \textbf{blade root bending moment} prediction
        (Top right) Histogram of the difference between the true and predicted peak absolute \textbf{blade root bending moment}.
        (Bottom left) Trace corresponding to the lowest RMSE. 
        (Bottom right) Trace corresponding to the highest RMSE. }
    }
    \label{fig:blade_moment_results_lstm}
\end{figure}

An important observation that complements this performance comparison of mNARX\textsuperscript{+} and LSTMs, is that the two approaches are not mutually exclusive. 
Indeed, the predictive performance of mNARX\textsuperscript{+} in this study is strongly influenced by the choice of polynomial NARX submodels, which are arguably one of the simplest available learners, known to have comparatively limited expressive power.  
Nonetheless, the mNARX\textsuperscript{+} layer elevates their performance enough to make them outperform a much more powerful learner like LSTMs in such a complex engineering scenario. 
In practice, the mNARX\textsuperscript{+} framework itself can be seen as an algorithmic tool to incorporate additional information in classical time-series emulation.
In particular, it is not restricted to, nor does it synergize particularly well with polynomial NARX submodels, and it can easily be coupled with more advanced learners such as Gaussian process-based NARX models [19] or even LSTM models, by simply substituting them in place of the polynomial NARX model in the \textit{Model update} step in Figure~\ref{fig:auto_mnarx}.
}

\section{Discussion and Conclusion}\label{sec:discussion_and_conclusion}
In this paper, we introduced mNARX\textsuperscript{+}, an automatic manifold nonlinear autoregressive with exogenous inputs (mNARX) modeling approach, which integrates the strengths of both mNARX modeling \citep{schaer_mnarx_2024}, and functional-NARX ($\mathcal{F}$-NARX) modeling \citep{schaer_fnarx_2025}, while reducing the prior knowledge required by the original mNARX approach. 

The key strength of mNARX\textsuperscript{+} is the automation of the mNARX construction. 
By employing a data-driven feature selection step based on residual correlation, the algorithm reduces the dependence on prior expert knowledge for identifying crucial auxiliary quantities and their causal relationships, which can be a bottleneck in manual mNARX modeling. 
The recursive nature of the algorithm ensures that all necessary intermediate models are constructed causally before their outputs (or features thereof) are used, respecting potential dependencies. 

The feature selection at each step in the algorithm is greedy and chooses the feature most correlated with the current residuals. 
While inspired by effective methods like LASSO and LARS, it cannot guarantee finding the globally optimal set of features or model structure. 
However, it offers a pragmatic approach to progressively improve the model by focusing on the currently unexplained output variance. 

The performance of the algorithm depends on several hyperparameters, including the choice of correlation measure, the complexity of the individual polynomial $\mathcal{F}$-NARX models (\eg polynomial degree), and the termination criteria. 
Appropriate tuning of these parameters is necessary for optimal results. 
Furthermore, while the algorithm automates the feature selection from a pool of candidate features, its success still relies on the user providing a sufficiently rich set of candidate auxiliary quantities during initialization from which temporal features can be extracted.

To evaluate the effectiveness of the \replace{automatic mNARX}{mNARX\textsuperscript{+}} modeling algorithm, we applied it to two case studies: a simple Bouc-Wen oscillator with hysteresis, and a highly complex aero-servo-elastic wind turbine simulator.
In the Bouc-Wen oscillator case study, the algorithm successfully identified the hysteretic displacement as an intermediate response, which improves prediction accuracy while maintaining forecast stability. 
Using only simple polynomial NARX base models the mNARX\textsuperscript{+} showed performance comparable to a powerful learning in the form of an LSTM neural network.
\addnew{
    In the wind turbine case study, the algorithm identified the blade pitch and rotor speed as intermediate responses which helped to predict the main blade moment as the main quantitiy of interest.
    By leveraging the intermediate responses the mNARX\textsuperscript{+} model can even significantly outperformed powerful learners, such as LSTM neural networks, even when relying on relatively weak learners.
}

The results demonstrate that \replace{automatic mNARX}{mNARX\textsuperscript{+}} modeling provides a systematic and data-driven approach to surrogate modeling for complex dynamical systems. 
The algorithm preserves most of the interpretability and efficiency of mNARX models, while alleviating the burden of manual feature selection and model configuration. 
Moreover, the correlation-based feature selection ensures that only the most relevant features are retained, further improving model stability and accuracy. 
By partially automating the mNARX model construction, we boost its potential to accelerate research in surrogate modeling and facilitate its broader adoption in industry applications.

\section*{Acknowledgments}
This project is part of the HIghly advanced Probabilistic design and Enhanced Reliability methods
for high-value, cost-efficient offshore WIND (HIPERWIND) project and has received funding from the European Union's Horizon 2020 Research and Innovation Programme under Grant Agreement No. 101006689.

\appendix

\section{Implementation details}\label{app:implementation_details}
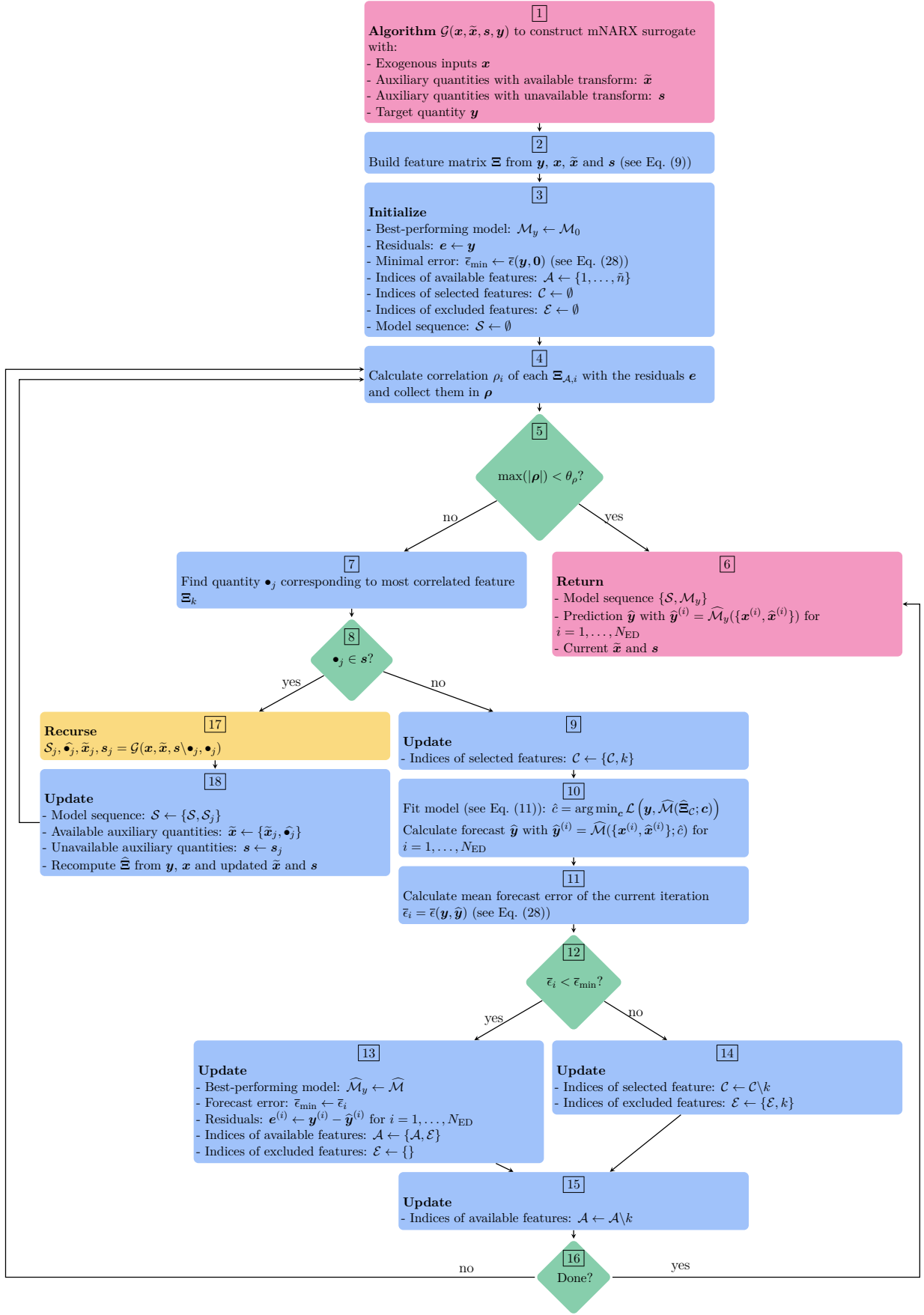
\begin{figure}[htb!]
    \centering
    \def\nodedistance{0.25cm}
    \def\boxwidth{9.7cm}
    \scalebox{0.60}{
        \begin{tikzpicture}[node distance=2.5cm]
        
        \node[autonumbered node] (start) [startstop, text width=\boxwidth, anchor=north, align=flush left] {\nodelabel{node:start}\\[4mm]
            \textbf{Algorithm} $\mathcal{G}(\vec{x}, \widetilde{\vec{x}}, \vec{s}, \vec{y})$ to construct mNARX surrogate with: \\
            - Exogenous inputs $\vec{x}$ \\
            - Auxiliary quantities with available transform: $\widetilde{\vec{x}}$ \\
            - Auxiliary quantities with unavailable transform: $\vec{s}$ \\
            - Target quantity $\vec{y}$ \\
        };
        
        \node[autonumbered node] (assemble_xi) [process, below=\nodedistance of start.south, anchor=north, text width=\boxwidth, align=flush left] {\nodelabel{node:assemble_xi}\\[4mm]
            Build feature matrix $\vec{\Xi}$ from $\vec{y}$, $\vec{x}$, $\widetilde{\vec{x}}$ and $\vec{s}$ (see Eq.~\eqref{eq:feature_matrix})
        };
        \draw [arrow] (start) -- (assemble_xi);
        
        \node[autonumbered node] (init) [process, below=\nodedistance of assemble_xi.south, anchor=north, text width=\boxwidth, align=flush left] {\nodelabel{node:init}\\[4mm]
            \textbf{Initialize} \\
            - Best-performing model: $\mathcal{M}_y \leftarrow \mathcal{M}_0$ \\ 
            - Residuals: $\vec{e} \leftarrow \vec{y}$ \\
            - Minimal error: $\overline{\epsilon}_\text{min} \leftarrow \overline{\epsilon}(\vec{y}, \vec{0})$ (see Eq.~\eqref{eq:mean_forecast_error})\\
            - Indices of available features: $\mathcal{A} \leftarrow \{1, \dots, \tilde{n}\}$ \\
            - Indices of selected features: $\mathcal{C} \leftarrow \emptyset$ \\
            - Indices of excluded features: $\mathcal{E} \leftarrow \emptyset$ \\
            - Model sequence: $\mathcal{S} \leftarrow \emptyset$ \\
        };
        \draw [arrow] (assemble_xi) -- (init);
        
        \node[autonumbered node] (calculate_rho) [process, below=\nodedistance of init.south, anchor=north, text width=\boxwidth, align=flush left] {\nodelabel{node:calculate_rho}\\[4mm]
            Calculate correlation $\rho_i$ of each $\vec{\Xi}_{\mathcal{A}, i}$ with the residuals $\vec{e}$ and collect them in $\vec{\rho}$
        };
        \draw [arrow] (init) -- (calculate_rho);
        
        \node[autonumbered decision node] (algo_rho_cond) [decision, below=\nodedistance of calculate_rho.south, anchor=north, align=center] {\nodelabel{node:algo_rho_cond}
            $\max(|\vec{\rho}|) < \theta_\rho$?
        };
        \draw [arrow] (calculate_rho) -- (algo_rho_cond);
        
        \node[autonumbered node] (algo_break) [startstop, below=\nodedistance of algo_rho_cond.south, anchor=north, text width=\boxwidth, align=flush left, xshift=5.5cm] {\nodelabel{node:algo_break}\\[4mm]
            \textbf{Return} \\
            - Model sequence $\{ \mathcal{S}, \mathcal{M}_y \}$ \\
            - Prediction $\widehat{\vec{y}}$ with $\widehat{\vec{y}}^{(i)}=\widehat{\mathcal{M}}_y(\{ \vec{x}^{(i)}, \widehat{\vec{x}}^{(i)} \})$ for $i=1, \dots, N_\text{ED}$ \\
            - Current $\widetilde{\vec{x}}$ and $\vec{s}$
        };
        \draw [arrow] (algo_rho_cond) -- node[anchor=south] {\large yes} (algo_break);
        
        \node[autonumbered node] (find_x) [process, below=\nodedistance of algo_rho_cond.south, anchor=north, text width=\boxwidth, align=flush left, xshift=-5.5cm] {\nodelabel{node:find_x}\\[4mm]
            Find quantity $\bullet_j$ corresponding to most correlated feature ${\vec\Xi}_{k}$
        };
        \draw [arrow] (algo_rho_cond) -- node[anchor=south] {\large no} (find_x);
        
        \node[autonumbered decision node] (check_dependent) [decision, below=\nodedistance of find_x.south, anchor=north, align=center] {\nodelabel{node:check_dependent}
            $\bullet_j \in \vec{s}$?
        };
        \draw [arrow] (find_x) -- (check_dependent);
        
        \node[autonumbered node] (update) [process, below=\nodedistance of check_dependent.south, text width=\boxwidth, align=flush left, xshift=6.5cm, anchor=north] {\nodelabel{node:update}\\[4mm]
            \textbf{Update} \\
            - Indices of selected features: $\mathcal{C} \leftarrow \{\mathcal{C}, k\}$ \\
        };
        \draw [arrow] (check_dependent) -- node[anchor=south] {\large no} (update);
        
        \node[autonumbered node] (eval_surrogate) [process, below=\nodedistance of update.south, anchor=north, text width=\boxwidth, align=flush left, xshift=0cm] {\nodelabel{node:eval_surrogate}\\[4mm]
            Fit model (see Eq.~\eqref{eq:loss_minimization}):
            $\hat{c} = \mathop{\arg\min}_{\vec{c}} \mathcal{L}\left(\vec{y}, \widehat{\mathcal{M}}(\widehat{\vec{\Xi}}_\mathcal{C}; \vec{c})\right)$ \\
            Calculate forecast $\widehat{\vec{y}}$ with
            $\widehat{\vec{y}}^{(i)}=\widehat{\mathcal{M}}(\{ \vec{x}^{(i)}, \widehat{\vec{x}}^{(i)} \}; \hat{c})$ for $i=1, \dots, N_\text{ED}$
        };
        \draw [arrow] (update) -- (eval_surrogate);
        
        \node[autonumbered node] (calc_error) [process, below=\nodedistance of eval_surrogate.south, anchor=north, text width=\boxwidth, align=flush left, xshift=0cm] {\nodelabel{node:calc_error}\\[4mm]
            Calculate mean forecast error of the current iteration $\overline{\epsilon}_{i} = \overline{\epsilon}(\vec{y}, \widehat{\vec{y}})$ (see Eq.~\eqref{eq:mean_forecast_error})
        };
        \draw [arrow] (eval_surrogate) -- (calc_error);
        
        \node[autonumbered decision node] (check_forecast_error) [decision, below=\nodedistance of calc_error.south, xshift=0cm, anchor=north, align=center] {\nodelabel{node:check_forecast_error}
            $\overline{\epsilon}_{i} < \overline{\epsilon}_\text{min}$?
        };
        \draw [arrow] (calc_error) -- (check_forecast_error);
        
        \node[autonumbered node] (reduced_error) [process, below=\nodedistance of check_forecast_error.south, text width=\boxwidth, align=flush left, xshift=-6cm, anchor=north] {\nodelabel{node:reduced_error}\\[4mm]
            \textbf{Update} \\
            - Best-performing model: $\widehat{\mathcal{M}}_{y} \leftarrow \widehat{\mathcal{M}}$ \\
            - Forecast error: $\overline{\epsilon}_\text{min} \leftarrow \overline{\epsilon}_{i}$ \\
            - Residuals: $\vec{e}^{(i)} \leftarrow \vec{y}^{(i)}-\widehat{\vec{y}}^{(i)}$ for $i=1, \dots, N_\text{ED}$ \\
            - Indices of available features: $\mathcal{A} \leftarrow \{ \mathcal{A}, \mathcal{E} \}$\\
            - Indices of excluded features: $\mathcal{E} \leftarrow \{\}$
        };
        \draw [arrow] (check_forecast_error) -- node[anchor=east] {\large yes} (reduced_error);
        
        \node[autonumbered node] (increased_error) [process, below=\nodedistance of check_forecast_error.south, text width=\boxwidth, align=flush left, xshift=4.5cm, anchor=north] {\nodelabel{node:increased_error}\\[4mm]
            \textbf{Update} \\
            - Indices of selected feature: $\mathcal{C} \leftarrow \mathcal{C} \backslash k$ \\
            - Indices of excluded features: $\mathcal{E} \leftarrow \{\mathcal{E}, k\}$
        };
        \draw [arrow] (check_forecast_error) -- node[anchor=south] {\large no} (increased_error);
        
        \node[autonumbered node] (available_features) [process, below=\nodedistance of reduced_error.south, text width=\boxwidth, align=flush left, xshift=6cm, anchor=north] {\nodelabel{node:available_features}\\[4mm]
            \textbf{Update} \\
            - Indices of available features: $\mathcal{A} \leftarrow \mathcal{A} \backslash k$
        };
        \draw [arrow] (reduced_error) -- (available_features);
        \draw [arrow] (increased_error) -- (available_features);
        
        \node[autonumbered decision node] (stopping_crit) [decision, below=\nodedistance of available_features.south, xshift=0cm, anchor=north, align=center] {\nodelabel{node:stopping_crit}
            Done?
        };
        \draw [arrow] (available_features) -- (stopping_crit);
        \draw [arrow] 
            (stopping_crit.east) -- ++ (9,0) 
            node[pos=0.222, above] {\large yes} 
            |- (algo_break.east);
        
        \draw [arrow] 
            (stopping_crit.west) -- ++ (-15.5,0) 
            node[pos=0.129, above] {\large no} 
            |- ($(calculate_rho.west) + (0,0.15cm)$);

        \node[autonumbered node] (recurse) [recursion, below=\nodedistance of check_dependent.south, text width=\boxwidth, align=flush left, xshift=-4cm, anchor=north] {\nodelabel{node:recurse}\\[1mm]
            \textbf{Recurse} \\
            $
            \mathcal{S}_{j}, 
            \widehat{\bullet_j}, 
            \widetilde{\vec{x}}_j, 
            \vec{s}_j 
            = \mathcal{G}(
                \vec{x}, 
                \widetilde{\vec{x}},
                \vec{s} \backslash \bullet_j, 
                \bullet_j
            )
            $
        };
        \draw [arrow] (check_dependent) -- node[anchor=south] {\large yes} (recurse);
        
        \node[autonumbered node] (recurse_update) [process, below=\nodedistance of recurse.south, text width=\boxwidth, align=flush left, xshift=0cm, anchor=north] {\nodelabel{node:recurse_update}\\[4mm]
            \textbf{Update} \\
            - Model sequence: $\mathcal{S} \leftarrow \{ \mathcal{S}, \mathcal{S}_{j} \}$ \\
            - Available auxiliary quantities: $\widetilde{\vec{x}} \leftarrow \{ \widetilde{\vec{x}}_j, \widehat{\bullet_{j}} \}$ \\
            - Unavailable auxiliary quantities: $\vec{s} \leftarrow \vec{s}_j$ \\
            - Recompute $\widehat{\vec{\Xi}}$ from $\vec{y}$, $\vec{x}$ and updated $\widetilde{\vec{x}}$ and $\vec{s}$
        };
        \draw [arrow] (recurse) -- (recurse_update);
        \draw [arrow] (recurse_update.west) -- ++ (-0.6,0) |- ($(calculate_rho.west) + (0,-0.15cm)$);
        
        \end{tikzpicture}
    }
    \caption{Implementation details of the data-driven mNARX construction algorithm.}\label{fig:auto_mnarx_details}
\end{figure}

The implementation of the \emph{recursive}, \replace{data-driven automatic mNARX}{mNARX\textsuperscript{+}} construction algorithm described in Section~\ref{sec:automatic_mnarx_modeling} requires careful handling of causality and bookkeeping to avoid redundant calculations, circular dependencies and prevent infinite recursion.
An extended version of Figure~\ref{fig:auto_mnarx}, intended as a visual representation of the complete mNARX\textsuperscript{+} construction procedure designed to avoid these issues, is shown in the flowchart in Figure~\ref{fig:auto_mnarx_details}, with a detailed step-by-step explanation of each node provided below.

\begin{itemize}
    \item \emph{Node~\ref{node:start}}:
    This initial node establishes the objective: to construct an mNARX surrogate model for a specified target quantity. To execute the algorithm, an experimental design (ED) must be available, comprising the following time series data:
    \begin{itemize}
        \item The exogenous inputs of the system, denoted as $\vec{x}=\{ \vec{x}^{(i)}, \dots, \vec{x}^{(N_\text{ED})} \}$.
        \item The auxiliary quantities with predefined transform $\mathcal{F}_i$ (see Section~\ref{sec:mnarx_modeling}), denoted as $\widetilde{\vec{x}}=\{ \widetilde{\vec{x}}^{(i)}, \dots, \widetilde{\vec{x}}^{(N_\text{ED})} \}$.
        \item The intermediate responses, \ie, auxiliary quantities requiring a predictive model for $\mathcal{F}_i$, denoted as $\vec{s}=\{ \vec{s}^{(i)}, \dots, \vec{s}^{(N_\text{ED})} \}$.
        \item The target quantity, denoted as $\vec{y}=\{ \vec{y}^{(i)}, \dots, \vec{y}^{(N_\text{ED})} \}$.
    \end{itemize}
    To describe the recursion of the algorithm, the notation $\bullet = \mathcal{G}(\vec{x}, \widetilde{\vec{x}}, \vec{s}, \vec{y})$ is used to indicate that the algorithm is invoked with arguments $\vec{x}$, $\widetilde{\vec{x}}$, $\vec{s}$, and $\vec{y}$, and its returned values are stored in $\bullet$.

    \item \emph{Node~\ref{node:assemble_xi}}:
    In this initial primary step of the algorithm, a comprehensive set of temporal features is extracted from the experimental design  $\{ \vec{x}^{(i)}, \widetilde{\vec{x}}^{(i)}, \vec{s}^{(i)}, \vec{y}^{(i)} \}_{i=1}^{N_\text{ED}}$. 
    These features are gathered into the feature matrix $\vec{\Xi}$ (see Eq.~\eqref{eq:feature_matrix_exp_design}). 
    For the construction of $\vec{\Xi}$, the number of retained principal components (PCs) (see Section~\ref{sec:fnarx_modeling}), $\tilde{n}$, can be set to a large value, or even equal to the total number of PCs, $n$. 
    This is because the objective of the algorithm is to greedily select only the most predictive PCs, meaning those that are expected to most effectively improve the surrogate forecast. 
    However, by choosing $\tilde{n} < n$, the overall number of candidate PCs is reduced, thereby decreasing the runtime of the algorithm. This optimization is particularly beneficial when the $\mathcal{F}$-NARX model memory is relatively large (\eg, $\mathcal{O}(10^{2-3})$ time steps), leading to an extensive initial pool of candidate PCs.

    \item \emph{Node~\ref{node:init}}:
    During algorithm execution, precise tracking of model performance and causal relationships and dependencies is essential. For this purpose, a comprehensive set of bookkeeping variables is initialized and continuously updated throughout the process. These include:
    \begin{itemize}
        \item The best-performing model, denoted as $\mathcal{M}_y$. This model is initially set to a trivial model $\mathcal{M}_0$, which predicts $\widehat{y}(t) = 0 \; \forall \; t \in \mathcal{T}$.
        \item The prediction residuals, $\vec{e} = \{ \vec{e}^{(1)}, \dots, \vec{e}^{N_\text{ED}} \}$. These quantify the discrepancy between the target quantity and the corresponding surrogate forecast, calculated as $\vec{e}^{(i)} = \vec{y}^{(i)} - \widehat{\vec{y}}^{(i)}$. Due to the trivial initial model, the residuals are initially equal to $\vec{y}^{(i)}$.
        \item An error metric used to track the surrogate performance. The mean forecast error $\overline{\epsilon}$, as defined in Eq.~\eqref{eq:mean_forecast_error}, is employed for this purpose. The algorithm continuously tracks the best mean forecast error, $\overline{\epsilon}_\text{min}$, which is initially set to $\overline{\epsilon}(\vec{y}, \vec{0})$.
        \item A set of indices $\mathcal{A}$ that refers to the available features (\ie columns of $\vec{\Xi}$) from which the algorithm can select at any given iteration. Initially, this set contains the indices of all features, $1,\dots,\tilde{n}$, thereby referring to $\vec{\Xi}_1,\dots,\vec{\Xi}_{\tilde{n}}$.
        \item A set of indices $\mathcal{C}$ that refers to features actively selected by the algorithm. This set is initially empty and is populated as the algorithm progresses.
        \item A set of indices $\mathcal{E}$ for features that are temporarily excluded from selection by the algorithm. This set is initialized as empty.
        \item A set $\mathcal{S}$ of predictive models $\mathcal{F}_i$ for the intermediate responses. This set is initially empty.
    \end{itemize}
    
    \item \emph{Node~\ref{node:calculate_rho}}:
    After initializing all variables, the algorithm proceeds to evaluate the correlation $\rho_i \in \mathbb{R}$ between each candidate feature $\vec{\Xi}_{\mathcal{A}, i}$ and the current prediction residuals $\vec{e}$. To assess this correlation, an estimator $\mathcal{H}$ is employed, such as Kendall's tau \citet{kendall_1938} or Pearson's correlation coefficient \citep{pearson_1895}:
    \begin{equation}
        \rho_i = \mathcal{H}(\vec{\Xi}_i, \vec{e}).
    \end{equation}
    All computed $\rho_i$ values are then gathered into a vector $\vec{\rho}$.

    \item \emph{Node~\ref{node:algo_rho_cond}}:
    The feature $\vec{\Xi}_{k}$ with the maximum correlation $\rho_k$ to the residual, is identified as the most important in the current iteration. 
    In this decision node, this value is compared against a predefined threshold $\theta_\rho$. 
    If $\max(|\vec{\rho}|) < \theta_\rho$, it is assumed that no further features will significantly improve the surrogate forecast. In this scenario, the algorithm terminates and proceeds to Node~\ref{node:algo_break}. 
    Conversely, if a sufficiently correlated feature is identified, the algorithm continues its execution and transitions to Node~\ref{node:find_x}.
    
    \item \emph{Node~\ref{node:algo_break}}:
    Upon reaching Node~\ref{node:algo_break}, the algorithm stops. 
    It then returns the final $\mathcal{F}$-NARX model $\widehat{\mathcal{M}}_{y}$ for the target quantity, along with the complete model sequence $\mathcal{S}$ for all intermediate responses. 
    Additionally, the algorithm provides the prediction $\widehat{\vec{y}}^{(i)} = \widehat{\mathcal{M}}_y(\vec{x}^{(i)}, \vec{y}^{(i)})$ for each realization $i = 1, \dots, N_\text{ED}$, as well as the current states of $\widetilde{\vec{x}}$ and $\vec{s}$. 
    It should be noted that $\widetilde{\vec{x}}$ and $\vec{s}$ may have been modified during the algorithm execution, a process explained in later nodes. 
    
    \item \emph{Node~\ref{node:find_x}}:
    In this step, the quantity $\bullet_j$ from which the most correlated feature $\vec{\Xi}_k$ originates is identified. Following this identification, the algorithm continues to Node~\ref{node:check_dependent}.

    \item \emph{Node~\ref{node:check_dependent}}:
    Depending on whether the identified quantity $\bullet_j$ is a not-yet-available intermediate response, the algorithm branches. 
    If it is unavailable, the process moves to Node~\ref{node:recurse}, which initiates a recursion of the algorithm to make $\bullet_j$ available. 
    The algorithm proceeds to Node~\ref{node:update} in the case that the most correlated feature $\vec{\Xi}_k$ corresponds to an already available auxiliary quantity $\widetilde{\vec{x}}$, an exogenous input $\vec{x}$, or the autoregressive component $\vec{y}$

    \item \emph{Node~\ref{node:update}}:
    In this node the index $k$ corresponding to the most correlated feature $\vec{\Xi}_k$ is added to the set of indices of selected features $\mathcal{C}$.

    \item \emph{Node~\ref{node:eval_surrogate}}:
    With the newly included feature, an $\mathcal{F}$-NARX model, denoted as $\widehat{\mathcal{M}}$, is trained on the expanded feature set for the target quantity. 
    Subsequently, its forecast is evaluated using the experimental design data. These resulting predictions are denoted as $\widehat{\vec{y}}$.

    \item \emph{Node~\ref{node:calc_error}}:
    With the obtained predictions, the accuracy of the surrogate model is assessed by computing the mean forecast error, $\overline{\epsilon}$, as defined in Eq.~\eqref{eq:mean_forecast_error}, on the experimental design. This calculated error at the $i$-th iteration is denoted as $\overline{\epsilon}_{i}$.
    
    \item \emph{Node~\ref{node:check_forecast_error}}:
    Depending on whether an error reduction is achieved, specifically if $\overline{\epsilon}_{i} < \overline{\epsilon}_\text{min}$, the algorithm flow diverges. It either proceeds to Node~\ref{node:reduced_error} (to update the model with reduced error) or to Node~\ref{node:increased_error} (to handle an increased error).

    \item \emph{Node~\ref{node:reduced_error}}:
    In the event that an error reduction is achieved:
    \begin{itemize}
        \item The best-performing surrogate $\widehat{\mathcal{M}}_y$ is replaced by the most recently trained surrogate $\widehat{\mathcal{M}}$.
        \item The minimal achieved error $\overline{\epsilon}_\text{min}$ is updated to the current error $\overline{\epsilon}_{i}$.
        \item New forecast residuals $\vec{e}$ are computed.
        \item The indices of previously excluded features $\mathcal{E}$, if available, are cleared and re-added to the indices of available features $\mathcal{A}$. This is done because the surrogate model has received new information, which may now enable it to benefit from features previously found unhelpful.
    \end{itemize}

    \item \emph{Node~\ref{node:increased_error}}:
    In the case where the forecast error increased, the latest index $k$ is removed from the set of selected features $\mathcal{C}$, and added to the set of excluded features $\mathcal{E}$. 
    When a feature is excluded because it does not improve the model forecast, it is temporarily also kept out from the pool of available features to prevent its reselection in subsequent iterations. 
    However, excluded features become available for selection again after a subsequent suitable feature that improves the model forecast has been found (see Node~\ref{node:reduced_error}). 
    
    \item \emph{Node~\ref{node:available_features}}:
    In this step, the index $k$ of the most correlated feature $\vec{\Xi}_{k}$ that was just processed is removed from the set of available features $\mathcal{A}$. 
    At this point, the index $k$ resides either in the set of excluded features $\mathcal{E}$ or the set of selected features $\mathcal{C}$.

    \item \emph{Node~\ref{node:stopping_crit}}:
    After updating all bookkeeping variables, the algorithm checks for termination. 
    The algorithm will certainly stop if no more features are available (\ie $\text{card}(\mathcal{A})=0$) but also additional termination criteria can be employed. 
    These may include a maximum number of iterations, a maximal duration if a computational budget is limited, or an error threshold, where the algorithm stops if the error $\overline{\epsilon}_\text{min}$ is sufficiently small. 
    It is also important to recall that the algorithm can terminate earlier, specifically after Node~\ref{node:algo_rho_cond}, if no sufficiently correlated feature remains.
    
    \item \emph{Node~\ref{node:recurse}}:
    If this node is reached, it signifies a challenging scenario: the quantity $\bullet_j$, corresponding to the most correlated feature $\vec{\Xi}_k$, is a not-yet-available intermediate response. 
    This means that, unlike exogenous inputs or already-modeled auxiliary quantities, $\bullet_j$ cannot be directly obtained during the prediction phase for unseen data. 
    Therefore, to ensure that $\vec{\Xi}_k$ (and thus $\bullet_j$) can be computed when making predictions, a predictive model for $\bullet_j$ must first be constructed. 
    This predictive model will be an $\mathcal{F}$-NARX model.

    To achieve this, the algorithm initiates a recursive call, effectively treating $\bullet_j$ as a temporary target quantity for which its own surrogate model needs to be built. This recursive call is invoked with the following inputs for the sub-problem:
    \begin{itemize}
        \item The exogenous inputs $\vec{x}$.
        \item The available auxiliary quantities $\widetilde{\vec{x}}$.
        \item The set of intermediate responses $\vec{s}$, explicitly excluding $\bullet_j$. This exclusion is critical because $\bullet_j$ is now the target of the recursive call, and its model is currently being constructed.
        \item The quantity $\bullet_j$ itself, which becomes the new target quantity for this recursion.
    \end{itemize}
    The outputs returned from this recursive step are:
    \begin{itemize}
        \item The model sequence $\mathcal{S}_j$ that leads to the prediction of $\bullet_j$.
        \item The prediction of the intermediate response, denoted as $\widehat{\bullet}_j$.
        \item The auxiliary quantities that are available at the termination of this recursive step, $\widetilde{\vec{x}}_j$.
        \item The intermediate responses that remain unavailable at the end of the recursion, $\vec{s}_j$.
    \end{itemize}

    \item \emph{Node~\ref{node:recurse_update}}:
    Upon returning from the recursive call (from Node~\ref{node:recurse}), the algorithm performs several critical updates:
    \begin{itemize}
        \item The parent model sequence $\mathcal{S}$ is updated by appending the child sequence $\mathcal{S}_j$ obtained from the recursion.
        \item The predicted quantity $\widehat{\bullet}_j$ is added to the set of available auxiliary quantities $\widetilde{\vec{x}}$. This is done because a model sequence $\mathcal{S}_j$ now exists to predict it, making it available for subsequent steps.
        \item The set of unavailable intermediate responses $\vec{s}$ is updated to $\vec{s}_j$, reflecting any changes or removals that occurred during the recursive process.
        \item The full feature matrix $\vec{\Xi}$ is recomputed by using the auxiliary response approximation in place of the true value. This recomputation is necessary because both $\widetilde{\vec{x}}$ and $\vec{s}$ have been altered. Previously, the true values of the intermediate response $\bullet_j$ were part of $\vec{s}$. However, after the recursive call, a predictive model for $\bullet_j$ has been constructed, and its \emph{predicted values} ($\widehat{\bullet}_j$) are now available. These predictions have been moved from the category of unavailable intermediate responses ($\vec{s}$) to the category of available auxiliary quantities ($\widetilde{\vec{x}}$). 
        This step significantly improves prediction stability on unseen data, as no \textit{true} auxiliary responses are available at that stage.
    \end{itemize}
    Due to the modification of $\vec{\Xi}$, the algorithm returns to Node~\ref{node:calculate_rho} to reassess the correlation of the features. This re-evaluation is crucial because, while initially feature $\vec{\Xi}_k$ (corresponding to $\bullet_j$) was correlated with the true $\bullet_j$, it is now associated with the prediction $\widehat{\bullet}_j$. Given that the prediction may be imperfect, the correlation with the residuals might have reduced and require a re-ranking of the features.
\end{itemize}

\bibliography{bibliography}

\begin{thebibliography}{}

\bibitem[\protect\citeauthoryear{Abbas, Zalkind, Pao, and Wright}{Abbas
  et~al.}{2022}]{rosco_2022}
Abbas, N.~J., D.~S. Zalkind, L.~Pao, and A.~Wright (2022).
\newblock A reference open-source controller for fixed and floating offshore
  wind turbines.
\newblock {\em Wind Energy Science\/}~{\em 7\/}(1), 53--73.

\bibitem[\protect\citeauthoryear{Atila and Spence}{Atila and
  Spence}{2025}]{atila2025}
Atila, H. and S.~M. Spence (2025).
\newblock Metamodeling of the response trajectories of nonlinear stochastic
  dynamic systems using physics-informed lstm networks.
\newblock {\em Journal of Building Engineering\/}~{\em 111}, 113447.

\bibitem[\protect\citeauthoryear{Azarhoosh and {Ilchi Ghazaan}}{Azarhoosh and
  {Ilchi Ghazaan}}{2025}]{azarhoosh_2025}
Azarhoosh, Z. and M.~{Ilchi Ghazaan} (2025).
\newblock A review of recent advances in surrogate models for uncertainty
  quantification of high-dimensional engineering applications.
\newblock {\em Computer Methods in Applied Mechanics and Engineering\/}~{\em
  433\/}(117508).

\bibitem[\protect\citeauthoryear{Bhattacharyya, Jacquelin, and
  Brizard}{Bhattacharyya et~al.}{2020}]{bhattacharyya_2020}
Bhattacharyya, B., E.~Jacquelin, and D.~Brizard (2020).
\newblock A {Kriging}–{NARX} model for uncertainty quantification of
  nonlinear stochastic dynamical systems in time domain.
\newblock {\em Journal of Engineering Mechanics\/}~{\em 146\/}(7).

\bibitem[\protect\citeauthoryear{Billings}{Billings}{2013}]{billings_2013}
Billings, S.~A. (2013).
\newblock {\em Nonlinear system identification: {NARMAX} methods in the time,
  frequency, and spatio-temporal domains}.
\newblock Chichester, West Sussex, United Kingdom: John Wiley \& Sons, Inc.

\bibitem[\protect\citeauthoryear{Blatman and Sudret}{Blatman and
  Sudret}{2010}]{Blatman_2010}
Blatman, G. and B.~Sudret (2010).
\newblock An adaptive algorithm to build up sparse polynomial chaos expansions
  for stochastic finite element analysis.
\newblock {\em Probabilistic Engineering Mechanics\/}~{\em 25\/}(2), 183--197.

\bibitem[\protect\citeauthoryear{Chandra, Matsagar, and Marburg}{Chandra
  et~al.}{2023}]{chandra_2025}
Chandra, S., V.~Matsagar, and S.~Marburg (2023).
\newblock Stochastic dynamic analysis of composite plates in thermal
  environments using nonlinear autoregressive model with exogenous input in
  polynomial chaos expansion surrogate.
\newblock {\em Computer Methods in Applied Mechanics and Engineering\/}~{\em
  416\/}(116303).

\bibitem[\protect\citeauthoryear{Cheng, Papaioannou, Lyu, and Straub}{Cheng
  et~al.}{2025}]{cheng_2025}
Cheng, K., I.~Papaioannou, M.~Lyu, and D.~Straub (2025).
\newblock State space {Kriging} model for emulating complex nonlinear dynamical
  systems under stochastic excitation.
\newblock {\em Computer Methods in Applied Mechanics and Engineering\/}~{\em
  442\/}(117987).

\bibitem[\protect\citeauthoryear{Chiras, Evans, and Rees}{Chiras
  et~al.}{2001}]{chiras_2001}
Chiras, N., C.~Evans, and D.~Rees (2001).
\newblock Nonlinear gas turbine modeling using {NARMAX} structures.
\newblock {\em IEEE Transactions on Instrumentation and Measurement\/}~{\em
  50\/}(4), 893--898.

\bibitem[\protect\citeauthoryear{Dassanayake, Mousa, Fowmes, Susilawati, and
  Zamara}{Dassanayake et~al.}{2023}]{Dassanayake_2023}
Dassanayake, S., A.~Mousa, G.~J. Fowmes, S.~Susilawati, and K.~Zamara (2023).
\newblock Forecasting the moisture dynamics of a landfill capping system
  comprising different geosynthetics: A {NARX} neural network approach.
\newblock {\em Geotextiles and Geomembranes\/}~{\em 51\/}(1), 282--292.

\bibitem[\protect\citeauthoryear{Deshmukh and Allison}{Deshmukh and
  Allison}{2017}]{deshmukh_2017}
Deshmukh, A.~P. and J.~T. Allison (2017).
\newblock Design of dynamic systems using surrogate models of derivative
  functions.
\newblock {\em Journal of Mechanical Design\/}~{\em 139\/}(10), 101402.

\bibitem[\protect\citeauthoryear{Efron, Hastie, Johnstone, and
  Tibshirani}{Efron et~al.}{2004}]{Efron_2004}
Efron, B., T.~Hastie, I.~Johnstone, and R.~Tibshirani (2004).
\newblock Least angle regression.
\newblock {\em The Annals of Statistics\/}~{\em 32\/}(2), 407 -- 499.

\bibitem[\protect\citeauthoryear{Gao, Liu, Li, and Liu}{Gao
  et~al.}{2016}]{Gao_2016}
Gao, Y., S.~Liu, F.~Li, and Z.~Liu (2016).
\newblock Fault detection and diagnosis method for cooling dehumidifier based
  on {LS-SVM} {NARX} model.
\newblock {\em International Journal of Refrigeration\/}~{\em 61}, 69--81.

\bibitem[\protect\citeauthoryear{Garg, Gupta, and Chakraborty}{Garg
  et~al.}{2022}]{garg_2022}
Garg, S., H.~Gupta, and S.~Chakraborty (2022).
\newblock Assessment of {DeepONet} for reliability analysis of stochastic
  nonlinear dynamical systems.
\newblock arXiv:2201.13145.

\bibitem[\protect\citeauthoryear{Hochreiter and Schmidhuber}{Hochreiter and
  Schmidhuber}{1997}]{hochreiter1997}
Hochreiter, S. and J.~Schmidhuber (1997).
\newblock {Long short-term memory}.
\newblock {\em Neural Computation\/}~{\em 9\/}(8), 1735--1780.

\bibitem[\protect\citeauthoryear{Hu, Fang, Zheng, Li, Gao, and Zhang}{Hu
  et~al.}{2024}]{Hu_2024}
Hu, Z., J.~Fang, R.~Zheng, M.~Li, B.~Gao, and L.~Zhang (2024).
\newblock Efficient model predictive control of boiler coal combustion based on
  {NARX} neutral network.
\newblock {\em Journal of Process Control\/}~{\em 134}, 103158.

\bibitem[\protect\citeauthoryear{IEC}{IEC}{2019}]{iec_2005}
IEC (2019).
\newblock International standard {IEC} 61400-1 ed. 4. wind turbines -- part 1:
  Design requirements.
\newblock {\em IEC 61400-1 Ed. 3\/}.

\bibitem[\protect\citeauthoryear{Jolliffe}{Jolliffe}{2002}]{Jolliffe_2002}
Jolliffe, I.~T. (2002).
\newblock {\em Principal component analysis}.
\newblock Springer-Verlag.

\bibitem[\protect\citeauthoryear{Jonkman}{Jonkman}{2009}]{turbsim_2009}
Jonkman, J. (2009).
\newblock Turbsim user's guide: Version 1.50.
\newblock Technical report, National Renewable Energy Laboratory.

\bibitem[\protect\citeauthoryear{Jonkman, Butterfield, Musial, and
  Scott}{Jonkman et~al.}{2009}]{nrel_onshore_2009}
Jonkman, J., S.~Butterfield, W.~Musial, and G.~Scott (2009).
\newblock Definition of a 5-{MW} reference wind turbine for offshore system
  development.
\newblock Technical report, National Renewable Energy Laboratory.

\bibitem[\protect\citeauthoryear{Kendall}{Kendall}{1938}]{kendall_1938}
Kendall, M.~G. (1938).
\newblock A new measure of rank correlation.
\newblock {\em Biometrika\/}~{\em 30}, 81--93.

\bibitem[\protect\citeauthoryear{Kim}{Kim}{2015}]{Kim_2015}
Kim, Y. (2015).
\newblock Prediction of the dynamic response of a slender marine structure
  under an irregular ocean wave using the {NARX}-based quadratic {Volterra}
  series.
\newblock {\em Applied Ocean Research\/}~{\em 49}, 42--56.

\bibitem[\protect\citeauthoryear{Langeron, Huynh, and Grall}{Langeron
  et~al.}{2021}]{langeron_2021}
Langeron, Y., K.~T. Huynh, and A.~Grall (2021).
\newblock A root location-based framework for degradation modeling of dynamic
  systems with predictive maintenance perspective.
\newblock {\em Proceedings of the Institution of Mechanical Engineers, Part O:
  Journal of Risk and Reliability\/}~{\em 235\/}(2), 253--267.

\bibitem[\protect\citeauthoryear{Levin and Narendra}{Levin and
  Narendra}{1996}]{levin_1996}
Levin, A. and K.~Narendra (1996).
\newblock Control of nonlinear dynamical systems using neural networks. {II}.
  {Observability}, identification, and control.
\newblock {\em IEEE Transactions on Neural Networks\/}~{\em 7\/}(1), 30--42.

\bibitem[\protect\citeauthoryear{Li, Chuang, and Spence}{Li
  et~al.}{2021}]{Li_2021b}
Li, B., W.-C. Chuang, and S.~M.~J. Spence (2021).
\newblock Response estimation of multi-degree-of-freedom nonlinear stochastic
  structural systems through metamodeling.
\newblock {\em Journal of Engineering Mechanics\/}~{\em 147\/}(11).

\bibitem[\protect\citeauthoryear{Mai}{Mai}{2016}]{mai_2016b}
Mai, C.~V. (2016).
\newblock {\em Polynomial chaos expansions for uncertain dynamical systems:
  Applications in earthquake engineering}.
\newblock Doctoral thesis, ETH Zurich.

\bibitem[\protect\citeauthoryear{Mai, Spiridonakos, Chatzi, and Sudret}{Mai
  et~al.}{2016}]{mai_2016}
Mai, C.-V., M.~D. Spiridonakos, E.~Chatzi, and B.~Sudret (2016).
\newblock Surrogate modeling for stochastic dynamical systems by combining
  nonlinear autoregressive with exogeneous input models and polynomial chaos
  expansions.
\newblock {\em International Journal for Uncertainty Quantification\/}~{\em
  6\/}(4), 313--339.

\bibitem[\protect\citeauthoryear{Mattson and Pandit}{Mattson and
  Pandit}{2006}]{mattson_2006}
Mattson, S.~G. and S.~M. Pandit (2006).
\newblock Statistical moments of autoregressive model residuals for damage
  localisation.
\newblock {\em Mechanical Systems and Signal Processing\/}~{\em 20\/}(3),
  627--645.

\bibitem[\protect\citeauthoryear{{NREL}}{{NREL}}{2021}]{openfast_2021}
{NREL} (2021).
\newblock Openfast documentation, release v2.5.0.
\newblock Technical report, National Renewable Energy Laboratory.

\bibitem[\protect\citeauthoryear{Pearson}{Pearson}{1895}]{pearson_1895}
Pearson, K. (1895).
\newblock Note on regression and inheritance in the case of two parents.
\newblock {\em Proceedings of the Royal Society of London\/}~{\em
  58\/}(347-352), 240--242.

\bibitem[\protect\citeauthoryear{Ramin, Amin, and Alireza}{Ramin
  et~al.}{2023}]{Poursorkhabi_2023}
Ramin, V.~P., H.~G. Amin, and N.~Alireza (2023).
\newblock An investigation of the performance of the {ANN} method for
  predicting the base shear and overturning moment time-series datasets of an
  offshore jacket structure.
\newblock {\em International Journal of Sustainable Construction Engineering
  and Technology\/}~{\em 14\/}(4), 79–93.

\bibitem[\protect\citeauthoryear{Rezaeian and Kiureghian}{Rezaeian and
  Kiureghian}{2010}]{rezaeian_2010}
Rezaeian, S. and A.~D. Kiureghian (2010).
\newblock Simulation of synthetic ground motions for specified earthquake and
  site characteristics.
\newblock {\em Earthquake Engineering \& Structural Dynamics\/}~{\em 39\/}(10),
  1155--1180.

\bibitem[\protect\citeauthoryear{Samsuri, Raman, and Tuan~Ya}{Samsuri
  et~al.}{2023}]{Samsuri_2023}
Samsuri, N.~A., S.~A. Raman, and T.~M. Y.~S. Tuan~Ya (2023).
\newblock Evaluation of {NARX} network performance on the maintenance
  application of rotating machines.
\newblock In {\em ICPER 2020}, pp.\  593--609. Springer Nature Singapore.

\bibitem[\protect\citeauthoryear{Schär, Marelli, and Sudret}{Schär
  et~al.}{2024}]{schaer_mnarx_2024}
Schär, S., S.~Marelli, and B.~Sudret (2024).
\newblock Emulating the dynamics of complex systems using autoregressive models
  on manifolds ({mNARX}).
\newblock {\em Mechanical Systems and Signal Processing\/}~{\em 208\/}(110956).

\bibitem[\protect\citeauthoryear{Schär, Marelli, and Sudret}{Schär
  et~al.}{2025}]{schaer_fnarx_2025}
Schär, S., S.~Marelli, and B.~Sudret (2025).
\newblock Surrogate modeling with functional nonlinear autoregressive models
  ({F-NARX}).
\newblock {\em Reliability Engineering \& System Safety\/}~{\em 264\/}(111276).

\bibitem[\protect\citeauthoryear{Spiridonakos and Chatzi}{Spiridonakos and
  Chatzi}{2015}]{spiridonakos_2015}
Spiridonakos, M.~D. and E.~N. Chatzi (2015).
\newblock Metamodeling of nonlinear structural systems with parametric
  uncertainty subject to stochastic dynamic excitation.
\newblock {\em Earthquakes and Structures\/}~{\em 8\/}(4), 915--934.

\bibitem[\protect\citeauthoryear{Tibshirani}{Tibshirani}{1996}]{Tibshirani_1996_LASSO}
Tibshirani, R. (1996).
\newblock Regression shrinkage and selection via the {Lasso}.
\newblock {\em Journal of the Royal Statistical Society. Series B
  (Methodological)\/}~{\em 58\/}(1), 267--288.

\bibitem[\protect\citeauthoryear{Wunsch, Liesch, and Broda}{Wunsch
  et~al.}{2018}]{Wunsch_2018}
Wunsch, A., T.~Liesch, and S.~Broda (2018).
\newblock Forecasting groundwater levels using nonlinear autoregressive
  networks with exogenous input ({NARX}).
\newblock {\em Journal of Hydrology\/}~{\em 567}, 743--758.

\bibitem[\protect\citeauthoryear{Yetkin and Kim}{Yetkin and
  Kim}{2019}]{Yetkin_2017}
Yetkin, M. and Y.~Kim (2019).
\newblock Time series prediction of mooring line top tension by the {NARX} and
  volterra model.
\newblock {\em Applied Ocean Research\/}~{\em 88}, 170--186.

\bibitem[\protect\citeauthoryear{Yu, Zhu, Luo, Luo, and Li}{Yu
  et~al.}{2023}]{Yu_2023}
Yu, C., Y.-P. Zhu, H.~Luo, Z.~Luo, and L.~Li (2023).
\newblock Design assessments of complex systems based on design oriented
  modelling and uncertainty analysis.
\newblock {\em Mechanical Systems and Signal Processing\/}~{\em 188}, 109988.

\bibitem[\protect\citeauthoryear{Zhang, Draycott, and Stansby}{Zhang
  et~al.}{2024}]{Zhang_2024}
Zhang, L., S.~Draycott, and P.~Stansby (2024).
\newblock System identification and generalisation of elastic mooring line
  forces on a multi-float wave energy converter platform in steep irregular
  waves.
\newblock {\em Mechanical Systems and Signal Processing\/}~{\em 214}, 111259.

\bibitem[\protect\citeauthoryear{Zhang, Dong, and Beer}{Zhang
  et~al.}{2024}]{Zhang2_2024}
Zhang, Y., Y.~Dong, and M.~Beer (2024).
\newblock {rLSTM-AE} for dimension reduction and its application to active
  learning-based dynamic reliability analysis.
\newblock {\em Mechanical Systems and Signal Processing\/}~{\em 215\/}(111426).

\bibitem[\protect\citeauthoryear{Zhao, Jiang, Vega, Todd, and Hu}{Zhao
  et~al.}{2023}]{zhao2023}
Zhao, Y., C.~Jiang, M.~A. Vega, M.~D. Todd, and Z.~Hu (2023).
\newblock Surrogate modeling of nonlinear dynamic systems: A comparative study.
\newblock {\em Journal of Computing and Information Science in
  Engineering\/}~{\em 23\/}(1), 011001.

\bibitem[\protect\citeauthoryear{Zhou and Li}{Zhou and Li}{2023}]{Zhou_2023}
Zhou, J. and J.~Li (2023).
\newblock An efficient time-variant reliability analysis strategy embedding the
  {NARX} neural network of response characteristics prediction into probability
  density evolution method.
\newblock {\em Mechanical Systems and Signal Processing\/}~{\em 200\/}(110516).

\end{thebibliography}

\end{document}